\documentclass[12pt]{article}
\usepackage{epsfig,rotating}
\usepackage{graphicx} 
\usepackage{bm}
\newcommand{\beq}{\begin{equation}}
\newcommand{\eeq}{\end{equation}}
\newcommand{\ba}{\begin{array}}
\newcommand{\ea}{\end{array}}
\newcommand{\lsim}   {\mathrel{\mathop{\kern 0pt \rlap
  {\raise.2ex\hbox{$<$}}}
  \lower.9ex\hbox{\kern-.190em $\sim$}}}
\newcommand{\gsim}   {\mathrel{\mathop{\kern 0pt \rlap
  {\raise.2ex\hbox{$>$}}}
\lower.9ex\hbox{\kern-.190em $\sim$}}}

\begin{document}
\title{{\bf Analytic calculations of the spectra of ultra-high energy cosmic ray nuclei.  \\
I. The case of CMB radiation.}}
     
\author{R. Aloisio$^{1,2}$, V. Berezinsky$^{2}$ and S. Grigorieva$^{3}$\\
        {\it\small $^1$INAF, Osservatorio Astrofisico di Arcetri,
                   I--50125 Arcetri (FI), Italy} \\
        {\it\small $^2$INFN, Laboratori Nazionali del Gran Sasso,
                   I--67010 Assergi (AQ), Italy} \\
        {\it\small $^3$Institute for Nuclear Research, 60th October Revolution
	           Prospect 7A, 117312 Moscow, Russia} }

\maketitle

\abstract{
We present a systematic study of different methods for the analytic 
calculation of ultra-high energy nuclei diffuse spectra. Nuclei 
propagating in the intergalactic space are photo-disintegrated and 
decrease their Lorentz factor due to the interaction with cosmic 
microwave background and extragalactic background light. 
We calculate the evolution trajectories in the backward time, that describe how 
atomic mass number $A$ and Lorentz factor $\Gamma$ change with 
redshift $z$. Three methods of spectra calculations are investigated 
and compared: {\it (i)} trajectory method, {\it(ii)} kinetic equation 
combined with trajectory calculations and {\it (iii)} coupled kinetic 
equations. We believe that these three methods exhaust at least the
principal possibilities for any analytic solution of the problem. In 
the most straightforward method {\it(i)} only  trajectory calculations 
are used to connect the observed nuclei flux with the production rate 
of primary (accelerated) nuclei $A_0$. In the second method {\it (ii)} 
the flux (space density) of primary nuclei, and secondary nuclei and 
protons are calculated with the help of kinetic equation and trajectories 
are used only to determine the generation rates of these nuclei. The  
third method {\it (iii)} consists in solving the complete set of coupled  
kinetic equations, written starting with primary nuclei $A_0$, then for 
$A_0-1$ etc down to the $A$ of interest. The solution of the preceding 
equation gives the generation rate for the one which follows. An important 
element of the calculations for all methods is the systematic use of Lorentz 
factor instead of energy. We consider here the interaction of nuclei only 
with the cosmic microwave background, this case is particularly suitable for 
understanding the physical results. In paper (II) of this series the extragalactic 
background light will be also included. Estimating the uncertainties of all 
methods discussed above, we conclude that the method of coupled kinetic 
equations gives the most reliable results. 
}

\section{Introduction}
\label{introduction}

The observation of particles with ultra high energies
($E>10^{18}$ eV) has a  fundamental importance in high energy 
astrophysics and, maybe, more generally in physics. These particles,
hereafter referred to as Ultra High Energy Cosmic Rays (UHECR), 
attract much attention in the recent years as the most energetic 
particles ever observed and because they are
messengers from cosmic accelerators to extreme energies. 

The observational data on  UHECR can be divided into three 
categories: spectra, mass composition and correlation with
astrophysical sources.

Several theoretically well established features are predicted for the  
energy spectra of UHE {\em protons} interacting with the Cosmic Microwave
Background (CMB) radiation. Most noticeably these are: (i) the 
Greisen-Zatsepin-Kuzmin (GZK) feature \cite{GZK}, a sharp steepening
of the spectrum at $E\simeq 5\times 10^{19}$ eV due to photo-pion
production \cite{GZK}; (ii) a rather faint feature, the dip, 
at energies $1\times 10^{18} - 4\times 10^{19}$~eV, caused by the 
$e^+e^-$ pair production \cite{BG88}, and (iii) an even more 
faint and sharp peak at energy $6.3 \times 10^{19}$~eV produced 
by an interference effect in the proton interaction with CMB \cite{BGK}. 

Two of these features, the GZK cutoff and the dip, most probably 
are already detected. The observation of the GZK feature 
in the differential spectrum of HiRes \cite{HiResGZK} is strengthened  
by the measured value of $E_{1/2}$ in the integral spectrum, which 
precisely coincides with the theoretical prediction. 
The recent observations of the Telescope Array detector also confirm the presence
of a spectrum steepening consistent with the GZK cut-off \cite{TAGZK}
The Auger data
\cite{WatsonICRC,Kampert08,AugerFlux,Auger2010} also show a sharp steepening
of the spectrum, which however, according to our calculations, 
does not fit well the theoretical shape of the GZK feature, especially in the data 
of 2010 \cite{Auger2010} and 2011 \cite{Auger2011}.

The dip is very well confirmed by the data of Hires, AGASA and Yakutsk
detectors \cite{BGGpl,BGGprd,Aloisioetal}, 
and by the Auger data  of 2007 \cite{VB08} though with a larger $\chi^2$.
The agreement becomes even worse taking the Auger data released in 2010
\cite{Auger2010} and the latest release of 2011 \cite{Auger2011}. 

One may notice that both features discussed above are signatures of a 
pure proton composition. Therefore, agreement of the data 
with the dip and GZK cutoff, most noticeably in HiRes data, can be considered
as an indirect evidence for a proton-dominated composition.

The direct observation of UHECR mass composition is a very
difficult task (for a review see \cite{NW} and \cite{Watson04}). 
At present, observations are contradictory. 
While HiRes \cite{HiRes}, Telescope Array \cite{TA}, HiRes-MIA \cite{HiRes-MIA} 
and Yakutsk \cite{Yak} detectors favor a proton-dominated flux at energies
$E>10^{18}$ eV, Fly's Eye \cite{FlysEye}, Haverah
Park \cite{HavPark}, and, recently, 
Auger \cite{Auger-mass2010} indicate a mixed composition with 
a substantial fraction of heavy nuclei, in particular at the highest energies.

Nowadays the most serious conflict exists between Auger \cite{Auger-mass2010} 
and HiRes \cite{HiRes-mass2010} data. Both 
detectors measure the mass composition through the fluorescence emission
produced in the atmosphere by UHECR interactions. In particular the quantity 
that characterizes the mass composition is the position of the maximum 
of the fluorescence emission, $X_{max}$, in the atmosphere 
and its Root Mean Square (RMS) \cite{ABBO}. While HiRes data evidence a 
pure proton composition till the highest energies, the Auger data show a progressively 
heavier composition with increasing energy that approaches an almost pure Iron
composition at energy $\sim 3.5\times 10^{19}$~eV.   
 
The third important observable in UHECR is the possible  
correlation with astrophysical sources. The most energetic UHECR
particles can show the direction to nearby sources. This
expectation depends  
on the electrical charge of the particles due to the deflection in
surrounding magnetic fields, in particular in the galactic magnetic
field. In this case for protons with $E > 50$~ EeV in a microGauss 
field on homogeneous scale of Kpc order the deflection 
angle is around few degrees. With an angular resolution of the
same order it is possible to resolve UHECR sources on the Mpc scale. 
For larger electric charge at $E > 50$ EeV, the deflection angle 
increases up to 10 degrees in the case of Helium and 50 degrees for Iron. 
Therefore, in the case of nuclei it is difficult to  
observe any correlation with sources. However,  
there have been proposed methods to disentangle the effect of the galactic 
magnetic field on the observed flux, pursuing a correlation study 
with UHE nuclei \cite{Giacinti}.  

The observational evidences concerning correlations are  
controversial. The Auger collaboration performed a search for correlation 
of their events with AGN from the 12th edition of the VCV catalog 
\cite{VCVcat}, using the data 
collected between January 1 2004 and August 31 2007 the collaboration 
found that 20 out of 27 events with energy larger than 
$(5 - 6)\times 10^{19}$~eV correlated with at least one 
of the selected AGN inside an angle of $3.2^\circ$. 
The updated analysis of such correlation 
has been  performed by the Auger collaboration using the UHECR events
collected up to March 31 
2009 with the same selection criteria as before. This updated analysis
has not strengthened \cite{AugerCorrelation} the signal of correlation with
sources as should be expected from an increased statistics.

The Auger results on correlations, coupled with chemical composition, 
produce an experimental picture with both heavy nuclei and correlation with 
nearby astrophysical sources at the highest energies, while such correlation is 
possible only in the case of a large fraction of protons \cite{AloisioBoncioli}. 

On the other hand, the HiRes collaboration performed a similar
analysis for correlations between stereo UHECR events and AGN  from
the VCV catalog with no significant correlation found \cite{HiResCorrelation}. 

From a theoretical point of view, there are in literature three models with different 
predictions for the mass composition. The {\it dip model} 
predicts (almost) a pure proton composition starting from energies 
$E\ge 10^{18}$ eV \cite{ABBO}; the {\it mixed composition model} \cite{Allard06} 
favours a mixed composition at the lowest energies with still a proton 
dominated spectrum at the highest energies and finally the 
{\it disappointing model}  
explains the observed spectrum in terms of a pure proton composition 
at $E \sim (1 - 3)\times 10^{18}$~eV with steadily heavier mass composition  
at increasing energy \cite{disappointing}.  

As shown in \cite{Kampert08}, the dip and mixed composition models 
in their present form contradict the Auger data on mass composition. 
However, the mixed composition model has more power in the explanation of the
Auger mass composition: due to presence of several free parameters, 
the predictions can be, in principle, adjusted to explain a wide 
range of different mass compositions. In contrast the {\em disappointing
model} gives the best description of the Auger spectrum and mass composition 
\cite{disappointing}. On the other hand, the correlations observed by Auger 
can be accommodated only in the framework of a
proton-dominated  composition at the highest energies \cite{AloisioBoncioli}, 
i.e. only in the case of mixed or dip models \cite{AloisioBoncioli}. 

For recent reviews on UHECR observations and related models see
\cite{Stanev,Olinto,Felix}.

The discussion above demonstrates the importance of a theoretical study 
of nuclei as carriers of the UHE signal. This study has important consequences
not only on the interpretation of the observations on mass composition, but 
also on the spectrum and correlations. The theoretical interest on UHE
nuclei is also supported by the difficulties in reaching the highest energies 
by the acceleration of protons in astrophysical models. 
In the case of nuclei this problem is ameliorated because the
maximum acceleration energy increases by a factor $Z$, the charge 
number of the nucleus, which is up to $26$ in the case of Iron. 

Historically, the interest to UHE nuclei started from the hope  
to solve the UHECR puzzle, i.e.  the absence of the 
GZK cutoff in observations, with the help of nuclei as signal carrier. In 
this case the energy of a CMB photon in the rest system 
of a nucleus with atomic mass number $A$ is $A$ times lower 
than for a proton of the same energy, thus photo-pion production 
is suppressed.

The first calculations of the UHE nuclei photo-disintegration by the CMB
radiation have been performed in the works by Stecker \cite{Stecker69} and 
by Puget and Stecker \cite{PS75}, where infrared  radiation 
was also included. Berezinsky and Zatsepin \cite{BZ71} in 1971 have 
calculated the energy losses of nuclei with different A due to 
photo-disintegration and pair-production. It has been demonstrated 
that  steepening in the spectra of  heavy nuclei such as 
Fe, C etc occurs at a Lorentz--factor determined by the equality of 
adiabatic and pair-production energy losses, i.e. at energies lower 
than the GZK cutoff for protons.

The first calculations of the diffuse nuclei spectrum have 
been made by Berezinsky, Grigorieva, Zatsepin \cite{BGZ75} and 
Hillas \cite{Hillas75}, both works have been presented at the same
conference and published in the same volume of the proceedings. 

Propagating in intergalactic space, UHE nuclei interact with the CMB and
Extragalactic Background Light (EBL), i.e. Infrared, Visible and Ultra-violet 
radiation, experiencing two main processes: (i) photo-disintegration 
and (ii) $e^+e^-$ pair production, that reduce the nucleus kinetic energy.
The systematic study of the photo-disintegration started from  
the pioneering works by Stecker \cite{Stecker69,PS75,Stecker76,
Stecker99}. These works are based on a very convenient parameterization of 
the photo-nuclear cross-sections, widely used in the study of UHE 
nuclei propagation. Recently, a refined parameterization of this 
cross-section was presented in \cite{Allard05}.

Using the parameterization of the photo-nuclear cross section by 
Stecker \cite{Stecker76}, updated in 1999 \cite{Stecker99}, or
using other approximations as in \cite{Allard05}, many papers have 
been  published in the last decade, all (except \cite{Gelmini07,Kalashev,Sarkar,Ahlers}) 
based on a Monte Carlo (MC) approach to the study of UHE nuclei propagation 
in astrophysical backgrounds \cite{Allard06,Stecker99,Taylor,Elbert95,
Epele98,Bertone02,Yamamoto04,Ave05,Armengaud05,Sigl05,
Harari06,Anchor07,Allard08,SimProp}. 

The MC computation schemes, typically implemented in these studies, are 
all based on a statistical treatment only of the process of photo-disintegration, 
in the case of nuclei, or photo-pion production, in the case of protons, 
that takes into account the fluctuations associated to the interaction process. 
The other channel of interaction with astrophysical backgrounds, the pair production 
process, is usually treated, for both nuclei and nucleons, in the approximation 
of continuous energy losses, that neglects fluctuations in the interaction assuming 
that particles loose energy continuously (see later). In the case of photo-pion production 
for protons this approximation produces a disagreement with the MC results only at the 
highest energies and at the level of few percent \cite{MC-Comparison}. 

The attempts to solve analytically the problem of UHE nuclei propagation
are all based on a kinetic-equation approach with the hypothesis of continuous 
energy losses \cite{Gelmini07,Kalashev,Sarkar,Ahlers}. This approach is very well suited to 
determine the flux produced by any kind of sources distribution. 
The approach of \cite{Gelmini07,Kalashev} is based on a numerical solution of the kinetic 
(transport) equations for primaries and all secondaries. The results of \cite{Gelmini07,Kalashev} 
are particularly interesting for us because represent a numerical test of one of the analytic 
computation schemes presented here. We will come back to this comparison at the end of the 
present paper and, more accurately, in the accompanying paper II. 
The same kinetic-equation approach is used in the computations presented in \cite{Sarkar}. 
The authors in this case propose a perturbative solution of the transport equation 
that, at the leading order, takes into account only the process of one nucleon 
emission in the photo-disintegration process. Finally, in \cite{Ahlers} the authors 
present the complete set of coupled kinetic equations describing the 
propagation of primary and secondary nuclei; however, their approach 
is focused on the computation of the flux of UHE neutrinos produced by the 
propagation of CR and doesn't address an explicit computation of the UHE nuclei 
fluxes. 

We present here three new methods for the analytic 
calculation of UHE nuclei spectra: the evolution-trajectory method,
kinetic equation method combined with trajectory calculations and 
the method of coupled kinetic equations (CKE). In this paper (paper I) we 
include only CMB as the radiation with which UHE nuclei interact. 
The main emphasis of this paper is given to the theoretical issues of 
these methods and comparing them with each other. It is more plausible
to do it using the  CMB alone, because 
the Planck spectrum and the exact knowledge of the cosmological
evolution of this radiation make formulae simple and the results 
more transparent. The phenomenological predictions 
obtained with CMB only are always relevant for the highest energy 
part of the spectra.

In the accompanying paper II we will include also the interaction  
with EBL radiations, and focus more on the observational
applications of our new computation scheme.

The paper is organized as follows: in section \ref{sec:propagation} 
we calculate the energy losses of nuclei on the CMB radiation, 
and  compute the evolution trajectories. In section 
\ref{sec:comb} the combined method is presented, where space 
particle density is obtained from kinetic equation, and trajectories 
are used only for calculations of the generation rates. 
In section \ref{sec:coupled-kinetic} the method of coupled kinetic 
equations is developed. In section \ref{comparison} the fluxes
obtained with the CKE method are compared with available results in literature.
Finally, conclusions are presented in section 
\ref{sec:conclusions}. In the appendixes we give 
a detailed calculation and discussion of the following technical 
problems: generation rates calculated  from number of particles
conservation, in  appendix \ref{app:generation}; the ratio of energy 
intervals at different redshifts, in  appendix \ref{app:dgamma};    
 the analytic solutions of kinetic equations, in appendix 
\ref{app:solution}, and the comparison of the secondary nuclei and
proton fluxes, in appendix \ref{app:comparison}. 

\section{Energy losses and evolution trajectories}
\label{sec:propagation}
Propagating through background radiations, mainly CMB and EBL, 
nuclei decrease their Lorentz factor $\Gamma$, due to 
$e^+e^-$ pair production, and atomic number $A$, due to 
photo-disintegration. Lorentz factor decreases also due to 
the expansion of the universe (adiabatic energy losses). In this section
we study the energy losses and the evolution of nuclei due to their 
propagation, considering the evolution trajectories in time (or redshift $z$) 
along which $A$ and $\Gamma$ are changing. We will consider these trajectories 
in the  backward time, starting from the point of observation 
($\Gamma,A,z_0=0$) and increasing $z$, so that $\Gamma$ and $A$ increase
too. 

An important ingredient of our method consists in the use of the 
Lorentz factor of the particles instead of their energy. This approach 
gives many simplifications in the theoretical study, in particular the
approximate equality of Lorentz factors of all three particles participating
in photo-disintegration process $A \to (A-1) + N$, where $N$ is a nucleon.

\subsection{Energy losses on CMB}
\label{sec:losses}
In this paper we consider the interaction of nuclei only with CMB. 
Adiabatic energy losses and pair production change only the Lorentz factor
and photo-disintegration changes only the atomic mass number $A$. Then for 
$E=\Gamma A m_N$ we have  

\begin{equation}
\frac{1}{E} \frac{dE}{dt}=\frac{1}{\Gamma}\frac{d\Gamma}{dt} +
\frac{1}{A}\frac{dA}{dt}.
\label{eq:lossE}
\end{equation}

For any process the energy losses in continuous approximation for 
nuclei or protons interacting with CMB 
at $z=0$ can be written as \cite{book}: 
\begin{equation}
\beta_0(\Gamma)=-\frac{1}{E}\frac{dE}{dt}=\frac{T}{2\pi^2\Gamma^2}
\int_{\epsilon_0}^{\infty}d\epsilon 
\sigma(\epsilon)f(\epsilon)\epsilon
\left [-\ln\left (1-\exp[- \frac{\epsilon}{2\Gamma T}]\right)\right ]~,
\label{eq:lossE-CMB}
\end{equation}
where $\epsilon$ is the photon energy in the nucleus or proton rest frame, 
$\epsilon_0$ is the threshold of the considered reaction,   
$f(\epsilon)$ is the mean fraction of energy in the
laboratory system lost by a nucleus in a single interaction, 
i.e. inelasticity, $\sigma(\epsilon)$ is the cross-section 
and $T=2.726^{\circ}$K is the CMB temperature;
units $\hbar=c=k=1$ are used.  

At redshift $z$ the number of CMB photons is  $(1+z)^3$ larger than at
$z=0$ and their energies are $(1+z)$ times higher. Then 
the energy loss at arbitrary $z$ is given by
\beq
\beta(\Gamma,z)= (1+z)^3\beta_0[(1+z)\Gamma]
\label{eq:lossE-z}
\eeq
{\it In the following we will specify} the three processes relevant for our 
calculations: expansion of the universe (adiabatic energy losses), pair 
production and photo-disintegration.
\vskip 0.2cm
\noindent
{\em (i) Adiabatic energy losses}.\\
Adiabatic energy losses are given by 
\beq
\beta_{\rm ad}(z)=-\frac{1}{\Gamma}\frac{d\Gamma}{dt}=H(z),
\label{eq:ad}
\eeq
where $H(z)=H_0\sqrt{(1+z)^3\Omega_m+\Omega_\Lambda}$ is the Hubble
parameter at redshift $z$ with  $H_0=72$~km/sMpc,  $\Omega_m=0.238$
and $\Omega_{\Lambda}=0.716$ according to WMAP data \cite{WMAP}.
\vskip 0.2cm 
\noindent
{\em (ii) Electron-positron pair production }\\
This process can occur if the energy of the background photon is
larger than $1$~MeV in the rest system of the UHE particle. 
The inelasticity and cross-section for a nucleus are simply related
to the corresponding quantities for the proton pair-production:
\begin{equation}
f^A_{\rm pair}(\epsilon)=\frac{1}{A}f^p_{\rm pair}(\epsilon),\;\;  
\sigma^A_{\rm pair}(\epsilon)= Z^2 \sigma_{\rm pair}^{p} (\epsilon),
\;\;\beta_{\rm pair}^A(\Gamma)=\frac{Z^2}{A}\beta_{\rm pair}^p(\Gamma),
\end{equation}
where $Z$ is the nucleus charge number, $m_N$ is the nucleon mass,
$\sigma_{\rm pair}^{i}$ is the pair production cross section for protons
$(i=p)$ or nuclei $(i=A)$.
Using these equations we can rewrite Eq. (\ref{eq:lossE-CMB}) as
\begin{equation}
\beta_{\rm pair}^A(\Gamma)=-\frac{1}{\Gamma}\frac{d\Gamma}{dt}=
\frac{Z^2}{A}\frac{T}{2\pi^2\Gamma^2}
\int_{\epsilon_0}^{\infty}d\epsilon\; 
\sigma_{\rm pair}^{p}(\epsilon)f^p_{\rm pair}(\epsilon)\; 
\epsilon
\left [-\ln\left (1-\exp[-\frac{\epsilon}{2\Gamma T}]\right)\right ]~.
\label{eq:lospairs}
\end{equation}
The expression above is valid for $z=0$. For arbitrary $z$ one should
use Eq.~(\ref{eq:lossE-z}).
In our calculations we use the function $\beta_{\rm pair}^p(\Gamma)$ for
protons as computed in \cite{BGGprd}. The quantity 
$\tau_{\Gamma}=\beta_{\rm pair}^{-1}$ has an important physical 
meaning being the characteristic time of the Lorentz factor 
decreasing.\\  
\vskip 0.2cm
\noindent
{\em (ii) Photo-disintegration}.\\
The photo-disintegration energy losses on CMB at $z=0$ are given by   
\begin{equation}
\beta_{\rm dis}^A(\Gamma)=-\frac{1}{A}\frac{dA}{dt}=
\frac{T}{2\pi^2\Gamma^2}\frac{1}{A}
\int_{\epsilon_0(A)}^{\infty}d\epsilon 
\sigma_{\rm dis}(\epsilon,A)\nu(\epsilon)\epsilon
\left [-\ln\left (1-\exp[-\frac{\epsilon}{2\Gamma T}]\right)\right ]~,
\label{eq:losdis}
\end{equation}
where $\nu(\epsilon)$ is the average multiplicity of the ejected
nucleons.

The quantity $\beta^A_{\rm dis}$ determines the time-scale of the total 
photo-disintegration of a nucleus $\tau_{\rm tot} \sim 1/\beta_{\rm dis}^A$, 
while $\tau_A = (dA/dt)^{-1}$ gives the mean time for one nucleon
loss, or the time between two collisions under an assumption of 
one-nucleon photo-disintegration $\nu=1$.

At arbitrary $z$  the pair-production energy losses 
$\beta_{\rm pair}(\Gamma,z)$ and photo-disintegration energy 
losses  $\beta_{\rm dis}^A(\Gamma,z)$  
are given by Eq. (\ref{eq:lossE-z}).

In the present paper we use the remarkable collection of 
nuclear cross-sections and their parametrization from the works 
by Stecker et al \cite{Stecker76,Stecker99}. 

Depending on the photon energy $\epsilon$ in the
nucleus rest frame, one can distinguish in the general case 
two different regimes of
photo-disintegration, namely  the low-energy regime $\epsilon<30$~MeV 
and the high-energy regime: $30<\epsilon<150$~MeV. At energies 
$\epsilon>150$~MeV the photo-disintegration process is not
important \cite{Stecker76,Stecker99}  and we include 
at these energies the photo-pion production. 

In the low-energy regime the leading process of photo-disintegration
is represented by one and two nucleon emission,
$A+\gamma\to (A-1)+N~$, $A+\gamma\to (A-2)+2N~,$
and the photo-disintegration cross-section is dominated by the Giant
Dipole Resonance (GDR) \cite{Stecker76,Stecker99}. At higher 
energies a multi-nucleon emission regime takes place 
(see \cite{Stecker76,Stecker99}).
In this regime the photo-disintegration cross-section can be approximated 
as constant \cite{Stecker76,Stecker99}. 
The main contribution to photo-disintegration is given by the GDR 
cross-section with the number of emitted nucleons $\nu =1$ and $\nu=2$.

In the case of CMB radiation considered in the present paper we may limit ourselves 
by one-nucleon emission only. The typical threshold for two-nucleon 
emission is $\epsilon_{\rm th} \sim 20$~MeV to be compared with one-nucleon
emission threshold $\sim 10$~MeV. For the Lorentz-factor values of 
interest $\Gamma \le 10^{10}$ the photo-disintegration process takes place at 
the high-energy tail of the Planckian CMB spectrum and two-nucleon production 
is suppressed by the number of active photons. In fact two-nucleon emission 
processes are strongly suppressed (by one order of magnitude) at larger 
Lorentz-factors too by the smallness of the corresponding cross-section 
(see Fig.~2 from \cite{Stecker99}). At extremely large Lorentz-factors the 
evolution of nuclei enters an explosive regime, described below, for which the
difference between one-nucleon and many-nucleons regimes looses any significance.
Further on we assume in all formulae below the nucleon multiplicity 
$\nu=1$, though we keep this quantity in the formulae.

\begin{center}
\begin{figure}[!ht]
\includegraphics[width=0.9\textwidth,angle=0]{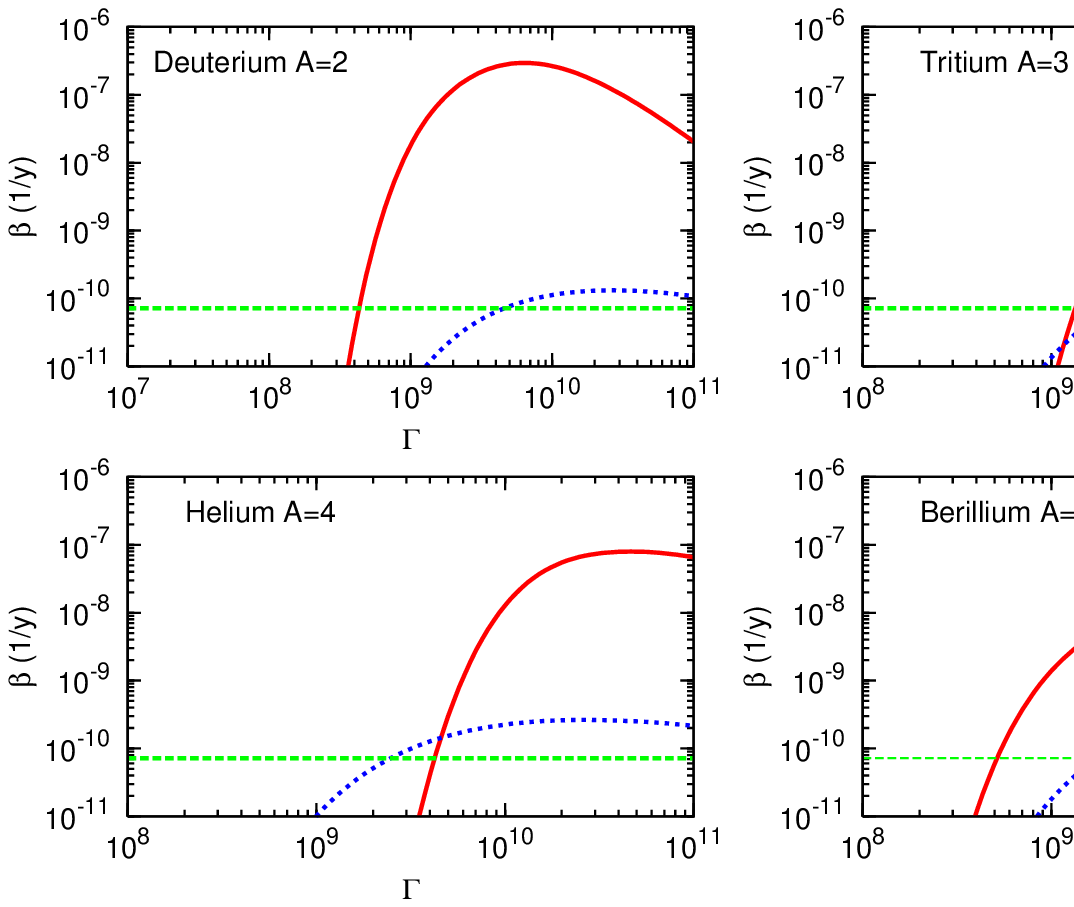}
\caption{ Energy losses for light nuclei due to photo-disintegration 
and pair production on CMB (red full line
and blue dotted line, respectively) and adiabatic energy losses  
given by $H_0$ (green dashed line).}
\label{fig:LosseLow}
\vspace{1.0 cm}
\includegraphics[width=0.9\textwidth,angle=0]{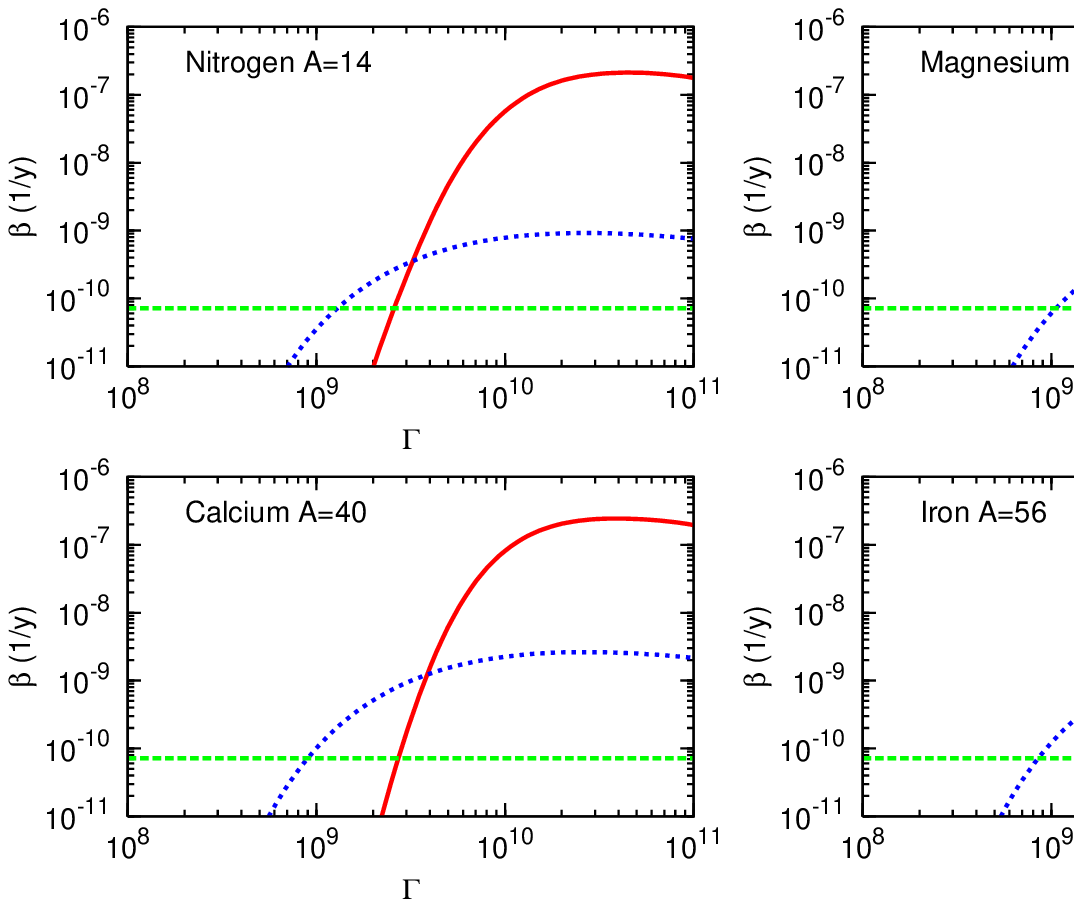}
\caption{The same as in Fig. \ref{fig:LosseLow} for heavy nuclei.}
\label{fig:LosseHigh}
\end{figure}
\end{center}

We have calculated the energy losses for all nuclei with existing data,
presenting some of them in Figs. \ref{fig:LosseLow} and 
\ref{fig:LosseHigh} as function of the Lorentz factor.

As discussed below, an important quantity that 
characterizes nuclei propagation and energy spectra is the 
{\it critical Lorentz factor} $\Gamma_c$, defined from the condition 
of equality between photo-disintegration lifetime $\tau_A = (dA/dt)^{-1}$ and   
the characteristic time $\tau_\Gamma$ for Lorentz factor changing, 
\beq
\tau_A (\Gamma_c) = \tau_{\Gamma}(\Gamma_c),
\label{eq:Gamma_c}
\eeq
where $\tau_{\Gamma}$ is given by the sum of adiabatic and 
pair-production losses as $\tau_{\Gamma}^{-1}=\tau^{-1}_{\rm ad}
+ \tau^{-1}_{\rm pair}$.

Eq. (\ref{eq:Gamma_c}) is written for $z=0$. The same condition 
for an epoch $z$ determines $\Gamma_c$ at redshift $z$, for which 
we will use the notation $\Gamma_c(z)$.

\begin{table}[!ht]
\vspace{-0.3cm}
\begin{center}
\begin{tabular}{|c|c|c|c|c|c|c|c|c|}
\hline\hline
Nucleus & $D$  & $^3He$ & $^4He$ & $^9Be$ & $^{14}N$ & $^{24}Mg$ 
& $^{40}Ca$ & $^{56}Fe$ \\
\hline
$\Gamma_c$ & $4\times 10^8$ & $1.2\times 10^9$ & $4\times 10^9$ &
$3.8\times 10^8$ & $2.2 \times 10^9$ & $2.1\times 10^9$ & $2.3\times 10^{9}$ 
& $1.9\times 10^9$ \\
\hline\hline
\end{tabular}
\caption{Values of $\Gamma_c$ for some selected nuclei}
\label{tab:Gamma_c}
\end{center}
\end{table}
The values of $\Gamma_c$ calculated for some nuclei from Eq. \ref{eq:Gamma_c} 
are listed in Table~\ref{tab:Gamma_c}.  From this Table and Fig. \ref{fig:LosseHigh}
one may notice that $\Gamma_c$ is almost the same for all heavy nuclei with a value 
around $(2 - 2.5)\times 10^{9}$. The values of $\Gamma_c$
calculated from the condition $\beta_A (\Gamma_c) = \beta_{\Gamma}(\Gamma_c)$
(see Figs. \ref{fig:LosseLow} and \ref{fig:LosseHigh}) differ
but little from the values of Table~\ref{tab:Gamma_c}. 
 
\subsection{Evolution trajectories}
\label{sec:trajectories}
A primary nucleus $A_0$ accelerated to large Lorentz factor $\Gamma_g$ 
at redshift $z_g$ is soon photo-disintegrated and then evolves 
as a secondary nucleus with decreasing  $A$ and $\Gamma$ as redshift 
$z$ decreases. 

Our study of the evolution goes in the backward time, whose role is played by 
redshift. We consider as initial 
state a secondary nucleus $A$ with Lorentz factor $\Gamma$ and evolve 
it to larger $z$ with increasing of $A(z)$ and $\Gamma(z)$,    
until $A(z)$ reaches $A_0$ at $z_g$ (index $g$ here and henceforth implies 
{\it generation}, i.e. acceleration, of nucleus $A_0$). Evolution of $\Gamma$ 
from $z_0=0$ to $z_g$ gives $\Gamma_g$ of nucleus $A_0$. This 
backward-time evolution is governed by two coupled  differential
equations, which for the variable $z$ reads
\begin{equation}
\frac{1}{A}\frac{dA}{dz}=\left | \frac{dt}{dz} \right |
\beta_{\rm dis}(\Gamma,A,z) \qquad , \qquad
\frac{1}{\Gamma}\frac{d\Gamma}{dz}=\left | \frac{dt}{dz}\right |
\beta_{\rm pair}(\Gamma,A,z)+\frac{1}{1+z},
\label{eq:evolve}
\end{equation}
being the relation between time and redshift $dt/dz$ 
\begin{equation}
\frac{dt}{dz}=-\frac{1}{(1+z) H(z)},\;\;\;\;
H(z) = H_0 \sqrt{(1+z)^3\Omega_m + \Omega_\Lambda}
\label{eq:dtdz}
\end{equation}
The term $1/(1+z)$ in Eq. (\ref{eq:evolve}) corresponds 
to adiabatic energy losses $d\Gamma/dt=-\Gamma H(z)$.

The numerical solution of the coupled equations 
(\ref{eq:evolve}) with initial values $A$ and $\Gamma$ at $z_0$ gives
the {\em evolution trajectories} 
$A(z)=\mathcal{A}(A,\Gamma,z_0,z)$ and 
$\Gamma (z)=\mathcal{G}(A,\Gamma,z_0,z)$, 
where the first three arguments describe the initial condition, in 
most cases with $z_0=0$.
\begin{figure}[!ht]
\begin{center}
\includegraphics[width=0.6\textwidth]{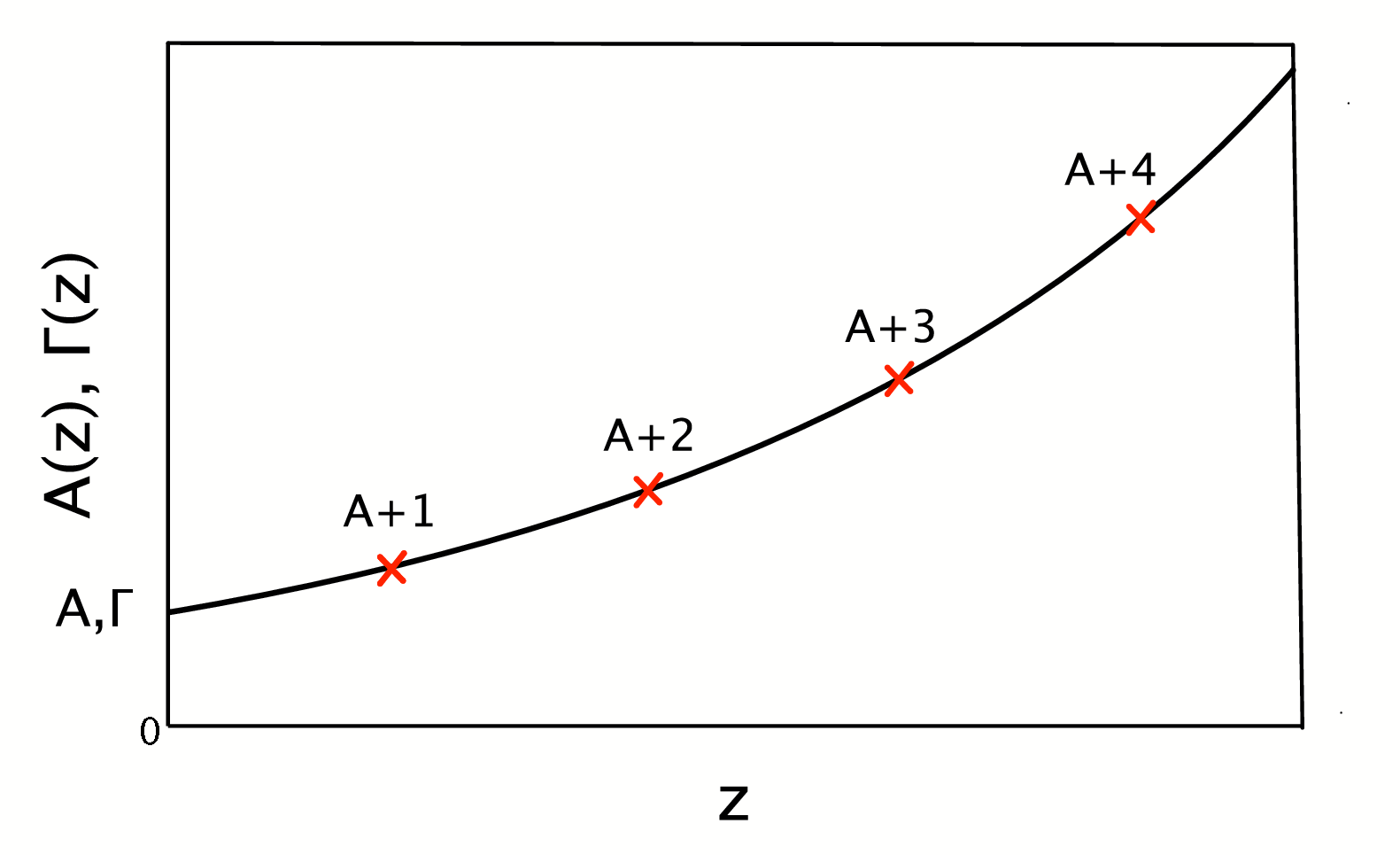}
\caption{A sketch of $\Gamma(z)$ and $A(z)$ backward-time evolution with 
$\Gamma(z)$ evolution shown by continuous curve and integer values of 
$A$ shown by crosses. A(z) is calculated as continuous quantity only
for determination of $z$ where integer values of A(z) are reached.   
The transition e.g. from $A$ to $A+1$  nuclei (or $A+2$), is assumed to 
occur instantaneously and marked by crosses. A nucleus between 
two crosses is considered as one with fixed $A$. Evolution of 
$\Gamma$ between two crosses is calculated for fixed $A$.
}
\label{fig:sketch}
\end{center}
\end{figure} 

\begin{figure}[!ht]
\begin{center}
\includegraphics[width=0.49\textwidth]{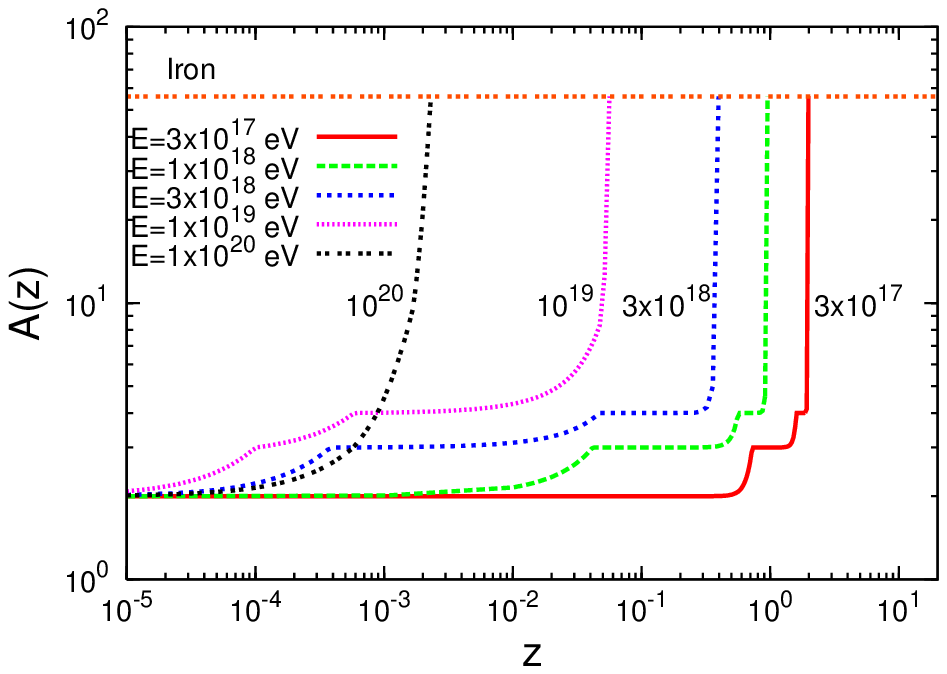}
\includegraphics[width=0.49\textwidth]{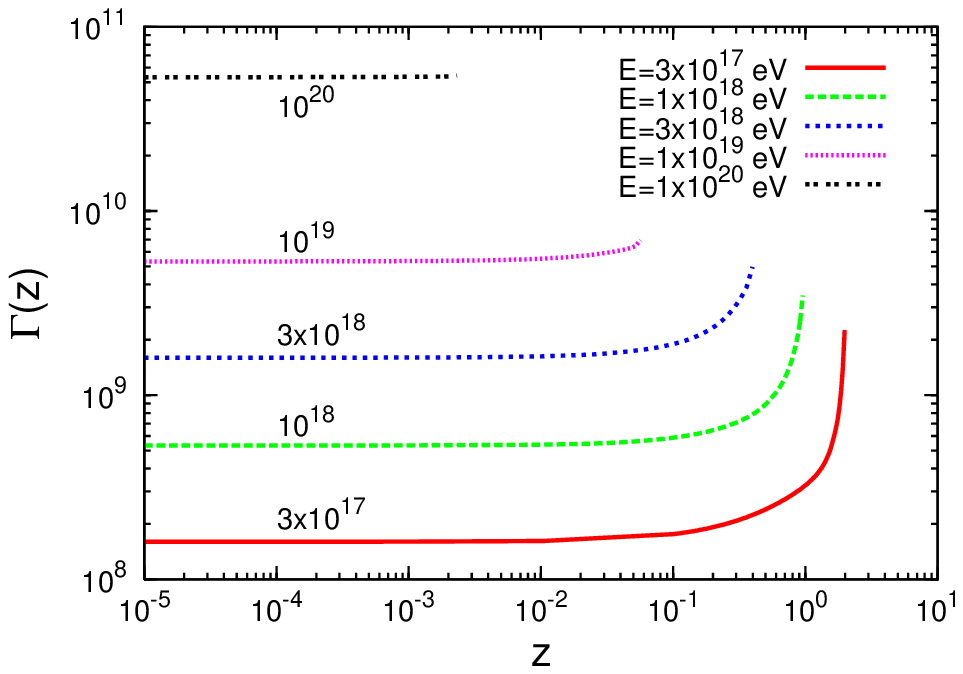}
\caption{Calculated evolution of the atomic mass number $A(z)$ 
(left panel) and Lorentz factor $\Gamma(z)$ (right panel) for 
a deuterium nucleus observed at $z=0$ with various energies as labeled.}
\label{fig:evolve}
\end{center}
\end{figure} 
In principle the solution $A(z)$ is given for continuous $A$.
We shall refer to the corresponding trajectories as 
{\em A-continuous}. In this solution we use $\beta^A_{\rm dis}$ 
smoothly changing between two integer values of $A$. 

In the kinetic-equation approach, which is used for calculation 
of particle density  $n_A(\Gamma,z)$, we assume that $A$ is constant 
until it reaches due to A-evolution $A+1$ (or $A+2$). The trajectories 
with this evolution of $A$ we call {\em A-jump trajectories}. They are
calculated assuming in Eq. (\ref{eq:evolve}) for $\Gamma$-evolution 
$A=const$, until $A$ reaches $A+1$ (or $A+2$).    
Accordingly, we use in this case the jump behaviour of 
$\beta^A_{\rm dis}$.

In fact, both methods give practically identical results.  

A calculated  trajectory is sketched in Fig. \ref{fig:sketch}  
as $\Gamma(z)=\mathcal{G}(\Gamma,A,z)$ with integer $A$ marked by crosses. 
In the kinetic equation approach we will interpret this trajectory 
assuming that the nucleus $A$ is produced in an instantaneous decay of the 
nucleus $A+1$, then it lives as nucleus $A$ and finally
instantaneously decays to $A-1$. The
intervals between crosses are determined by continuously changing $A(z)$
setting $\Delta A=1$  (or $\Delta A=2$). This is the most natural 
interpretation, because a nucleus with continuously changing $A$ just does 
not exist in nature, while instantaneous production of the nucleus $A$ from the 
decay of $A+1$ is a realistic photo-disintegration assumption, as well as an 
assumption that after decay of $A+1$ in the forward-running time a nucleus 
$A$ propagates with fixed $A$ until the instantaneous $A \to (A-1)$ decay (we always 
imply below the case of $A+2$ too). 

In Fig. \ref{fig:evolve} the calculated evolution trajectories 
$\mathcal{A}(A,\Gamma,z_0,z)$ and $\mathcal{G}(A,\Gamma,z_0,z)$ are 
displayed for the initial condition $A=2$ (Deuterium), $z_0=0$ and 
different values 
of $\Gamma$, corresponding to energy $E$ indicated in the figure. 
$A(z)$ is evolved to $A_0=56$ (Iron) and $z_g$ is determined. 
Then $\Gamma_g=\mathcal{G}(A,\Gamma,z_0,z_g)$ is calculated as shown 
by the end points in the right panel of Fig. \ref{fig:evolve}. 
The trajectories for other values of $A$, e.g. $A=$ 9, 14,~ 40, are
very similar. The characteristic feature for all of them is the 
{\em explosive regime} in $A(z)$ evolution at the end of the trajectory.
We will discuss it in subsection \ref{sec:explos}.

In the  calculations of the evolution trajectories 
we always include two additional conditions: maximum energy of acceleration 
$E_{\rm max}^{\rm acc}$ (in terms of $\Gamma_{\rm max}^{\rm acc}$)
and nucleus stability restriction.  The first restriction is imposed
as follows. With $z_g$ determined from 
$\mathcal{A}(A,\Gamma,z_g)=A_0$ we calculate 
$\Gamma_g=\mathcal{G}(A,\Gamma,z_g)$. If 
$\Gamma_g \geq \Gamma_{\rm max}^{\rm acc}$
this trajectory is forbidden, and the contribution to the flux along this 
trajectory is set to zero. The second compulsory restriction, caused by 
nucleus stability, is considered in the next subsection. 

From the point of view of the  final result of our work as 
calculation of UHE nuclei {\em spectra}, the evolution trajectories 
are needed to different extent. For {\em trajectory method} they are 
the most essential  component of calculations. For the  
{\em combined method} the trajectories are needed 
for calculation of two quantities, $z_g$ and $\Gamma_g$. 
The former is calculated from $\mathcal{A}(\Gamma,A,z_g)=A_0$
(left panel of Fig. \ref{fig:evolve}). The latter is calculated 
as $\Gamma_g=\mathcal{G}(\Gamma,A,z_g)$ (the end-points of trajectories in 
the right panel of Fig. \ref{fig:evolve}). In the {\em coupled
kinetic equations} the  $\mathcal{A}(\Gamma,A,z_g)$
trajectory formally is not included, but it presents implicitly,
providing the evolution from $A_0$ to $A$.   

\subsection{Nucleus stability restriction}
\label{sec:stability}
\begin{figure}[!ht]
\begin{center}
\includegraphics[width=0.44\textwidth,angle=0]{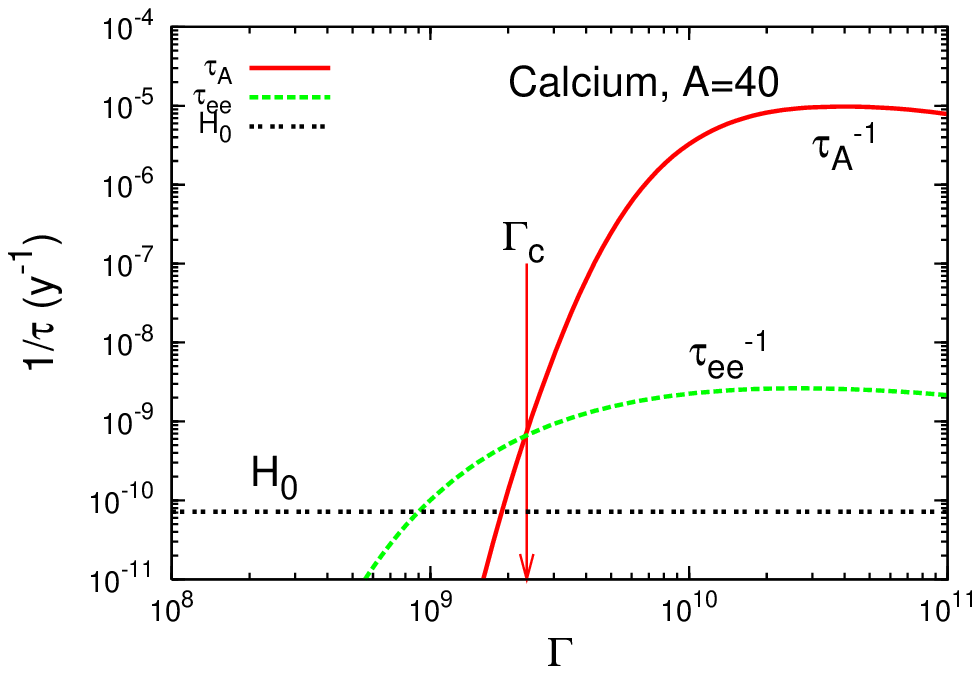}          
\includegraphics[width=0.44\textwidth,angle=0]{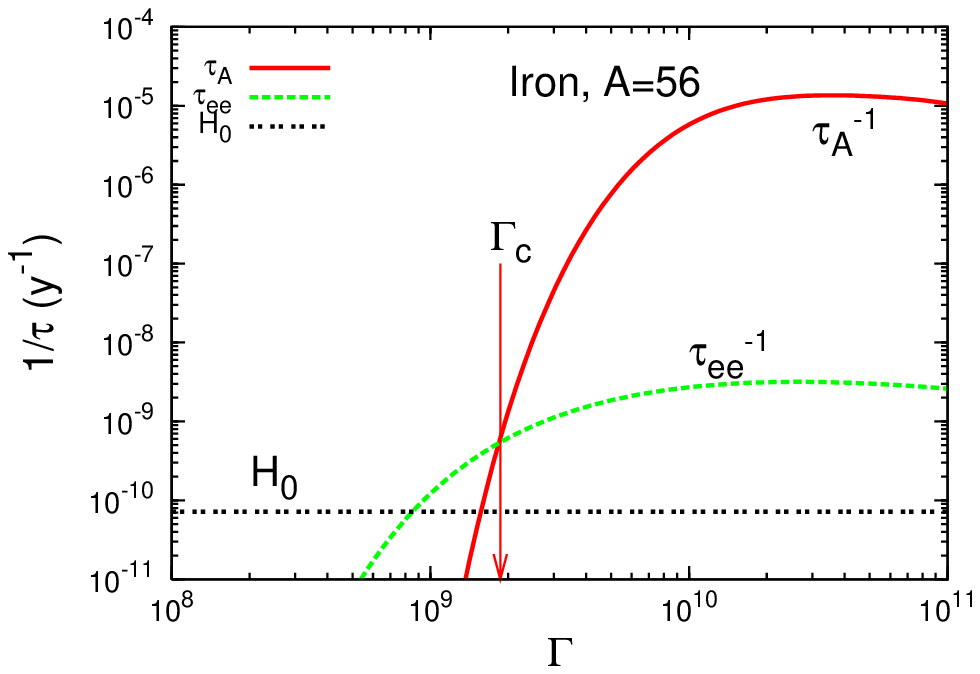}
\caption{Stability condition for $A=40$ and $A=56$ nuclei.
The nucleus $A$ is stable if the decay rate $\tau_A^{-1}$ is less than 
the rate of diminishing the Lorentz factor $\tau^{-1}_{ee} +H_0$, 
where $\tau_{ee}$ is the characteristic time of the pair-production  
energy losses (see Eq. \ref{eq:Gamma_c}). The critical Lorentz
factor given by Eq. (\ref{eq:Gamma_c}) coincides approximately with 
the intersection of $\tau_A^{-1}$ and $\tau^{-1}_{ee}$ curves, as shown 
in the figure.
}
\label{fig:stability}
\end{center}
\end{figure}

A nucleus $A$ with Lorentz factor $\Gamma < \Gamma_c^A$  at $z=0$ is stable, 
i.e. it is not photo-disintegrated during the Hubble time. This is easy to
understand from the plots in Fig. \ref{fig:stability}.

The critical Lorentz factor is defined by Eq. (\ref{eq:Gamma_c}) 
equating the photo-disintegration lifetime $\tau_A = (dA/dt)^{-1}$ and   
the characteristic time $\tau_{\Gamma}$ for Lorentz factor changing. 
Consider as an example the forward-in-time evolution of $^{40}$Ca nucleus 
with the Lorentz factor $\Gamma < \Gamma_c$. From Fig. \ref{fig:stability} it
is clear that such nucleus is stable: 
decreasing of the Lorentz factor goes faster than changing of $A$ and 
the nucleus  slows down, with a further increase of $\tau_A(\Gamma)$. 
Thus, nuclei $A-1$ with $\Gamma < \Gamma_c^A$ cannot be produced by
the decay $A \to (A-1) + N$ at $z=0$. However, they can be produced at
larger $z$. 

Consider now the backward-time evolution of the nucleus $(A-1)$ with Lorentz
factor $\Gamma < \Gamma_c^A$. The Lorentz factor of this
nucleus increases with $z$ quite fast  
as $\Gamma(z)={\mathcal G}(A-1,\Gamma,z_0,z)$, while $A-1$ remains
practically unchanged,  
because $\tau_A^{-1} \ll \tau_{\rm pair}^{-1}$. Simultaneously 
$\Gamma_c^A(z)$ decreases with increasing $z$, because the intersection
point of $\tau_A^{-1}$ and $\tau_{\rm pair}^{-1}$ is shifted to lower energies
by a factor $(1+z)$. As a result at some {\it critical redshift} 
$z_c(\Gamma)$,  
determined by the condition $\tau_A(\Gamma,z_c) = \tau_{\Gamma}(\Gamma,z_c)$
the considered nucleus becomes unstable.

In case of nucleus $A$ produced in $(A+1) \to A+N$ decay the equation for
critical redshift $z_c$ becomes
\begin{equation}
\mathcal{G}(A,\Gamma,z_0,z_c)= \Gamma_c^{A+1}(z_c),
\label{eq:z_c}
\end{equation}
where $z_0=0$. Therefore, to produce the secondary nucleus A with Lorentz 
factor $\Gamma <\Gamma_c^{A+1}$ at $z_0=0$, the parent nucleus $A+1$  
must decay  at $z \geq z_c$ determined by the Eq. (\ref{eq:z_c}). 
Thus, $z_c(\Gamma)$ can be considered as the minimum red-shift $z_{\rm min}$ 
for the production of nucleus A with Lorentz factor $\Gamma < \Gamma_c^{A+1}$.

A trajectory for A-nucleus can be blocked not only by $A+1$ nucleus, 
but also by $A' > A+1$, if $A'$-nucleus has an anomalously high $\Gamma_c$. 
But in this case the nucleus $A'-1$ increases its Lorentz factor with 
fixed $A'-1$ until it reaches  $\Gamma_c$. Numerical calculations show 
that stability condition for the considered case puts no restrictions 
to the calculated trajectories. 

\subsection{Explosive regime in $A(z)$ evolution}
\label{sec:explos}
The most striking and important feature of $A(z)$ evolution,
seen in the left panel of Fig. \ref{fig:evolve}, is its explosive 
behaviour at the end of the evolution. It is explained by the short 
lifetime $\tau_A$ of nucleus $A$ relative to the transition $A \to (A+1)$. 
This lifetime is connected with $\beta_A^0$, calculated for $z=0$ and 
plotted in Fig. \ref{fig:LosseHigh}, see also  Eq. (\ref{eq:lossE-z}), as 
\beq
\tau_A(\Gamma,z)= (1+z)^{-3} A^{-1}/\beta_A^0[(1+z)\Gamma].    
\label{eq:tau(z)}
\eeq
This equation demonstrates that $\tau_A(\Gamma,z)$ is short and
diminishes with $z$. Taking into account that $\sigma_{\rm dis}$ 
in Eq. (\ref{eq:losdis}) is approximately proportional to $A$ one can
understand that $A(t)$ increases exponentially with time. The 
Lorentz factor at this stage remains constant because the ratio 
$\tau_A/\tau_{\Gamma}$ is less than 100 (see Figs. \ref{fig:LosseLow} 
and \ref{fig:LosseHigh}).

At large Lorentz factors $\Gamma > \Gamma_c$ the explosive stage
starts at very small $z$ (see the curves $10^{19}$ and $10^{20}$ 
in the left panel of Fig. \ref{fig:evolve}), and the Lorentz factor during
this stage remains constant (see right panel of the figure). 
\begin{figure}[!ht]
\begin{center}
\includegraphics[width=0.49\textwidth]{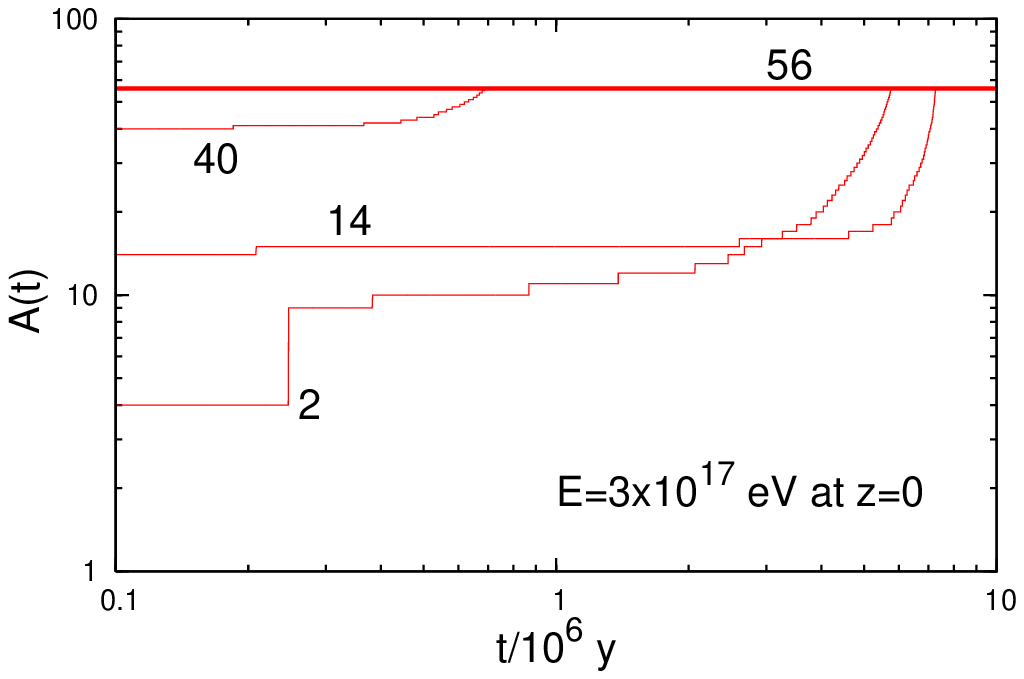}
\includegraphics[width=0.49\textwidth]{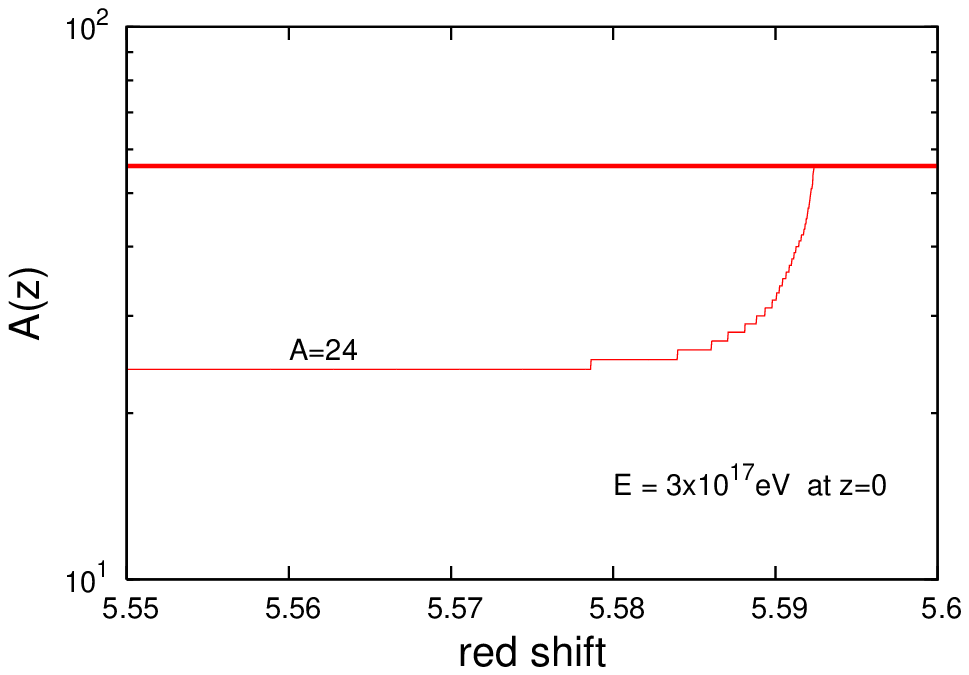}
\caption{The explosive regimes in the zoomed scale. 
{\em Left panel:} The explosive regime
for nuclei with different $A$ and with initial energy 
$3\times 10^{17}$~eV at $z_0=0$, 
shown in a natural time scale. The numbers give $A$.  
The duration of explosive regime is seen to be short. 
{\em Right panel:}
The explosive part of $A(z)$-evolution for $A=24$ and 
$E=3\times 10^{17}$~eV at $z=0$. It starts from $A \rightarrow (A+1)$
transition at $z_A=5.58$ and finishes at $z_g=5.59$ reaching $A_0=56$.  
}
\label{fig:explos}
\end{center}
\end{figure} 

Let us consider now a nucleus $A$ at $z_0=0$ with 
$\Gamma_A \ll \Gamma_c^{A+1}$. 
This case is shown for A=2 in Fig. \ref{fig:evolve} (left panel) for 
$\Gamma = 1.5\times 10^8$, i.e. $E=3\times 10^{17}$~eV. From $z_0=0$ to 
$z \sim 0.5$ a nucleus evolves not changing $A$ and changing $\Gamma$  
only due to adiabatic energy losses. The critical moment $z_c$ occurs when 
Lorentz factor reaches $\Gamma_c$ for $^3He$ according to 
Eq. (\ref{eq:z_c}). Then photo-disintegration starts and almost
immediately enters the exponential regime with $\tau_A$ given by 
Eq. (\ref{eq:tau(z)}), which in terms of redshifts is an explosive regime.

The typical cases of explosive trajectories with $\Gamma_A \ll \Gamma_c^{A+1}$
at  $z_0=0$ are shown in Fig. \ref{fig:explos}. In the left panel the 
explosive regimes for $A= 2,~14$ and 40 are shown in a natural time
scale to demonstrate the short time of evolution. In the right panel
the evolution of nucleus $A=24$ is shown for $E=3\times 10^{17}$ at
$z=0$. Until $z=5.58$ $A$ is not changed, and then during $\Delta z= 0.01$
$A$ jumps to $A_0=56$. It illustrates a typical case when the
redshift of $A \rightarrow (A+1)$ transition approximately coincides
with the redshift of $A_0$ appearance, $z_A \approx z_g$. In this example 
$z_A=5.58$ and $z_g=5.59$. Considering then the part of trajectory 
$A \rightarrow A_0$, one obtains 
$\Gamma_g= \mathcal{G}(A,\Gamma_A,z_A,z_g) \approx \Gamma_A$, due to 
$z_A \approx z_g$. The relations 
\beq 
z_A \approx z_g, \;\;\; \Gamma_A \approx \Gamma_g .
\label{eq:explos}
\eeq
are important feature of the explosive trajectories. 

For the large Lorentz factors $\Gamma > \Gamma_c$,  Eq. (\ref{eq:explos})
is naturally fulfilled, too. 

\subsection{Fluxes of nuclei and protons in the trajectory method}
\label{sec:flux-trajectory}
In principle trajectory method is similar to MC simulation: 
both are following trajectory of a nucleus propagating through CMB,
calculating the flux of secondary nuclei with Lorentz factor 
$\Gamma$ from the flux of primary nuclei $A_0$ with Lorentz factor
$\Gamma_g$. As advantage MC includes fluctuations in the interactions 
and does not need the introduction of any integration limit.  

In the trajectory method the space density $n_a (\Gamma,z_0)$, 
of the components $a=A_0, A, p$ is found from the conservation
of the number of particles as 
$n_a(\Gamma)d\Gamma = \int dt' Q_a(\Gamma',t') d\Gamma'$, where 
$Q_a(\Gamma',t')$ is the generation rate of these particles (see 
appendix \ref{app:generation}). In a way convenient for calculations 
this formula has the form 
\beq
n_a(\Gamma,z_0) = \int_{z_{\rm min}} ^{z_{\rm max}}dz'
\left |\frac{dt}{dz} \right | Q_a(\Gamma',z')\frac{d\Gamma'}{d\Gamma},
\label{eq:n-traj}
\eeq
where $z_0=0$, $Q_a(\Gamma',z')$ is the generation rate of $a$
particles given in appendix \ref{app:generation} and 
$d\Gamma'/d\Gamma$ is given in appendix \ref{app:dgamma}.  
The limits of integration are as follows.

For primary nuclei ($a=A_0$) $z_{\rm min}=0$ and for $z_{\rm max}$ we have
three options: $z(\tau_{A_0})$, i.e. redshift corresponding to lifetime 
$\tau_{A_0}(\Gamma)$, $z_c(\Gamma)$ as given by Eq. (\ref{eq:z_c}), and 
$z_g(A_0+1)$. These three cases give somewhat different results. 

For secondary nuclei ($a=A$) $z_{\rm min}=0$ for $\Gamma \geq \Gamma_c$,
and is $z_c(\Gamma)$ for $\Gamma \leq \Gamma_c$. The maximum limit is
given by $z_{A+1}(\Gamma)$. 

For secondary protons produced by $(A+1)$ nuclei the limits are the
same as for secondary $A$-nuclei, since the production of both is identical 
(see appendixes \ref{app:generation} and \ref{app:comparison}). 

As numerical calculations show, the trajectory method gives the worst  
accuracy for flux calculations, especially for secondary nuclei, and
this is most probably connected with rather rough estimate of the
integration limits, which affect strongly the results. 
\section{Kinetic equation combined with trajectory calculations}
\label{sec:comb}
In this section we calculate the fluxes (space densities) of primary 
nuclei,  secondary nuclei and protons using the kinetic equation 
and trajectory calculations. The latter are used in a limited way, 
only for calculation of generation rate of nuclei and protons, using 
the number of particles conservation (see 
appendix \ref{app:generation}). In the case of secondary nuclei,   
the trajectories are used only to 
determine $\Gamma_g$ and $z_g$ associated to the primary nucleus $A_0$, 
these calculations may be greatly simplified taking into account that    
in all practical cases the trajectories are explosive. 

We start with the calculations of the secondary-nuclei flux as the 
most general and technically most interesting part of  our calculations. 

\subsection{Secondary nuclei }
\label{sec:secondary}
We consider an expanding universe homogeneously filled by the sources
of accelerated primary nuclei $A_0$ with a generation rate per unit
of comoving volume $Q_{A_0}(\Gamma,z)$ given by 
\begin{equation}
Q_{A_0}(\Gamma,z)=\frac{(\gamma_g-2)}{m_N A_0}
{\mathcal L}_0 \Gamma^{-\gamma_g},
\label{eq:inj1}
\end{equation}
where $\gamma_g>2$ is the generation index, $m_N$ is the nucleon mass and 
${\mathcal L}_0$ is the source {\em emissivity}. i.e. the energy
generated per unit
of comoving volume and per unit time at $z=0$. In Eq. (\ref{eq:inj1}) 
$\Gamma_{\rm min} \sim 1$ is assumed. In all calculations below 
we assume also a maximum energy of acceleration 
$E^{\rm acc}_{\rm max}=Z_0\times 10^{21}$~eV 
(or $\Gamma_{\rm max}^{\rm acc}= (Z_0/A_0)\times 10^{12}$) with 
the condition $Q_{A_0}(\Gamma_g)=0$~ 
at $\Gamma_g \geq \Gamma_{\rm max}^{\rm acc}$.    

The trajectory $A(z)$ is calculated as a continuous quantity
only to determine the redshifts of $(A+1) \to A$ and 
$A \to (A-1)$ transitions. Between these values we assume $A=const$
calculating the evolution $\Gamma(z)$ with fixed $A$.
Thus, we assume that both of the above-mentioned transitions are  
instantaneous photo-disintegration, e.g. $\gamma_{\rm CMB}+(A+1) \to A+N$.  
Since a recoil momentum in these processes is negligibly small, one 
has approximate equality of Lorentz factors of all three particles 
\beq
\Gamma_{A+1} \approx \Gamma_A \approx \Gamma_N .
\label{eq:equality1}
\eeq

The generation rate of secondary nuclei (and secondary nucleons)
is derived in  appendix \ref{app:generation} using the trajectory 
calculation combined with the conservation of the number of particles 

\begin{equation}
Q_A(\Gamma_A,z)=Q_{A_0}(\Gamma_g,z_g) \frac{1+z}{1+z_g} 
\frac{d\Gamma_g}{d\Gamma_A}
\label{eq:QA1}
\end{equation}
with $d\Gamma_g/d\Gamma_A$ given in appendix \ref{app:dgamma}. 

The kinetic equation for the comoving space density of secondary nuclei 
$A$,~ $n_A(\Gamma_A, t)$, reads 
\begin{equation}
\frac{\partial n_A(\Gamma_A,t)}{\partial t} - 
\frac{\partial }{\partial \Gamma_A}\left [ b_A(\Gamma_A,t) n_A(\Gamma_A,t) 
\right ] + \frac{n_A(\Gamma_A,t)}{\tau_A(\Gamma_A,t)} = Q_A(\Gamma_A,t),
\label{eq:kin}
\end{equation}
where $b_A=-d\Gamma/dt=(\beta_{\rm pair}+ \beta_{\rm ad})\Gamma_A $ is 
the rate of the Lorentz-factor loss and $Q_A(\Gamma_A,t)$ is the generation 
rate given by Eq. (\ref{eq:QA1}), the nucleus $A$ lifetime is 

\begin{equation}
\tau_{A}^{-1}(\Gamma_{A},z)=- dA/dt~=\frac{T}{2\pi^2\Gamma^2}
\int_{\epsilon_0(A)}^{\infty}d\epsilon 
\sigma_{\rm dis}(\epsilon,A)\nu(\epsilon)\epsilon
\left [-\ln\left (1-\exp[-\frac{\epsilon}{2\Gamma T}]\right)\right ]~,
\label{eq:tau}
\end{equation}
where $\nu(\epsilon)$ is the average multiplicity of the emitted
nucleons. 

\begin{figure}[!ht]
\begin{center}
\includegraphics[width=0.9\textwidth,angle=0]{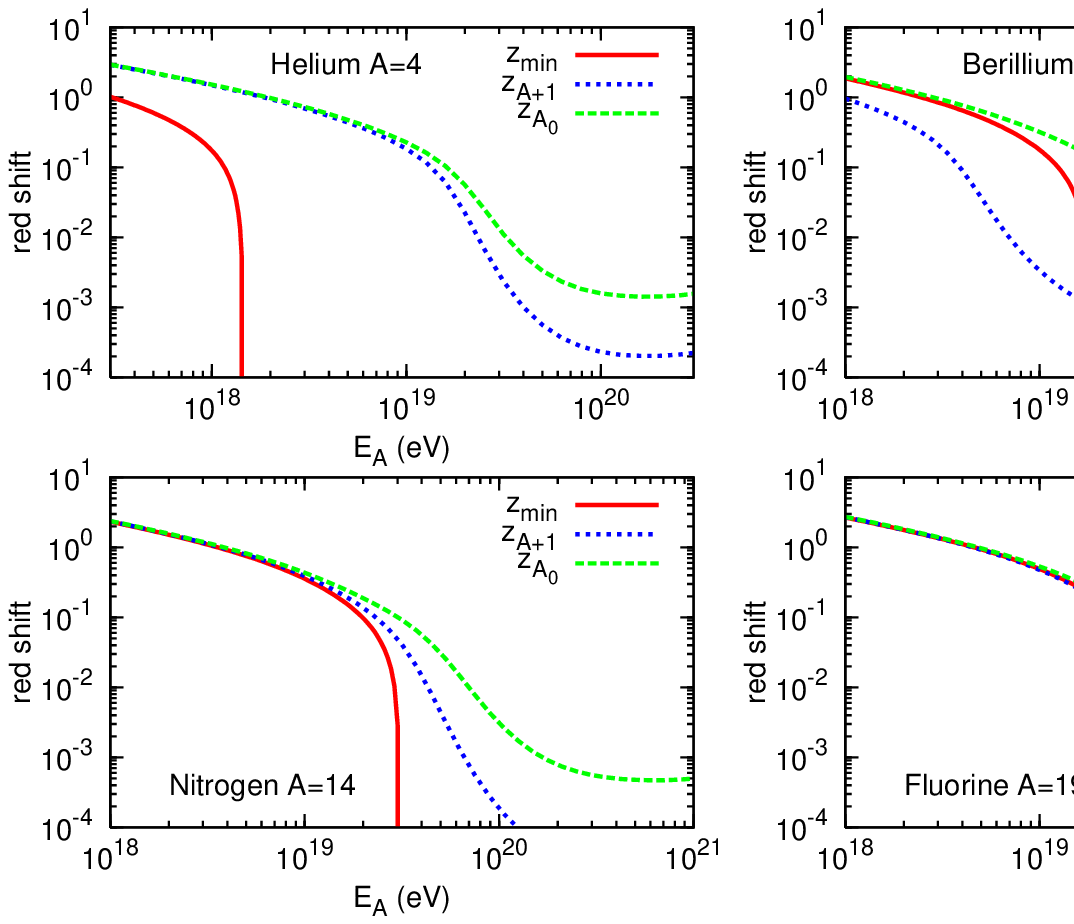}
\caption{Redshift limits in Eq. (\ref{eq:fluxA}) for nuclei with 
small mass number $A$. The values $z_{A+1}$ and $z_{A_0}$ correspond to
redshifts when the running $A(z)$ reaches $A+1$ and $A_0$, respectively.}
\label{fig:z_limitLow}
\vskip 1 cm
\includegraphics[width=0.9\textwidth,angle=0]{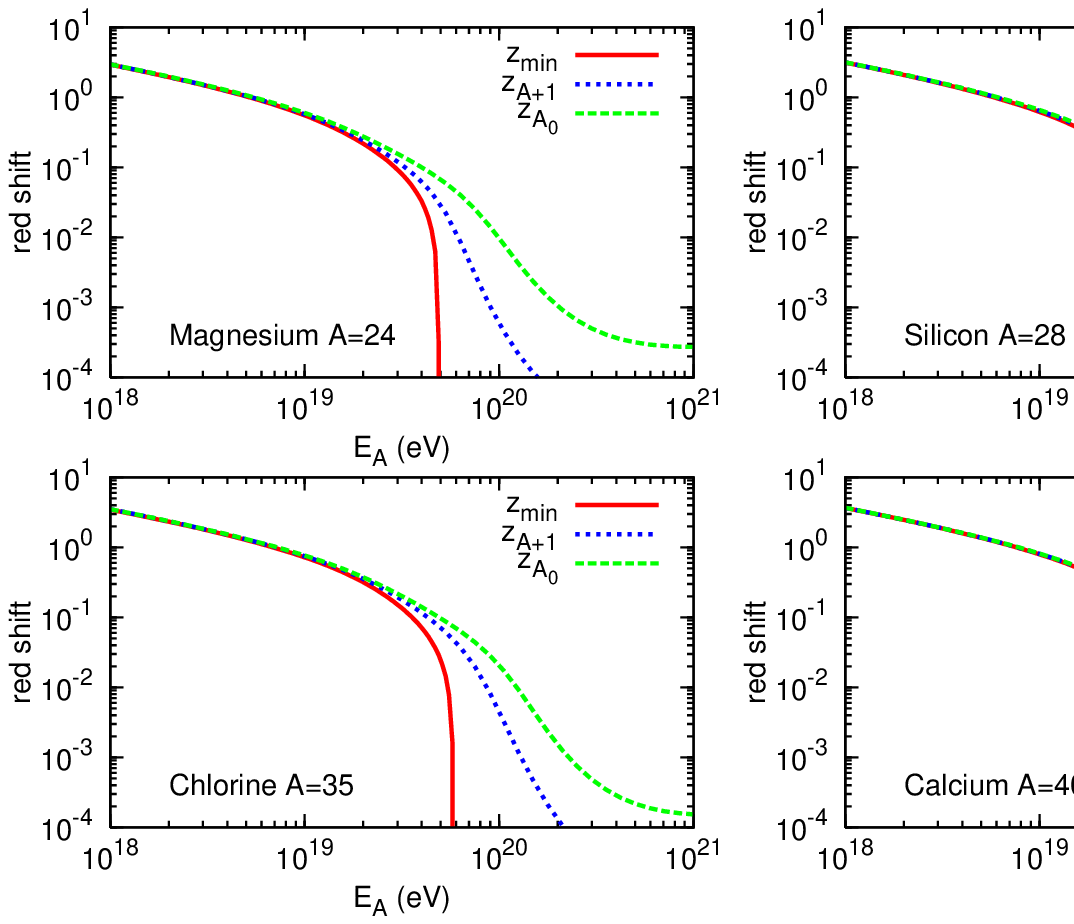}
\caption{Redshift limits for nuclei with large $A$.
Notation is the same as in Fig. \ref{fig:z_limitLow}. }
\label{fig:z_limitHigh}
\end{center}
\end{figure}

The solution of the kinetic equation (\ref{eq:kin})
is found in appendix \ref{app:solution} and can be presented as 

\begin{equation}
n_A(\Gamma)=\int_{z_{\rm min}}^{z_{\rm max}} dz_A \left | \frac{dt_A}{dz_A} 
\right | Q_A(\Gamma_A,z_A) 
\frac{d\Gamma_A}{d\Gamma} e^{-\eta(\Gamma_A,z_A)}
\label{eq:fluxA}
\end{equation}
where $n_A(\Gamma)$ is given for $z_0=0$, $z_A$ is the redshift of 
$A$-nuclei production in $A+1 \to A+N$ decay, 
$\Gamma_A={\mathcal G}(A,\Gamma,z_0,z_A)$ is the Lorentz factor of the 
nucleus $A$ at the moment of production, calculated using $A=const$.
The quantity $\eta(\Gamma_A,z)$ is given by 
\begin{equation}
\eta(\Gamma_A,z)=\int_0^z dz' \left | \frac{dt'}{dz'} \right | 
\frac{1}{\tau_A(\Gamma_A,z')}~~,
\label{eq:tauA}
\end{equation} 

The physical meaning of the factor $\exp(-\eta)$ becomes clear 
from Eq. (\ref{eq:app-solut2}) of appendix \ref{app:solution}.
Assuming $ \tau_A=const$ and introducing as variable a propagation 
time $t=t_0-t'$, one obtains this factor as $\exp(-t/\tau_A)$, which
gives the survival probability for the nucleus A during the propagation 
time $t$. 

The factor $\exp(-\eta)$ provides an upper limit of integration
in Eq. (\ref{eq:fluxA}). Without it the trajectory of integration in
kinetic equation $\Gamma_A(z)$ extends to any large $z$ with the same 
$A=const$. Considering this trajectory as the one on which $A$ changes
with $z$ too, we can put the marks $z_{A+1}$ and $z_0$, which  
corresponds $A(z)$ reaching $(A+1)$ and $A_0$, respectively. 
From Eq. (\ref{eq:tauA}) one can see that at these redshifts the
suppression factor $\exp(-\eta)$ is equal to $\exp(-1)$ and
$\exp[-(A_0-A)]$. We shall use $z_{A+1}$ and $z_{A_0}$ as trial
upper limits in  Eq. (\ref{eq:fluxA}) for comparison with the basic case
when the upper limit is  set to infinity. 

The calculation of the lower limit of integration in Eq. (\ref{eq:fluxA}) 
is more complicated. The lower limit $z_{\rm min}(\Gamma)$ is provided 
by vanishing of the generation rate $Q_A(z)$ at $z \leq z_{\rm min}$. 
As has been discussed in section \ref{sec:stability} at 
z=0 there is a critical Lorentz factor $\Gamma_c^{A+1}$ below which 
the photo-disintegration of the parent nucleus $A+1$ is absent 
(see Fig. \ref{fig:stability}). With increasing $z$ this critical Lorentz 
factor $\Gamma_c^{A+1}(z)$ slightly diminishes because both the 
pair production and photo-disintegration curves are shifted in 
Fig. \ref{fig:stability} by factor $(1+z)$ to lower energies.
If Lorentz factor of a considered $A$-nucleus  $\Gamma_A < \Gamma_c^{A+1}$, 
the transition $A \to (A+1)$ at $z=0$ is forbidden 
and only $\Gamma$ increases with $z$ ($\tau_A^{-1}$ in
Fig. \ref{fig:stability} is very small). At $z_c$ when the 
Lorentz factor of $A$-nucleus ${\mathcal G}(A,\Gamma,0,z_c)$
reaches $\Gamma_c^{A+1}(z_c)$ photo-disintegration starts
(see Eq. \ref{eq:z_c}). Thus, for $\Gamma < \Gamma_c^{A+1}$  
$z_{\rm min}(\Gamma)=z_c(\Gamma)$, and for $\Gamma > \Gamma_c$ 
$z_{\rm min}(\Gamma)=0$.

This effect of $z_{\rm min}$-appearance can be explained in the 
evolution-trajectory approach as follows. The production rate of 
$A$-nuclei, $Q_A(z)$, is caused by the decay of $A+1$ nuclei. 
However, at all $z$ for which $\Gamma_{A+1}(z)<\Gamma_c^{A+1}(z)$ 
a nucleus $A+1$ does not decay and $A+1=const$, as we 
observe indeed in the calculation of the trajectories forward 
in time, and hence $Q_A(z)=0$ at $z \leq z_c$.

The lower limit $z_{\rm min}(\Gamma)$ and two trial upper limits 
are presented in Figs. \ref{fig:z_limitLow} and \ref{fig:z_limitHigh}.
The upper limit $z_{A+1}$ gives the redshift when $A$ formally evolves 
to $A+1$. The upper limit $z_{A_0}$ corresponds to the evolution
of $A$ to $A_0$. The actual upper limit in this case is provided
by the factor $e^{-\eta}$. One may recognize in these 
figures the regions in the energy-redshift plane which
contribute significantly to the flux of the secondary nuclei. 
In particular, at low energies the interval $z_{\rm max}-z_{\rm min}$ 
tends to zero and the flux vanishes. At the highest energies 
$E> 3\times 10^{19}$ eV, the interval is widening, but the absolute 
values of $z_{\rm min}$ and $z_{\rm max}$ become vanishingly small, 
in the range of $10^{-3}$, and thus the fluxes are strongly suppressed. 

\begin{figure}[!ht]
\begin{center}
\includegraphics[width=0.49\textwidth]{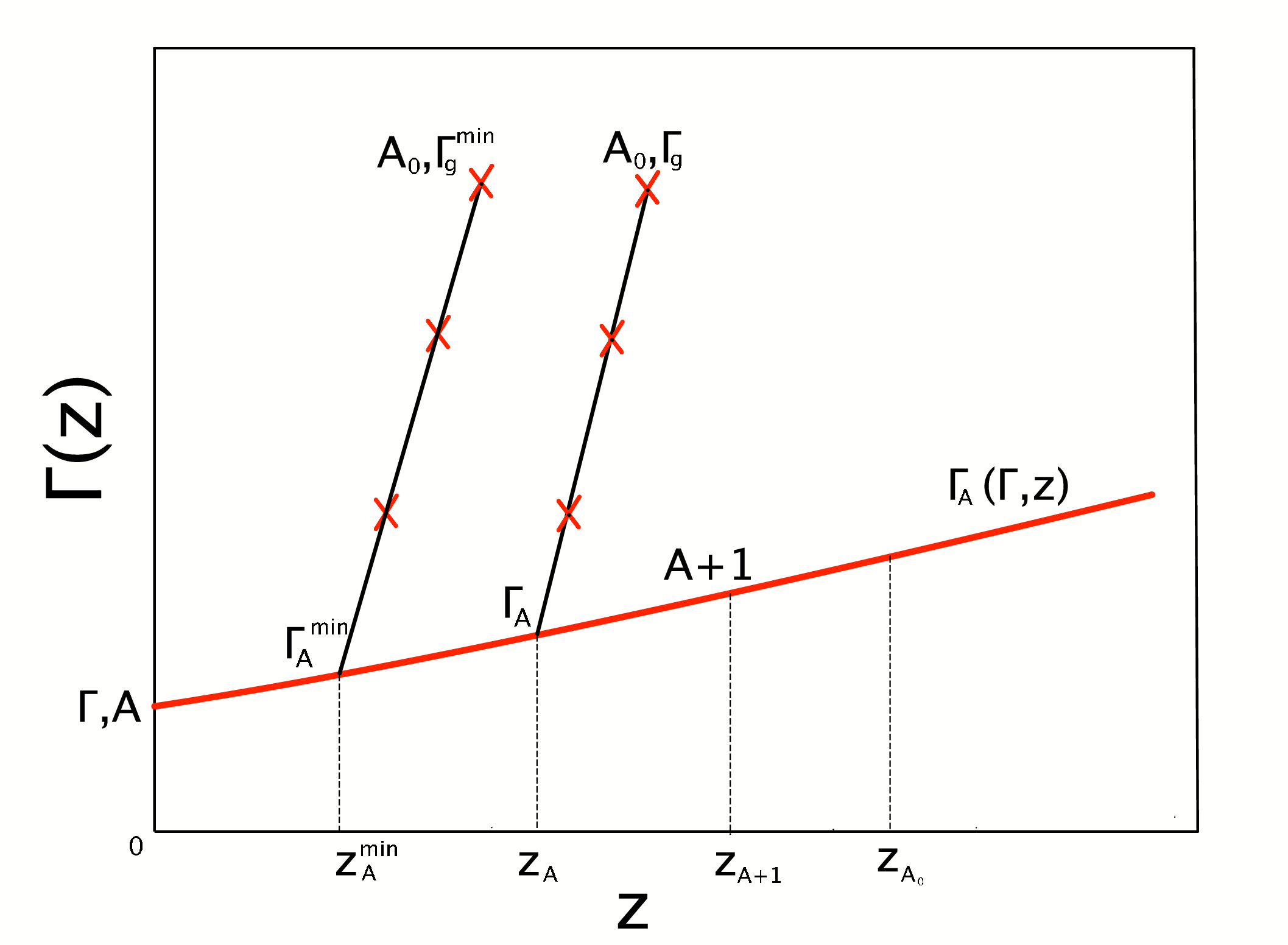}
\includegraphics[width=0.49\textwidth]{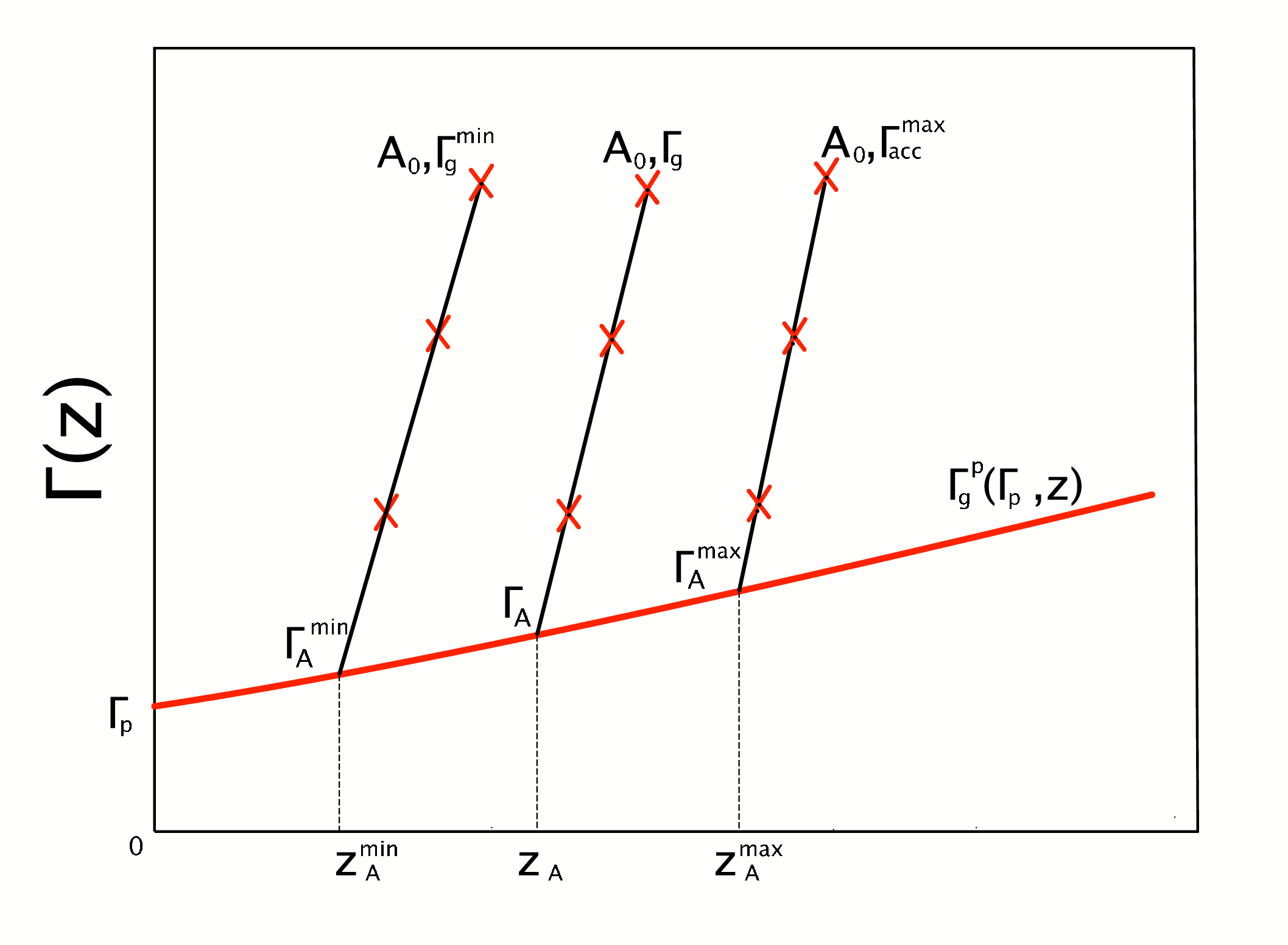}
\caption{A sketch of $\Gamma (z)$ evolution in kinetic 
equation approach. {\em Left panel}: The thick red line gives the
integration trajectory  in Eq. (\ref{eq:fluxA}) described by 
$\Gamma_A(\Gamma,z_0,z)$ with fixed $A=const$. The auxiliary  
$A$-evolution trajectories $(A+1) \to A_0$ are needed 
for calculation of generation rate of $A$-nuclei 
$Q_A(\Gamma_A,z_A)$. In practice all these trajectories are 
explosive, providing $z_A \approx z_g$ and 
$\Gamma_A \approx \Gamma_g$. A point $(z_A,\Gamma_A)$ represents 
the running variables in the integral given by   
Eq. (\ref{eq:fluxA}). The lower limit of integration is given by 
$z_A^{\rm min}$, while $z_{A+1}$ and $z_{A_0}$ are the trial upper 
limits (see text). {\em Right panel} related to section 
\ref{sec:prot}: The thick red line gives the integration trajectory in 
Eq. (\ref{eq:flux-pA}) for secondary protons. 
The trajectory describes the Lorentz-factor evolution of a proton 
$\Gamma_g^p(\Gamma_p,z)$. Proton is produced at a running point 
$(z,\Gamma)$ of a proton trajectory in a decay 
$(A+1) \rightarrow A+N$, and we use in the figure the variables 
of the ´brother´ nucleus $A$:~ $z_A=z$ and $\Gamma_A=\Gamma$.  
The $A$-evolution sub-trajectories 
$A \to A_0$ are needed for calculation of generation rate 
$Q_p^A(\Gamma_A,z_A)$ given by Eq. (\ref{eq:QpA}). All these
trajectories are explosive, providing $z_A \approx z_g$ and 
$\Gamma_A \approx \Gamma_g$. } 
\label{fig:trajAP}
\end{center}
\end{figure}

\begin{figure}[!ht]
\begin{center}
\includegraphics[width=0.49\textwidth]{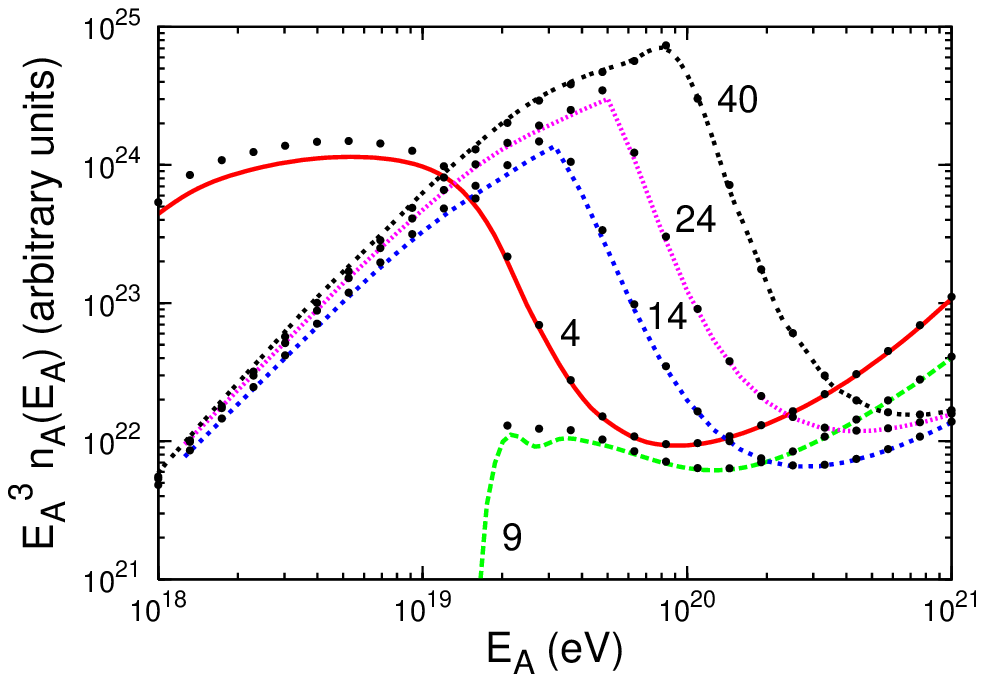}
\includegraphics[width=0.49\textwidth]{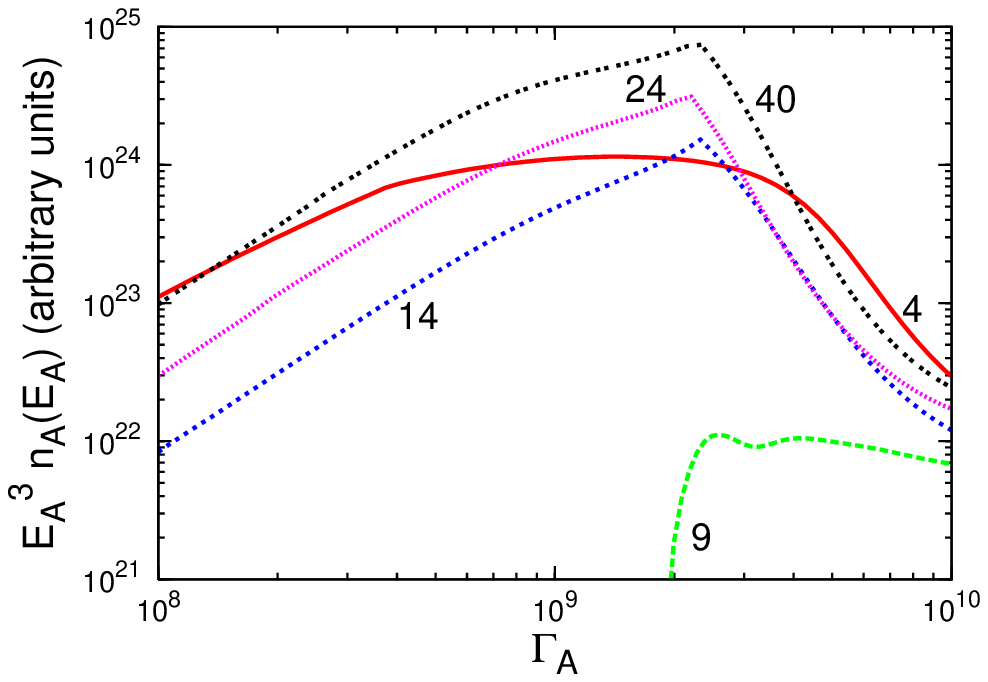}
\caption{ Fluxes of secondary nuclei, as function of energy 
(left panel) and of the Lorentz factor (right panel). The numbers on
the curves show A. The small filled circles present calculations in 
the explosive trajectory approximation (see the text).}
\label{fig:fluxA}
\end{center}
\end{figure}

We will now come over to calculation of fluxes in the form of particle
densities $n_A(\Gamma,z)$. For this we calculate the generation rates 
$Q_A(\Gamma_A,z_A)$ using Eq. (\ref{eq:QA1}) and put it into 
Eq. (\ref{eq:fluxA}). The left panel in Fig. \ref{fig:trajAP} explains the
procedure of integration. The fluxes are shown in Fig. \ref{fig:fluxA} 
in terms of energy (left panel) and Lorentz factor (right panel). 

One should realize the following feature of calculations: the time
scale of $A$ evolution from $(A+1)$ to $A_0$ needed for calculation 
of the rate $Q_A(\Gamma_A,z_A)$, and the time scale of the kinetic-equation
trajectory $\Gamma_A(z)$, which contribute the integral in 
Eq. (\ref{eq:fluxA}), is the same and very short $\sim \tau_A$. 
In case of $A$-evolution this is the explosive part of the trajectory,  
in the case of kinetic-equation trajectory it is regulated by 
$\exp(-t/\tau_A)$. For low energies $E \lsim 10^{18}$~eV 
the lower limit of integration $z_{\rm min}=z_c \approx 2-3$,
and $A$ evolution at this $z$ occurs in explosive regime with 
short evolution time given by Eq. (\ref{eq:tau(z)}). The region 
of integration in Eq. (\ref{eq:fluxA}) is controlled by the same 
$\tau_A(\Gamma,z)$. As it is easy to understand the situation 
is the same at $\Gamma \geq \Gamma_c$. 
In both cases we can use an explosive trajectory for calculation 
of generation rate, which implies $z_g \approx z_A$ and $\Gamma_g
\approx \Gamma_A$, and thus Eq. (\ref{eq:fluxA}) results in
\beq
n_A(\Gamma)=\frac{\gamma_g-2}{m_N A_0}\mathcal{L}_0 
\int_{z_{\rm min}(\Gamma)}^{z_{\rm max}} \frac{dz_A}{(1+z_A)H(z_A)}
\Gamma_A^{-\gamma_g}\frac{d\Gamma_A}{d\Gamma}e^{-\eta(\Gamma_A,z_A)},
\label{eq:fluxA-inst}
\eeq
where $\Gamma_A = \Gamma_A(\Gamma, z_0, z_A)$ with $A=const$ and $z_0=0$.

In Fig. \ref{fig:fluxA} the fluxes calculated in this explosive trajectory 
approximation are shown by small filled circles. One may observe,
as expected, a good agreement with the exact calculations. 

The basic parameter which determines the spectra of nuclei $A$  
is given by the critical Lorentz factor $\Gamma_c^{A+1}$ at which 
the photo-disintegration process of the parent nucleus  (A+1, or 
Beryllium in the case of Helium) is allowed at $z=0$. One may  expect that  
the peaks seen in the spectra $E_A^3 n_A(E)$ set at
the critical Lorentz factors and the right panel of
Fig. \ref{fig:fluxA} confirms this expectation with a good precision
(see Table~\ref{tab:Gamma_c} of section~\ref{sec:losses}).

The calculated spectra at low $\Gamma$ exhibit the $A$ hierarchy:
the higher $A$ the larger the flux. The exceptional case of low Be flux 
is explained by large $\tau_A$, which determines the upper limit 
in Eq. (\ref{eq:fluxA}). In effect, the upper limit $z_{A+1}$ is below  
$z_{\rm min}$ (see upper-right panel in Fig. \ref{fig:z_limitLow}),
and factor $e^{-\eta}$ in Eq. (\ref{eq:fluxA}) operates exclusively 
at large $\eta$ suppressing strongly the Beryllium flux.     

The explosive-trajectory approximation allows us  to obtain the 
exact formula for the asymptotic high-energy flux and to compare 
it with calculations presented in Fig. \ref{fig:fluxA}.

In the asymptotic regime one can use in  
Eq. (\ref{eq:fluxA-inst}) $\Gamma_A \approx \Gamma$ (Lorentz
factor is not changing in the pure photo-disintegration process),
$z_{\rm min}=0$ and  $z_g \ll 1$. Then after simple calculations we 
obtain 
\begin{equation}
n_A (\Gamma)= (\gamma_g-2)\frac{\mathcal{L}_0}{A_0 m_N}
\Gamma^{-\gamma_g} \tau_A(\Gamma).
\label{eq:n_Aasympt}
\end{equation}
The asymptotic formula (\ref{eq:n_Aasympt}) follows also directly 
from the kinetic equation (\ref{eq:kin}) if one neglects there
$\partial n_A/\partial t$ and $\partial (b_A n_A)/\partial \Gamma_A$.
Eq. (\ref{eq:n_Aasympt}) predicts the asymptotic ratio as 
\begin{equation}
n_{A_1} : n_{A_2} : n_{A_3} .. = \tau_{A_1} : \tau_{A_2} : \tau_{A_3} ..
\label{eq:ratio}
\end{equation}
Numerical calculations confirm that asymptotic relations 
(\ref{eq:n_Aasympt}) and (\ref{eq:ratio}) are fulfilled for the curves 
in Fig. \ref{fig:fluxA}. 

For all secondary nuclei at the highest energies the contributing 
universe is vanishingly small, when the relevant redshifts are of the 
order of $10^{-3}$ or less,  as it follows from curves $z_{A+1}(E)$ in  
Figs. \ref{fig:z_limitLow} and \ref{fig:z_limitHigh}. The fluxes 
in Fig. \ref{fig:fluxA} are non-vanishing only because 
in the present paper we  consider a  homogeneous 
distribution of sources for primary nuclei $A_0$ with constant 
density at any redshift, 
while in a realistic situation the presence of UHECR sources 
at these low redshifts $(z<10^{-3})$ is unlikely. 

The energy spectrum at high energies follows  
the corresponding energy losses, with a recovery at the highest 
energy due to an increase of the nucleus lifetime $\tau_A$ 
(i.e. a decrease in the photo-disintegration energy losses, see 
Figs. \ref{fig:LosseLow} and \ref{fig:LosseHigh}). 

\subsection{Secondary protons}
\label{sec:prot}
In this section we discuss the flux of secondary nucleons produced 
in the photo-disintegration of the heavier nuclei. We do not need to 
distinguish 
neutrons and protons, because at the characteristic lengths involved
here neutrons decay fast at all energies of interest. For this reason 
we will often refer to protons, instead of nucleons. 

Production of secondary protons accompanies the production of 
secondary nuclei, but in contrast to them, protons are not destroyed
and can arrive from any redshift, being suppressed only by high energy
of generation.  

Following the same approach as for secondary nuclei one can write a kinetic 
equation that describes the propagation of protons: 
\begin{equation}
\frac{\partial n_p(\Gamma_p,t)}{\partial t} - \frac{\partial}{\partial 
\Gamma_p} \left [ b_p(\Gamma_p,t)n_p(\Gamma_p,t) \right ] = Q_p(\Gamma_p,t)
\label{eq:kin_p}
\end{equation}
where $n_p$ is the secondary-proton density, $b_p=-d\Gamma/dt$
describes the loss of proton Lorentz factor (energy loss) 
due to adiabatic energy losses, 
pair-production and photo-pion production on the CMB radiation, and  
$Q_p$ is the generation rate for secondary protons, produced by  
photo-disintegration of secondary or primary nuclei. 
The secondary nucleons $N$ which accompany production of $A$-nuclei in 
the process $A+1 \rightarrow A+N$ have the same Lorentz factor and
generation rate as A-nuclei. 
We refer to these nucleons as ``$A$-associating 
protons'' with notation $n_p^A$ for their space density, and use for
their generation rate $Q_p^A(\Gamma,t)=Q_A(\Gamma,t)$.   
The solution of Eq. (\ref{eq:kin_p}) for $A$-associating protons   
similarly to \cite{AB04}, reads: 
\begin{equation}
n^A_p(\Gamma_p)=\int_{z_{\rm min}}^{z_{\rm max}} 
dz' \left | \frac{dt'}{dz'} \right | Q^A_p(\Gamma',z')
\left (\frac{d\Gamma'}{d\Gamma_p}\right )_p , 
\label{eq:flux-pA}
\end{equation}
where $\Gamma'(z')=G_p(\Gamma_p,0,z')$ and $d\Gamma'/d\Gamma_p$ is
taken along the proton trajectory ( $\Gamma_g^p(\Gamma_p,z)$ line 
in the right panel of Fig. \ref{fig:trajAP} ).

As in Eq. (\ref{eq:QA1}) we have from the conservation of particles
number: 
\beq
Q^A_p(\Gamma',z')=Q_{A_0}(\Gamma_g,z_g) \frac{1+z'}{1+z_g} 
\left (\frac{d\Gamma_g}{d\Gamma'}\right )_A ,
\label{eq:QpA}
\eeq
where the redshift of $A+1$ decay $z_A=z'$ and $\Gamma_A=\Gamma'$, 
index $A$ at derivative $d\Gamma_g/d\Gamma'$ indicates the
$A$-variable subtrajectory $A \rightarrow A_0$ for which derivative is 
given by Eq. (\ref{eq:dE_g/dE}), and $Q_{A_0}$ is the  
generation (acceleration) rate of primaries $A_0$ at redshift $z_g$ with 
Lorentz factor $\Gamma_g$, given by Eq. (\ref{eq:inj1}). 
The generation redshift $z_g$ is determined from 
$\mathcal{A}(A,\Gamma',z',z_g)=A_0$ and the generation Lorentz factor 
is calculated as $\Gamma_g=\mathcal{G}(A,\Gamma',z',z_g)$.
Finally, we obtain for $n_p^A(\Gamma_p,z_0)$ at $z_0=0$:
\begin{equation}
n_p^A(\Gamma_p)=\frac{\gamma_g-2}{A_0 m_N}{\mathcal L}_0
\int_{z_{\rm min}}^{z_{\rm max}} 
dz' \left | \frac{dt'}{dz'}\right |
\Gamma_g^{-\gamma_g}\frac{1+z'}{1+z_g}
\left (\frac{d\Gamma_g}{d\Gamma'}\right )_A
\left (\frac{d\Gamma'}{d\Gamma_p}\right )_p,
\label{eq:n-pA}
\end{equation}
the integration in Eq. (\ref{eq:n-pA}) goes along 
the proton trajectory $\Gamma_g^p(\Gamma_p,z)$, i.e $G_p(\Gamma_p,0,z)$, 
as sketched in the right panel of Fig. \ref{fig:trajAP}. A proton can 
be produced at any running point of this trajectory  $(z_A,\Gamma_A)$, 
with $z_A=z'$ and $\Gamma_A=\Gamma'$, in a decay 
$(A+1) \rightarrow A +N$. The generation rate at such point 
$Q_p^A(z_A,\Gamma_A)$ is calculated using Eq. (\ref{eq:QpA})
along the $A$-evolution subtrajectory $A \rightarrow A_0$. For the 
different subtrajectories we have different values of  
$\Gamma_g$ for the primary nucleus $A_0$ and 
different redshifts of generation $z_g$. 

\begin{figure}[!ht]
\begin{center}
\includegraphics[width=0.49\textwidth,angle=0]{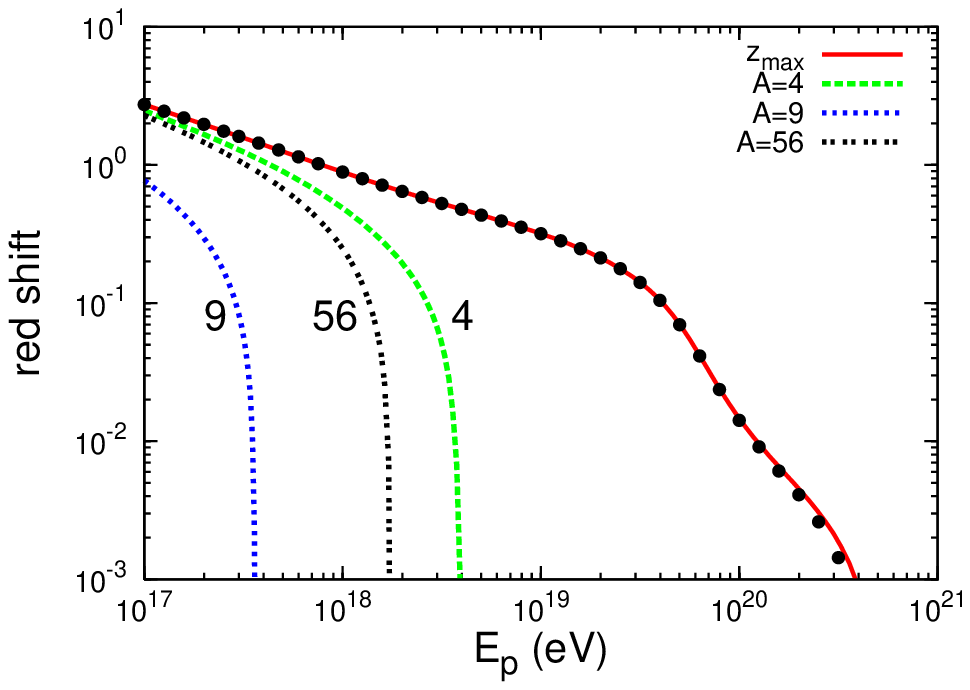}
\caption{Upper and lower redshift limits in Eq. (\ref{eq:n-pA}). 
The lower limits are shown for three proton-associating nuclei  
with A=4,~ 9 and 56 and labelled by these numbers. The highest 
limit for Helium and the lowest for Beryllium are explained by 
the highest and lowest critical Lorentz factors $\Gamma_c$ for 
these nuclei. The lower limits for A= 14,~24,~35 and 40 (not shown 
here) are grouped between A=4 and 56. The upper limit (see text) is 
given by full red curve. The small filled circles show the 
upper limit for the instantaneous approach.}
\label{fig:z-limits}
\end{center}
\end{figure}

The minimum redshift in Eq. (\ref{eq:n-pA}) is calculated as explained
in sections \ref{sec:stability} and \ref{sec:secondary}. If the proton 
Lorentz factor $\Gamma_p < \Gamma_c^{A+1}(z=0)$, the minimum redshift
corresponds to $z_c$ at which the proton Lorentz factor 
$\Gamma_g^p(z_c)$ reaches the critical Lorentz factor of the 
$(A+1)$-nucleus $\Gamma_c^{A+1}(z_c)$.  Note that $z_c=z_{\rm min}$ 
for protons are higher
than for secondary nuclei  (compare Fig. \ref{fig:z-limits} with 
Figs. \ref{fig:z_limitLow} and \ref{fig:z_limitHigh}). It happens
because $\beta_{\rm pair}^A(\Gamma)$ is $Z^2/A \approx A/4$ of that
for proton (see Eq. \ref{eq:lospairs}).
In the case $\Gamma_p > \Gamma_c$,~ $z_{\rm min}=0$.
The maximum upper limit $z_{\rm max}$ corresponds to $A(z)$ subtrajectory 
on which the Lorentz factor of the $A_0$-nucleus reaches 
$\Gamma_{\rm acc}^{\rm max}$ (see Fig. \ref{fig:trajAP}). 
These limits are plotted in Fig. \ref{fig:z-limits}, with $z_{\rm max}$ 
shown by solid red curve. 
The other upper limit shown by small filled circles is valid 
for the {\em instantaneous method} (see below). Here it is important to note
that both limits coincide with great accuracy.

The total space density of secondary protons is given by the summation 
over the primary $A_0$ and all the secondaries $A$ in Eq. (\ref{eq:n-pA}):
\begin{equation}
n_p(\Gamma_p)=\sum_{A\le A_0} n_p^A(\Gamma_p)~~.
\label{eq:fluxNtot}
\end{equation}
Before discussing the numerical results for the calculated flux, 
we present an alternative and more simple method of calculation,
which, being less precise, can be considered as a test at high 
energies for the calculations
described  above.

This approach will be referred to as {\it ``primary-nucleus 
instantaneous decay''}. It is based on the assumption that at the moment of
production $z_g$ the primary nucleus is instantaneously 
photo-disintegrated to $A_0$ nucleons. At large Lorentz factors this 
assumption is well justified because the nucleus lifetime 
$\tau_{A_0}(\Gamma_g)$
is much shorter than all other relevant time scales of the problem. 
We will demonstrate below why this approach works well (with some 
exceptions) at lower Lorentz factors $\Gamma < \Gamma_c$ and 
$\Gamma \ll \Gamma_c$.
To distinguish the previous more precise approach from the instantaneous 
one, we will refer to the former as {\it intermediate A-decay} method.

The proton space density in the {\em instantaneous} approach can be 
written as   
\begin{equation}
n^{\rm inst}_p(\Gamma_p)=A_0 \int_{z_g^{\rm min}(A_0)}^{z_g^{\rm max}}dz_g 
\left | \frac{dt_g}{dz_g} \right | Q_{A_0}(\Gamma_g^p,z_g)
\frac{d\Gamma_g^p}{d\Gamma_p}, 
\label{eq:flux-inst}
\end{equation}
where all notations are as before and $Q_{A_0}$ are given by 
Eq. (\ref{eq:inj1}). 

\begin{figure}[!ht]
\begin{center}
\includegraphics[width=0.9\textwidth,angle=0]{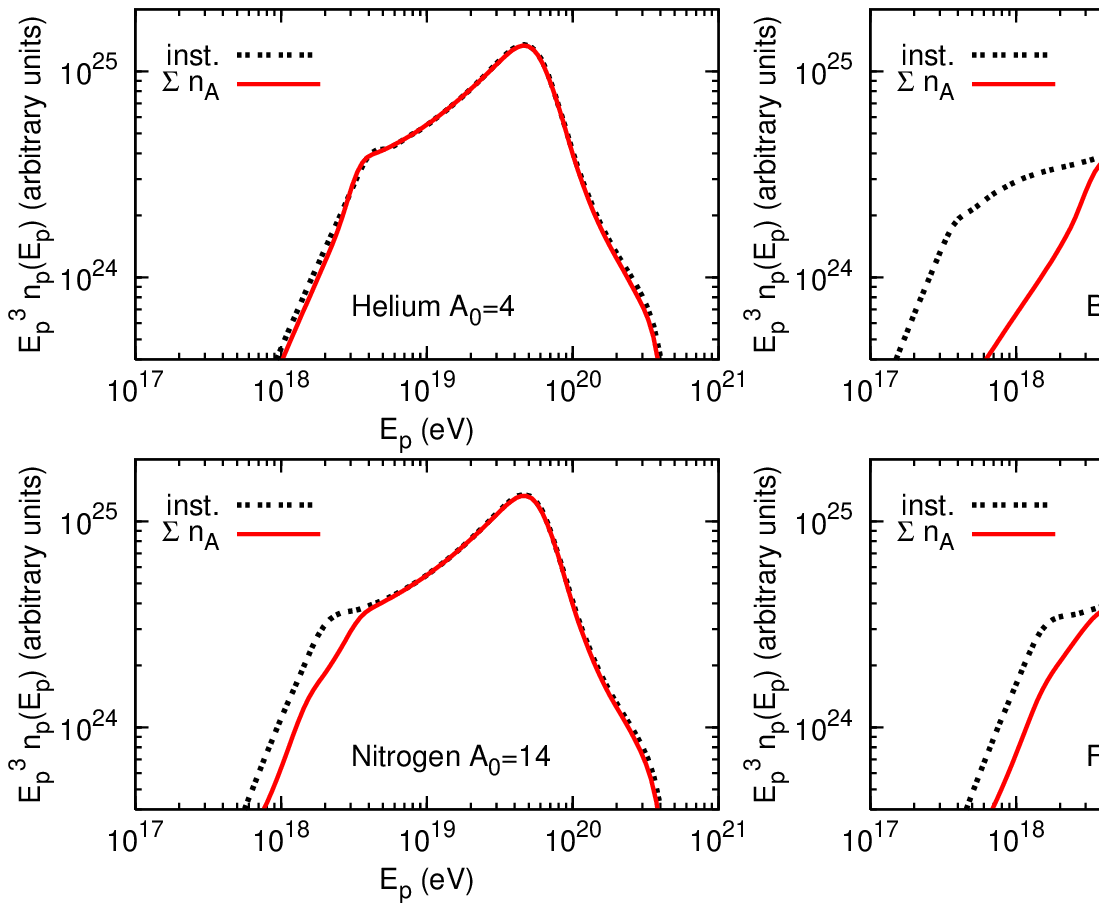}
\caption{Spectra of secondary protons computed by the 
A-intermediate method (continuous red curves) and in the 
instantaneous approximation (black dotted curves) for different $A_0$.
The generation spectral index is $\gamma_g=2.3$ in all cases. 
}
\label{fig:fluxN_Low}
\vspace{1.0 cm}
\includegraphics[width=0.9\textwidth,angle=0]{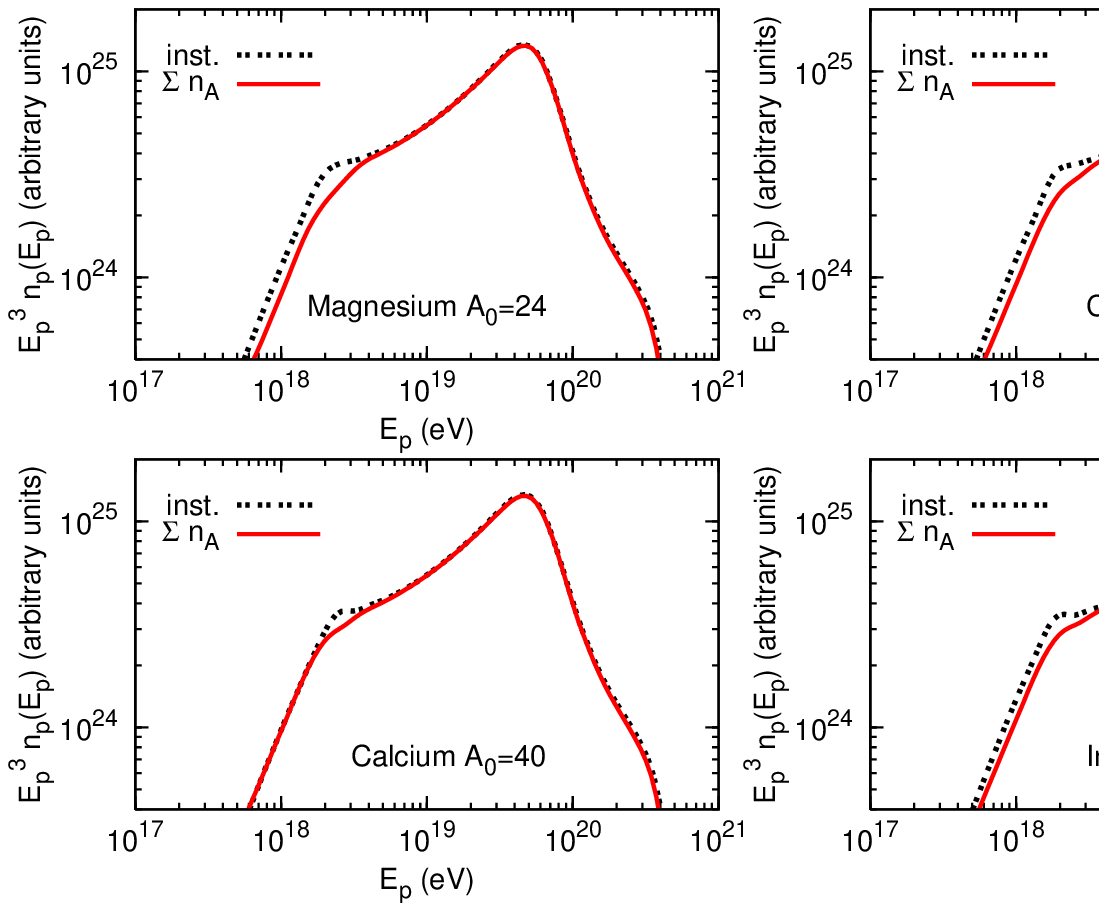}
\caption{The same as in Fig. \ref{fig:fluxN_Low} for heavy nuclei. 
} 
\label{fig:fluxN_High}
\end{center}
\end{figure}

Since $Q_{A_0} \propto 1/A_0$, the factor $A_0$ disappears from 
Eq. (\ref{eq:flux-inst}) and  $n^{\rm inst}_p(\Gamma_p)$ can depend on $A_0$
only through the limits of integration. 
 
After a simple rearrangement of Eq. (\ref{eq:flux-inst}) one obtains 
\beq
n_p^{\rm inst}(\Gamma_p)=\frac{{\mathcal L}_0}{H_0}\frac{\gamma_g-2}{m_N}
\int_{z_g^{\rm min}}^{z_g^{\rm max}} dz_g
\frac{[\Gamma_g^p(\Gamma_p,z_g)]^{-\gamma_g}}
{(1+z_g)\sqrt{\Omega_m(1+z_g)^3 +\Omega_{\Lambda}}} 
\frac{d\Gamma_g^p}{d\Gamma_p}.
\label{eq:flux-inst-f}
\eeq
Eq. (\ref{eq:flux-inst-f}) is very simple for calculations, because it 
involves only the {\em proton trajectory} in $\Gamma - z$ plane 
(subtrajectories $A \rightarrow A_0$ in the right panel of 
Fig. \ref{fig:trajAP} shrink to a point). 
The limits of integration are the same as those shown in 
Fig. \ref{fig:z-limits}. The basic observation involved in the calculation 
of these limits is that the $A_0$ nucleus is now located on the proton 
trajectory $\Gamma_g^p(\Gamma_p,z)$ in Fig. \ref{fig:trajAP}.  
Therefore, $z_g^{\rm min}$ is defined by the equation
$\Gamma_g^p(\Gamma_p,z_{\rm min})= \Gamma_c^{A_0}(z_{\rm min})$, 
which is the same as for nuclei $A$, considered now as $A_0$ in 
Fig. \ref{fig:z-limits}. The upper limit is defined by the condition 
$\Gamma_g^p(\Gamma_p,z_{\rm max})= \Gamma_{\rm acc}^{\rm max}$ and 
it is shown in  Fig. \ref{fig:z-limits} by small filled circles, which 
coincides with upper limit in the intermediate A method.  
We see, thus, that the upper limit $z_{\rm max}$ does not depend on 
$A_0$. The lower limit at $\Gamma_p \geq \Gamma_c^{A_0}$ 
is $z_{\rm min}=0$. At $\Gamma_p \leq \Gamma_c$ the lower limit
depends on $A_0$, but weakly, with the exceptional case of Beryllium. 

Thus, at high energy when 
$z_{\rm min}=0$ the flux has a universal form independent of the 
primary  nucleus, i.e. of  $A_0$. It is natural to expect that the
high energy regime starts from  $\Gamma_p > \Gamma_c^{A_0}$ 
when $z_{\rm min}=0$, and the calculations below confirm this expectation.

The high-energy regime is of particular interest as 
a test of the {\em intermediate A-decay} method. In this case  
the proton flux in {\em instantaneous approximation} $n_p^{\rm inst}$
must coincide with the intermediate $A$ method, when the flux is 
given by $\sum n_p^A$ , i.e. by Eq. (\ref{eq:fluxNtot}). 
It follows also from Figs. \ref{fig:LosseLow} and 
\ref{fig:LosseHigh}, which show that at $\Gamma > \Gamma_c$ the
photo-disintegration rate strongly dominates over the rate of the 
Lorentz-factor loss. 
Figs. \ref{fig:fluxN_Low} and \ref{fig:fluxN_High} reliably confirm 
the agreement of both calculations at $\Gamma > \Gamma_c$, 
demonstrating thus that the intermediate $A$ method has passed this test.

We address now the question of the agreement between A-intermediate and
instantaneous methods at $\Gamma < \Gamma_c$ and $\Gamma \ll \Gamma_c$. 
The explanation follows from the explosive character of $A(z)$ 
trajectories for 
$A \rightarrow A_0$ evolution. Using the explosive-trajectory result 
$z_A \approx z_g$ and $\Gamma_A \approx \Gamma_g$ in Eq. (\ref{eq:n-pA}),
one obtains after simple calculations 
\beq
n_p^A(\Gamma_p)=\frac{\gamma_g-2}{A_0 m_N}
\frac{{\mathcal L}_0}{H_0}
\int_{z_g^{\rm min}}^{z_g^{\rm max}} dz_g
\frac{[\Gamma_g^p(\Gamma_p,z_g)]^{-\gamma_g}}
{(1+z_g)\sqrt{\Omega_m(1+z_g)^3 +\Omega_{\Lambda}}} 
\frac{d\Gamma_g^p}{d\Gamma_p}.
\label{eq:npA-explos}
\eeq
The total flux of protons is obtained by summation in 
Eq. (\ref{eq:npA-explos}) over all $A$. Since limits of integration 
depend weakly on $A$, it means multiplication of 
Eq. (\ref{eq:npA-explos}) to $A_0$, which results in instantaneous 
flux given by Eq. (\ref{eq:flux-inst-f}).

We shall discuss now the calculated  spectra presented in 
Figs. \ref{fig:fluxN_Low} and \ref{fig:fluxN_High}.

One may observe that agreement between the exact 
($A$-intermediate) and instantaneous method of calculations is 
precise at high energies $(\Gamma > \Gamma_c)$ and approximate 
at low energies ( $(\Gamma < \Gamma_c)$ and $(\Gamma \ll \Gamma_c)$),  
as it is expected.

For small $A$ the excellent agreement at low energies is observed for  
$^4$He. This is an exceptional case: the energy threshold of 
photo-disintegration for $^4$He is much higher than for its daughter
nuclei: $^3$He and D. Hence, when $^4$He starts to photo-disintegrate,
its daughter-nuclei disintegrate too. Based on this argument we expect 
that in the Helium case the spectra of secondary protons 
calculated by both methods should be identical at all Lorentz factors,
including the low ones, and this expectation is confirmed by 
the upper-left panel of Fig. \ref{fig:fluxN_Low}. 

We  now come over to the "Beryllium excess" in the spectrum of 
the instantaneous decay, shown in the upper-right panel of 
Fig. \ref{fig:fluxN_Low}. The nature of this effect is simple:  
$\Gamma_c^{\rm Be} \ll \Gamma_c^{\rm He}$ and since $^4$He  
is born with Lorentz factor of $^9$Be, the former can be stable
and secondary protons are not produced. For $\Gamma > \Gamma_c^{\rm Be}$ 
it gives a suppression of the secondary proton flux. Under the assumption
of an instantaneous decay this effect is absent.  

Beryllium effect works for all $A_0 > 9$ and since it affects only 8 protons 
from two $^4$He nuclei, its influence on heavy nuclei is weaker, as one 
observes from Fig. \ref{fig:fluxN_High}.

The shape of the proton spectra in Figs. \ref{fig:fluxN_Low} and 
\ref{fig:fluxN_High} can be  naturally explained. They have two spectral 
breaks:  the high energy steepening, which  is the usual GZK cutoff at energy 
$E \sim 5\times 10^{19}$~eV, and the low-energy steepening which coincides 
with $\Gamma_c$:   
$\sim 4\times 10^{9}$ for $^4$He, $\sim 4\times 10^{8}$ for $^9$Be etc). 
Below $\Gamma_c$ the adiabatic energy losses dominate 
and it explains the flat spectrum $n_p(E_p)$ there,  i.e. steep 
$E^3 n_p(E)$ spectrum in Figs.~\ref{fig:fluxN_Low} and \ref{fig:fluxN_High}.
It is different from $\propto E^{-\gamma_g}$ because $z_{\rm min}$ in 
Eq. ({\ref{eq:n-pA}) and Eq. (\ref{eq:flux-inst-f}) depends on $\Gamma$, 
i.e. on $E_p$.

We complete this section with a remark about the universality of the
secondary-proton spectrum. Using Eq. (\ref{eq:flux-inst-f}) 
for the instantaneous spectrum we proved its universality. i.e. 
independence of $A_0$.  The intermediate A spectrum 
has the same property, as it was proved above. In other words 
the normalised spectrum calculated for one $A_0$ is
valid for any other $A_0$ with the same normalization. 

\subsection{Primary nuclei}
\label{sec:prim.nuc}

\begin{figure}[!ht]
\begin{center}
\includegraphics[width=0.49\textwidth,angle=0]{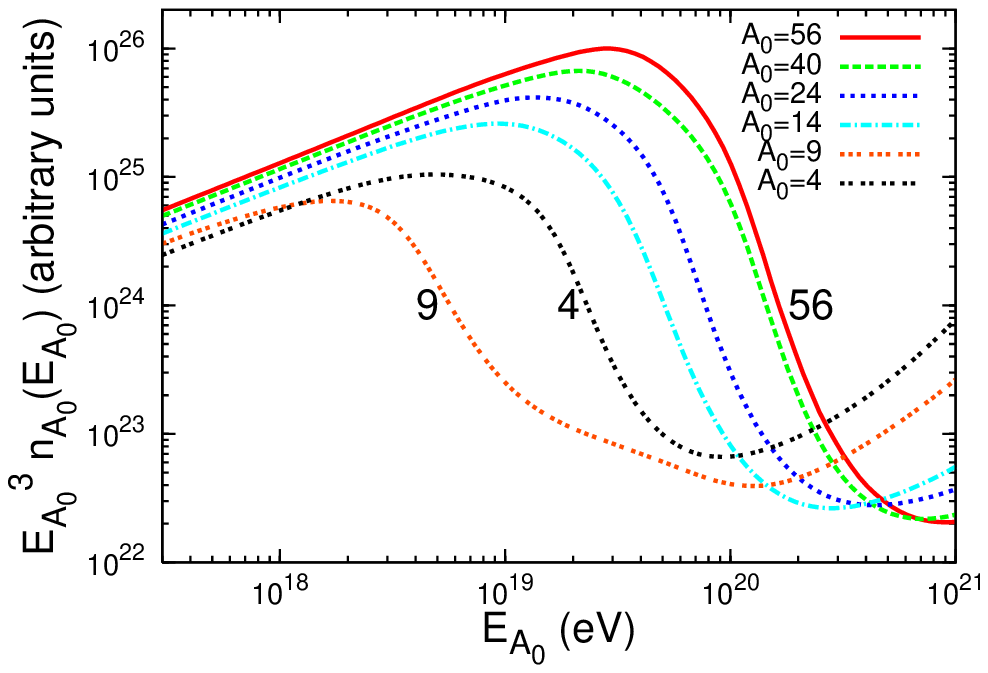}
\includegraphics[width=0.49\textwidth,angle=0]{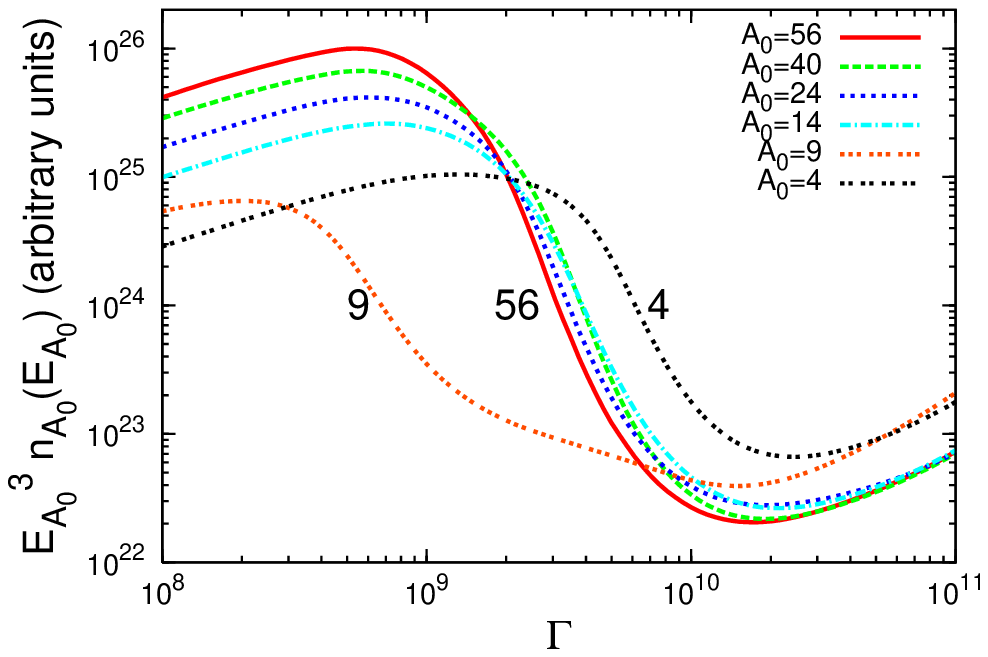}
\caption{Spectrum of primary nuclei with various $A_0$ 
as function of energy (left panel) and Lorentz factor (right panel).}
\label{fig:fluxA0}
\end{center}
\end{figure}

We calculate now the diffuse flux of primary nuclei $A_0$, which arrive 
undestroyed from the sources. The comoving space density of 
these nuclei $n_{A_0}(\Gamma,t)$ is described by the kinetic equation 
\begin{equation}
\frac{\partial n_{A_0}(\Gamma,t)}{\partial t} - 
\frac{\partial }{\partial \Gamma}\left [ b_{A_0}
(\Gamma,t) n_{A_0}(\Gamma,t) \right ] +
 \frac{n_{A_0}(\Gamma,t)}{\tau_{A_0}(\Gamma,t)} = 
Q_{A_0}(\Gamma,t) ~~, 
\label{eq:kin2}
\end{equation}
where $Q_{A_0}(\Gamma,t)$ is the generation rate of primary nuclei 
per unit time and comoving volume, and $\tau_{A_0}(\Gamma,t)$ is the 
photo-disintegration lifetime of the nucleus $A_0$ calculated as 
$[A_0\beta_{\rm dis}(A_0,\Gamma,z)]^{-1}$ with $\beta_{\rm dis}$ 
given by Eq. (\ref{eq:losdis}). 

As found in appendix \ref{app:solution} the solution of the kinetic equation  
(\ref{eq:kin2}) reads 
\begin{equation}
n_{A_0}(\Gamma)=\int_0^{\infty} dz_g \left | \frac{dt_g}{dz_g} 
\right | Q_{A_0}(\Gamma_g,z_g) 
\frac{d\Gamma_g}{d\Gamma} e^{-\eta(\Gamma_g,z_g)}
\label{eq:fluxA0}
\end{equation}
where $n_{A_0}(\Gamma)$ is the space density of nuclei $A_0$ at $z=0$, 
$\Gamma_g={\mathcal G}(A_0,\Gamma,z_0=0,z_g)$ is the Lorentz factor of 
nuclei $A_0$ at the moment of generation $z_g$ calculated by the evolution 
trajectory, with {\it fixed} $A_0$, which starts from $\Gamma$ at $z_0=0$. 
The ratio  $d\Gamma_g/d\Gamma$ is also calculated with fixed $A_0$.  
The quantity $\eta$ 
takes into account the photo-disintegration of the 
propagating nucleus: 
\beq
\eta(\Gamma_g,z_g)=\int_{t(z_g)}^{t_0}\frac{dt}{\tau_{A_0}(\Gamma(t),t)}~.
\label{eq:eta-t}
\eeq
The lower limit of integration in Eq. (\ref{eq:fluxA0}) reflects the 
assumption of an homogeneous distribution of the sources. The upper 
limit is imposed by the factor $\exp(-\eta)$ accompanied by the condition
of a maximum acceleration Lorentz factor: $Q_{A_0}(\Gamma_g)=0$,~ if 
$\Gamma_g \geq \Gamma_{\rm max}^{\rm acc}$. 

In Fig. \ref{fig:fluxA0} we plot the flux of primary nuclei for
various $A_0$, using the injection spectrum with $\gamma_g=2.3$. 
The steepening of the spectra in Fig. \ref{fig:fluxA0} starts 
from the Lorentz factor where pair-production energy losses are equal to 
adiabatic energy losses. Comparison of Figs. \ref{fig:LosseHigh}
and \ref{fig:fluxA0} confirms it. It is worth noting that this 
criterion for the beginning of the steepening was obtained first in 
\cite{BGZ75}.

\section{Coupled kinetic equations (CKE)}
\label{sec:coupled-kinetic}
In this section we present a straightforward method of
calculation for UHE nuclei fluxes, based on the analytic 
solution of a complete set of coupled kinetic equations. The great 
advantage of this method is a simple expression for the generation rates 
of secondary nuclei and protons, which do not include integration with 
independently determined lower and upper limits. We consider as before 
the expanding universe filled homogeneously by accelerated primary and 
secondary nuclei, including the secondary protons. 
As discussed in section \ref{sec:losses}, these secondary particles
are produced mainly by one-nucleon photo-disintegration process 
$(A+1) + \gamma \to A + p$. In this approximation one can 
write the generation rate for any secondary nucleus $A$ as a function 
only of the equilibrium distribution of the parent nucleus $(A+1)$, namely:
\beq
Q_A(\Gamma,z)=\frac{n_{A+1}(\Gamma,z)}{\tau_{A+1}(\Gamma,z)},
\label{eq:couplQ_A}
\eeq
where $\tau_A(\Gamma,z)$ is given by Eq. (\ref{eq:tau(z)}) and 
(\ref{eq:losdis}). Taking it into account one may write a set 
of coupled kinetic equations which describes the equilbrium of primary
nuclei $A_0$ with the products of their decay
\begin{eqnarray}
\nonumber
\frac{\partial n_{A_0}(\Gamma,t)}{\partial t} -
\frac{\partial}{\partial \Gamma}
\left [ n_{A_0}(\Gamma,t) b_{A_0}(\Gamma,t) \right ] +
\frac{n_{A_0}(\Gamma,t)}
{\tau_{A_0}(\Gamma,t)} & = & Q_{A_0}(\Gamma,t) \\
\nonumber
\frac{\partial n_{A_0-1}(\Gamma,t)}{\partial t} -
\frac{\partial}{\partial \Gamma}
\left [ n_{A_0-1}(\Gamma,t) b_{A_0-1}(\Gamma,t) \right ] +
\frac{n_{A_0-1}(\Gamma,t)}
{\tau_{A_0-1}(\Gamma,t)}  & = &
\frac{n_{A_0}(\Gamma,t)}{\tau_{A_0}(\Gamma,t)} \\
\nonumber
\\
& \vdots &  \label{eq:coupled}   \\
\nonumber
\\
\nonumber
\frac{\partial n_{A}(\Gamma,t)}{\partial t} - \frac{\partial}{\partial \Gamma}
\left [ n_{A}(\Gamma,t) b_{A}(\Gamma,t) \right ] + \frac{n_{A}(\Gamma,t)}
{\tau_{A}(\Gamma,t)}  & = & \frac{n_{A+1}(\Gamma,t)}{\tau_{A+1}(\Gamma,t)}\\
\nonumber
\end{eqnarray}
The solution of the kinetic equation for primaries $A_0$ is given by   
\begin{equation}
n_{A_0}(\Gamma,z)=\int_{z}^{z_{max}} \frac{dz'}{(1+z') H(z')}
Q_{A_0}(\Gamma ',z')
\frac{d\Gamma '}{d\Gamma} e^{-\eta_{A_0}(\Gamma ',z')}, 
\label{eq:n_A0}
\end{equation}
with 
\begin{equation}
\eta_{A_0}(\Gamma ',z ') = \int_{z}^{z'} \frac{dz''}{(1+z'')H(z'')}
\frac{1}{\tau_{A_0}(\Gamma '', z'')} ,
\label{eq:etaA0}
\end{equation}
and the solution for an arbitrary secondary nuclei $A$ is 
\begin{equation}
n_A(\Gamma,z)=\int_{z}^{z_{max}} \frac{dz'}{(1+z')H(z')} 
\frac{n_{A+1}(\Gamma',z')}{\tau_{A+1}(\Gamma',z')}
\frac{d\Gamma'}{d\Gamma} e^{-\eta_{A}(\Gamma',z')},
\label{eq:nA-solut}
\end{equation}
with
\begin{equation}
\eta_{A}(\Gamma ',z ') = \int_{z}^{z'} \frac{dz''}{(1+z'')H(z'')}
\frac{1}{\tau_{A}(\Gamma '', z'')}.
\label{eq:etaA}
\end{equation}
In each of these equations we assume $A=const$ and the Lorentz factor
changing due to adiabatic and pair-production energy losses. 
Accordingly, the ratio $d\Gamma'/d\Gamma$ is given by  
\begin{equation}
\frac{d\Gamma'}{d\Gamma}=\frac{1+z'}{1+z}
\exp\left [ \frac{Z^2}{A}\int_{z}^{z'} \frac{(1+z'')^2 dz''}{H(z'')} 
\left (\frac{d b_0^p (\tilde{\Gamma})}{d\tilde{\Gamma}} 
\right )_{\tilde{\Gamma}=(1+z'')\Gamma''} \right ] ,
\label{eq:ratio-dGamma}
\end{equation}
where 
$b_0^p(\Gamma)=d\Gamma /dt$ is the Lorentz-factor loss per unit time 
for protons at $z=0$ due to pair production. 

The procedure to solve the system of equations (\ref{eq:coupled}) 
consists in 
finding the analytic solution $n_{A_0}(\Gamma,t)$ for the first
equation in the form (\ref{eq:n_A0}), and putting it into the second 
equation. Taking the solution of this second equation, given by  
Eq. (\ref{eq:nA-solut}) with $A=A_0-1$, one finds $n_{A_0-1}(\Gamma,t)$.
Continuing this procedure one obtains the space density (fluxes) for  
all secondary nuclei $n_A(\Gamma,t)$. 

Note, that in this method there is no problem with the limits of integration:
the lower limit is $z$ by definition and the upper limit is regulated by the 
factor $e^{-\eta}$, while $z_{max}$ here plays only a symbolic role. 
The generation rate $Q_A(\Gamma,z)$ given by Eq. (\ref{eq:couplQ_A}) tends to
zero at small $\Gamma$ due to factor $\tau_{A+1} \to \infty$ at small 
Lorentz factors. 

For the calculations of {\em secondary protons} spectrum we use the 
$A$-associating proton approach developed in section \ref{sec:prot}.
The secondary nucleons $N$ which accompany production of $A$-nuclei 
in the process $(A+1) \rightarrow A+N$ have the same generation rate 
\beq
Q_p^A(\Gamma,z) = Q_A(\Gamma,z) = 
\frac{n_{A+1}(\Gamma,z)}{\tau_{A+1}(\Gamma,z)} .  
\label{eq:p-generation}
\eeq
The kinetic equation for secondary protons is given by Eq. (\ref{eq:kin_p})
with the solution described by Eq. (\ref{eq:flux-pA}). Using the
generation rate in the form of Eq. (\ref{eq:p-generation}) this solution reads 
\beq
n_p^A(\Gamma,z)=\int_z^{z_{\rm max}}\frac{dz'}{(1+z')H(z')}
\frac{n_{A+1}(\Gamma',z')}{\tau_{A+1}(\Gamma',z')}
\frac{d\Gamma'(z')}{d\Gamma} ,
\label{eq:p-coupled-solut}
\eeq
where $\Gamma'(z')=G_p(\Gamma,z,z')$ is the proton trajectory, and 
Eq. (\ref{eq:p-coupled-solut}) is valid for any arbitrary initial $z$
including $z=0$. In fact, the actual lower limit of integration 
$z_{\rm min}$ is automatically provided by $\tau_{A+1}(\Gamma',z')$, 
and this gives a great advantage of the CKE method in comparison
with the combined method, where $z_{\rm min}$ is calculated 
independently, using some particular features of the trajectory
evolution. As will be seen by the example of secondary $A$-nuclei, the 
brothers of secondary protons, the structure of the lower-limit  
cutoff exposed by Eq. (\ref{eq:low-limit-coupled}) is quite different
from the sharp cutoff at $z_{\rm min}$ introduced in the combined method. 

The upper limit of integration $z_{\rm max}$ in Eq. (\ref{eq:p-coupled-solut})
is quite different from secondary nuclei, where it is imposed by 
the life-time $\tau_A$ in the form of $e^{-\eta_A}$. In the case of secondary
protons the upper limit is provided by $\Gamma_{\rm max}$, which the brother 
$A$-nucleus (or parent $A+1$ nucleus) is allowed to have. Since at
these very large Lorentz factors the approximation of explosive
trajectories is fully justified, we can use $\Gamma_{\rm max}$ for 
primary nucleus $A_0$ and $z_{\rm max}$ is given (in case of initial 
$z=0$) by the equation for the proton trajectory 
$\Gamma_{\rm max}=G_p(\Gamma,0,z_{\rm max})$, where $\Gamma$ is the
proton Lorentz factor at $z=0$.

As we discuss below, the fluxes obtained by the CKE 
method for secondary nuclei and protons differ, most notably at low 
energies, from that in the combined method
(see section \ref{sec:comb}). This difference is due to the lower limit 
of integration in the Eq. (\ref{eq:nA-solut}) and (\ref{eq:p-coupled-solut}).
In the combined method it is sharply 
fixed at $z_{min}$ as discussed in section \ref{sec:secondary}, 
while in the CKE method the integrand is automatically suppressed 
by the term $\tau_{A+1}(\Gamma',z')$ (see discussion below). 

The fluxes of secondary nuclei and protons are displayed in 
Fig. \ref{fig:secAp-coupled}.

\begin{figure}[!ht]
\begin{center}
\includegraphics[width=0.49\textwidth]{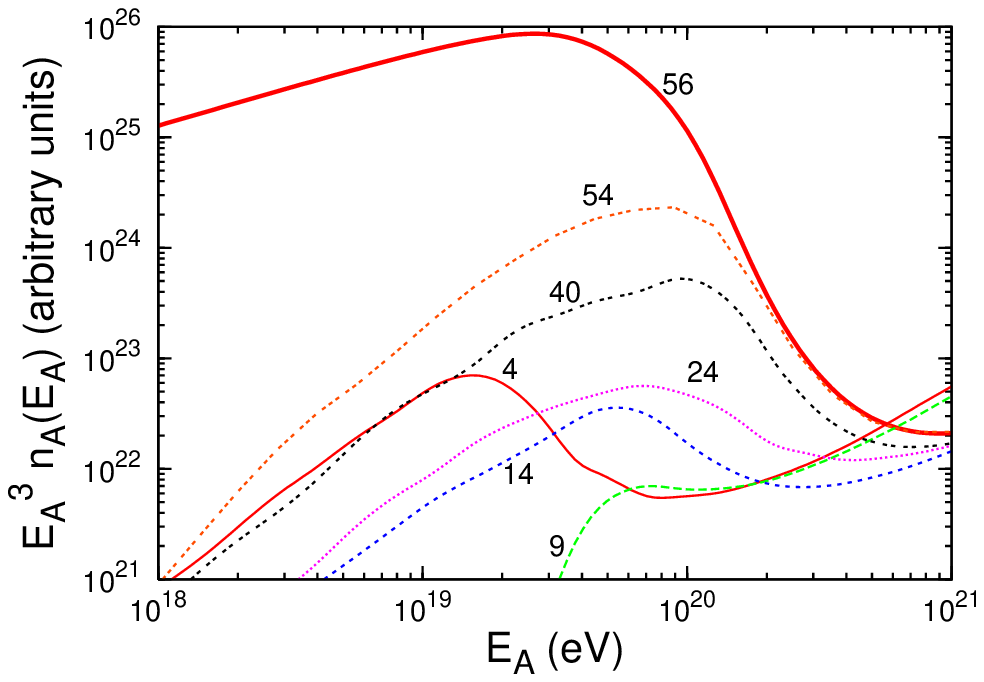}
\includegraphics[width=0.49\textwidth]{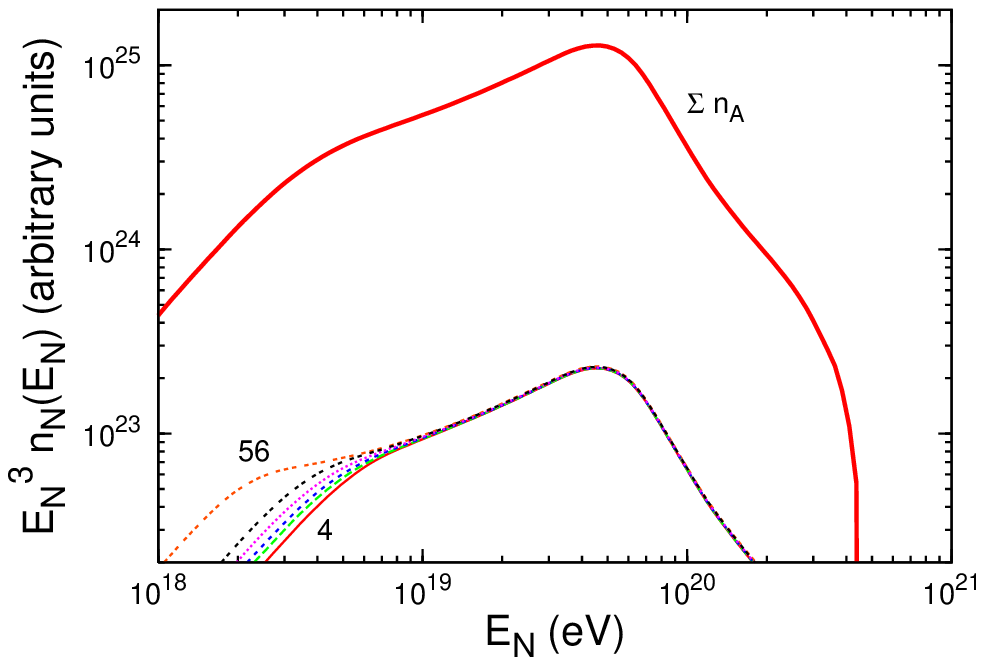}
\caption{Flux of secondary nuclei (the primary $A_0=56$ included)
(left panel) and of the secondary protons (right panel) as function of 
energy. The numbers on the curves show A. $\Sigma_A$ is related with
the total flux of protons produced by the primaries $A_0=56$, the
numbers indicate fluxes for each $A$-associating proton. The discussion
of these results and comparison with the combined method is given in
the next section.
}
\label{fig:secAp-coupled}
\end{center}
\end{figure} 

We will discuss now the solutions obtained for secondary nuclei 
and protons.
In particular we will study these solutions in their analytic form to compare
them with the combined method. 

First we obtain the analytic solutions of Eqs. (\ref{eq:coupled}) 
in the high-energy asymptotic regime, when all $\tau_A$ are very short. 
We re-write the solution (\ref{eq:n_A0}) in terms of the cosmological time 
using for $e^{-\eta}$ the expression of Eq. (\ref{eq:app-solut2}) 
(see appendix \ref{app:solution}):  
\beq
n_{A_0}(\Gamma,t)= \int_{t_g}^t dt'Q_{A_0}(\Gamma',t')\frac{d\Gamma'}
{d\Gamma}\exp \left [ -\int_{t'}^t\frac{dt''}{\tau_{A_0}(\Gamma'',t'')}
\right ].
\label{eq:nA0-time}
\eeq
Introducing the propagation time $t_{\rm prop}=t-t'$ as integration 
variable in Eq. (\ref{eq:nA0-time}) and taking $Q_{A_0}d\Gamma'/d\Gamma$ 
out of the integral as a slowly variable quantity, one obtains after simple 
calculations 
\beq
n_{A_0}(\Gamma,t)=Q_{A_0}(\Gamma,t)\tau_{A_0}(\Gamma,t) ,
\label{eq:nA0-asymp}
\eeq
and 
\beq
Q_{A_0-1}(\Gamma,t)=n_{A_0}(\Gamma,t)/\tau_{A_0}(\Gamma,t)= Q_{A_0}(\Gamma,t).
\label{eq:QA-asymp}
\eeq
Repeating these calculations for $A < A_0$ we obtain for the high-energy 
asymptotic regime 
\beq
Q_A(\Gamma,t)=Q_{A+1}(\Gamma,t)= ... =Q_{A_0}(\Gamma,t)
\label{eq:Q-equality}
\eeq
and 
\beq
n_A(\Gamma,t)=Q_A(\Gamma,t)\tau_A(\Gamma,t)= 
Q_{A_0}(\Gamma,t)\tau_A(\Gamma,t),
\label{eq:nA-asymp}
\eeq
i.e. the same high-energy regime that we obtained in
Eqs. (\ref{eq:n_Aasympt}) and (\ref{eq:ratio}) in the combined
method. In particular, the equality (\ref{eq:Q-equality}) is the same
obtained in the combined method for the explosive regime, which is provided
by short $\tau_A$ like in the case above. 

We will now come over to the general case valid also for low $\Gamma$. 
To obtain the general solution $n_A(\Gamma,t)$ in the analytic form 
we put into Eq. (\ref{eq:nA-solut}) the value $n_{A+1}(\Gamma',z')$  
determined from the preceding equation
\beq
n_{A+1}(\Gamma',z')=\int_{z'}^{z_{max}} \frac{dz''}{(1+z'')H(z'')} 
\frac{n_{A+2}(\Gamma'',z'')}{\tau_{A+2}(\Gamma'',z'')}
\frac{d\Gamma''}{d\Gamma'} e^{-\eta_{A}(\Gamma'',z'')}.
\label{eq:n(A+1)}
\eeq
Repeating this procedure with increasing $A$ until we reach $A_0$,
we obtain the general expression with many-fold integral 
\beq 
n_A(\Gamma,z) = \left [ \prod_{i} \int_{z_i}^{z_{max}} dz_i
\frac{e^{-\eta_{A_0-i}(\Gamma_i,z_i)}}{(1+z_i)H(z_i)\tau_{A_0+1-i}
(\Gamma_i,z_i)} \right ] \int_{z_1}^{z_{max}} dz' \frac{ Q_{A_0}
(\Gamma',z')}{(1+z')H(z')}
\frac{d\Gamma'}{d\Gamma}e^{-\eta_{A_0}(\Gamma',z')},
\label{eq:nA-analytic}
\eeq
where $d\Gamma'/d\Gamma$ is taken along the $A$-variable trajectory. 
Using Eq. (\ref{eq:inj1}) for $Q_{A_0}(\Gamma',z')$ and assuming that 
the external integration results in some functions $f(\Gamma,z)$ 
(we calculated these functions numerically) we obtain
\begin{equation}
n_A(\Gamma,0)=\int_{0}^{z_{max}} dz f(z) \int_z^{z_{\rm max}}
\frac{dz'}{(1+z')H(z')} {\cal L}_0 \frac{\gamma_{g}-2}{A_0 m_N} 
\left [ \Gamma_g(\Gamma,A,z') \right ]^{-\gamma_g} 
\frac{d\Gamma_g}{d\Gamma} e^{-\eta_{A_0}(\Gamma_g,z')},
\label{eq:low-limit-coupled}
\end{equation}
with $f(z)$ rapidly vanishing when $z \to 0$ and $z \to z_{max}$. }

The presentation of our solution as a manyfold integral allows 
us to discuss whether one should classify this solution as analytic 
or implicit-analytic, as in the first paper of \cite{Sarkar} is 
formulated. We follow the definition of the analytic solution used 
in many  monographs and text books: analytic solution is the one 
which can be presented by a {\em finite number} of quadratures,
or a solution which has the form of a {\em finite number} 
of successive integrations. With these definitions 
Eqs.~(\ref{eq:nA-analytic}) and (\ref{eq:low-limit-coupled}) 
demonstrate that  the solutions we obtain are analytic. 

Comparison of the solutions (\ref{eq:nA-analytic}) and 
(\ref{eq:low-limit-coupled}) with $n_A(\Gamma,0)$ given by 
Eq. (\ref{eq:fluxA-inst}) for the combined method shows 
similarity and differences. 
The crucial one consists in $z_{\rm min}$. While in the combined
method this is a well determined function of $\Gamma$, in the CKE
method the lower limit is given by different values
of $z$ weighted by the function $f(z)$. 

Numerical calculations show a very good agreement between the two
methods at the highest energies and for all primary nuclei $A_0$ at all 
energies. For secondary protons the disagreement is moderate, but 
for the secondary nuclei there is a strong disagreement at low energies. 
We see the main reason for this disagreement in the  
$z_{\rm min}(\Gamma)$ calculations. In the combined model $z_{\rm min}$
is calculated from reasonable physics connected with trajectories. 
In the CKE method the lower limit appears automatically, without any
assumption, directly from the kinetic equations. We think that as far as
numerical results are concerned, this last method should be trusted more.
  
\section{Comparison with other computation schemes}
\label{comparison}

In this section we will compare the CKE spectra with those present 
in literature and obtained 
by different methods. In paper II we will present 
such comparison in a more detailed way taking into account the total
background radiation, given by the sum of EBL and CMB, here we will 
restrict our study to the case of CMB only.
This comparison cannot be done straightforwardly, 
because in all available calculations EBL and CMB are not used separately. 

We solve this problem introducing three criterions of agreement. 

\begin{figure}[!ht]
\begin{center}
\includegraphics[width=0.49\textwidth]{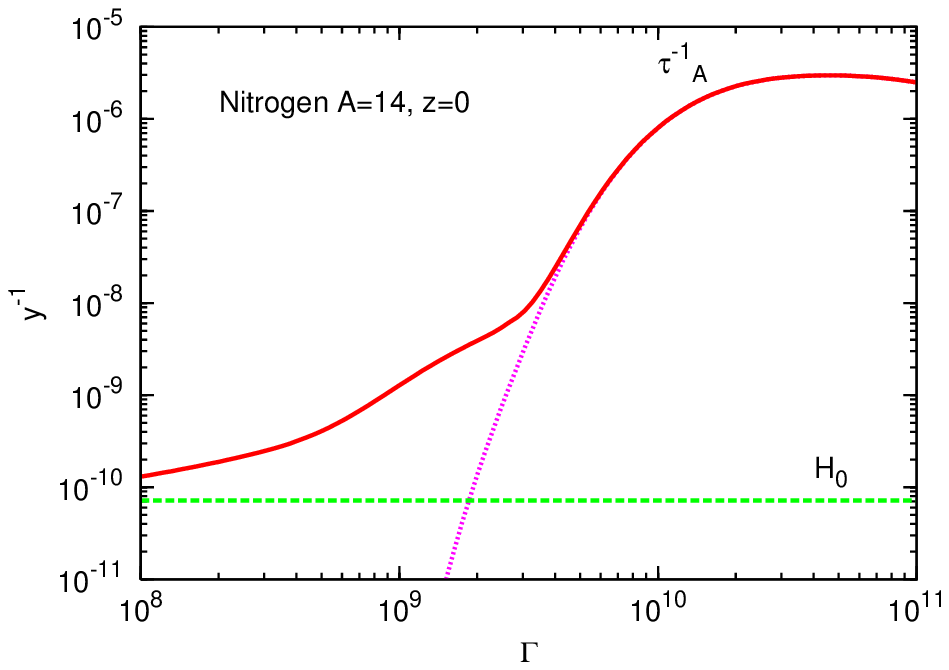}
\includegraphics[width=0.49\textwidth]{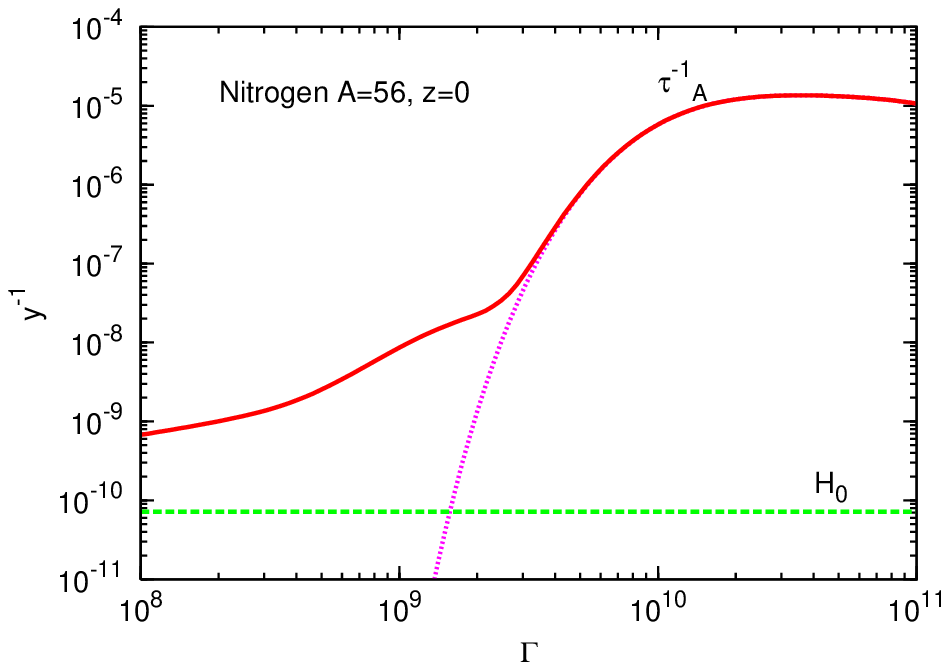}
\caption{Inverse photo-disintegration lifetime $\tau_A^{-1}$ for 
Nitrogen (left panel) and Iron (right panel) nuclei for CMB background 
only (dotted magenta curve) and for CMB+EBL (continuos red curve). The merging 
Lorentz-factor is defined as intersection of these two curves.
}
\label{fig16}
\end{center}
\end{figure} 

In Figs \ref{fig16} taken from paper II the photo-disintegration lifetimes 
$\tau_A$ are given for EBL and CMB separately (paper II and references therein). 
For convenience we 
present these graphs here in Fig.~\ref{fig16} for Iron (right panel) and 
Nitrogen (left panel). 
One can see that lifetimes $\tau_A$ for EBL and CMB are merged sharply 
at the critical Lorentz factor $\Gamma_m$ equal to $3\times 10^9$ and 
$4\times 10^9$ for Iron and Nitrogen, respectively; the corresponding 
energies are $1.7\times 10^{20}$~eV for Fe and $5.6 \times 10^{19}$~eV 
for N. Therefore, the {\em first criterion} of agreement between spectra 
for EBL+CMB (literature) and CMB only (our calculations) is given by the
merging energy $E_m=A\Gamma_m m_N$ taken from Fig.~\ref{fig16}.
The {\em second criterion} is the agreement of the spectra with CMB only and
with EBL+CMB above the merging energy, because at these energies 
photo-disintegration on CMB strongly dominates. 

The {\em third criterion} of agreement is given by the secondary proton 
spectrum.  EBL produces secondary protons with Lorentz-factors 
below $\Gamma_m$, i.e. with low energies 
$E_p \lsim \Gamma_m m_N \sim 3\times 10^{18}$~eV. Protons 
with higher energies are produced on CMB only. Therefore, the  agreement 
of secondary protons spectra on EBL+CMB (literature) and on CMB only 
(our calculations) at $E_p \gsim \Gamma_m m_N$ 
gives one more proof of the discussed agreement. 

Let us now come over to the explicit comparison of our spectra with those 
in literature. The most interesting case for us is the comparison with 
the results of \cite{Gelmini07,Kalashev}, where the same 
kinetic equations are used. The essential difference is that in 
our work the kinetic equations are solved analytically, while 
in \cite{Gelmini07,Kalashev} solutions are obtained numerically.
Thus this comparison can be considered as a numerical test of the CKE method.

\begin{figure}[!ht]
\begin{center}
\includegraphics[width=0.49\textwidth]{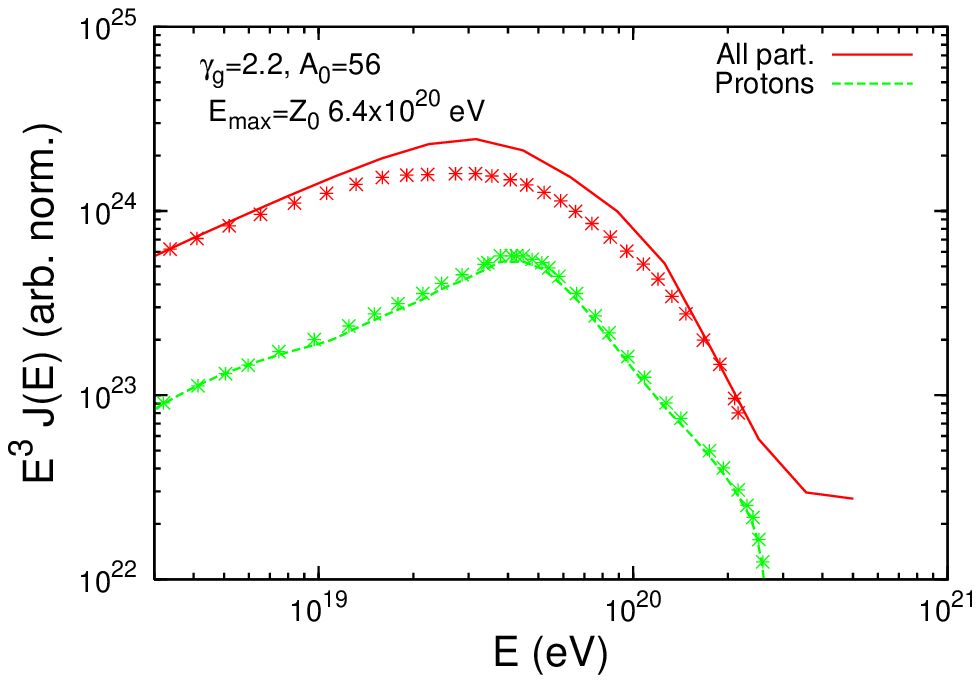}
\includegraphics[width=0.49\textwidth]{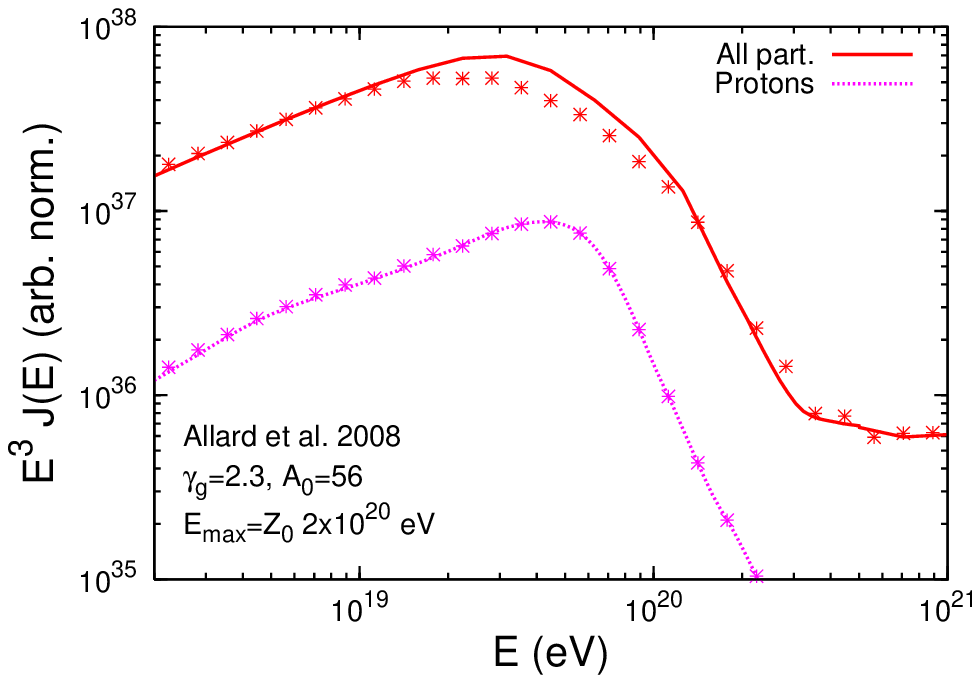}
\caption{{\em Left panel}:Comparison of energy spectra with 
CMB only (our calculations) with \cite{Gelmini07}, where 
both EBL and CMB are included. Primaries at the source are Iron nuclei.  
Solid red line shows all-particles spectrum from our calculations, 
the red asterisks present all-particles spectrum from \cite{Gelmini07}.
One may see the EBL suppression of the latter flux.
Both fluxes merge at $1.5\times 10^{20}$~ eV as expected. 
The secondary-proton spectra in the lower part of the figure 
coincide too. See the text for more details.
{\em Right panel}: Comparison of spectra with CMB only (our 
calculations) with MC simulation by  Allard et al \cite{Allard08} 
which includes EBL and CMB.  In the upper part of the figure the 
all-particles spectra are compared, in the lower part - that of
secondary protons. The solid lines show our calculations and asterisks
- calculations by Allard et al.
}
\label{fig17}
\end{center}
\end{figure}

In the left panel of Fig.~\ref{fig17} we show the comparison of our 
analytic calculations (using only CMB) with the numerical solution of the 
kinetic-equation as calculated in \cite{Gelmini07}, where EBL is also included. 
The parameters used in these calculations are: $\gamma_g=2.2$  
and $E_{max}=6.4\times 10^{20}~Z_0 $ eV, where $Z_0$ is the atomic charge 
number of the injected particles. Computations are performed assuming a 
homogeneous distribution of sources with a pure Iron injection.

The full red line shows the all-particles spectrum in our work (CMB only) 
and red asterisks show the spectrum \cite{Gelmini07} produced on EBL+CMB. 
One may notice the suppression of latter spectrum by photo-disintegration 
on EBL. These two spectra merge at $1.5\times 10^{20}$~ eV, as expected. 
The secondary protons spectrum shown by green dashed line (our calculations)
and by green asterisks \cite{Gelmini07} coincide well too.

We will compare now our spectra with MC simulation using the computation
by  Allard et al \cite{Allard08}. In this work the accelerated
particles are assumed to be Iron nuclei, the generation index is $\gamma_g=2.3$, 
the sources are assumed to be homogeneously distributed in space, and the maximum 
energy of acceleration is $E_{\max}= 2\times 10^{20} Z_0 $~eV. The results are 
shown in the right panel of Fig.~\ref{fig17}, where the computation by Allard 
et al \cite{Allard08} includes EBL and CMB, while our calculations - only CMB. 
All three 
criterions for a good agreement are present in these two spectra: the merging 
energy is in the right place, both spectra coincide above the merging energy 
and secondary-protons spectrum agree at all energies.  

It is interesting to comment on the agreement between all-particles 
spectra in both left and right panels of Fig.~\ref{fig17}. The
agreement above the merging energy is trivial and is explained by the
dominance of CMB photons in the formation of the highest energy part of the spectrum.
Agreement of spectra at low energies, where EBL photons dominate, is 
caused by an approximate conservation of the number of nuclei: in the process
$A+\gamma \to (A-1)+N$
the nucleon has energy A times lower than nucleus, thus it has
an energy out of the range we considered here. The suppression of the nuclei flux 
appears at the intermediate energies due to the full destruction of nuclei
by EBL photons.

We can conclude stating that the analysis performed shows a very good 
agreement of 
the fluxes computed with the three methods analyzed taking into
account only the CMB,  when the flux of target photons is rigidly fixed.

To conclude this section we will briefly discuss the comparison of
our theoretical results with the Auger observations. 
The attempt to compare the spectra obtained in this paper through a theoretical 
toy-model, based on the simplified assumptions of a pure Iron injection and only CMB photons as     
target, with the observed spectrum may be motivated as follows. 

\begin{figure}[!ht]
\begin{center}
\includegraphics[width=0.49\textwidth,angle=0]{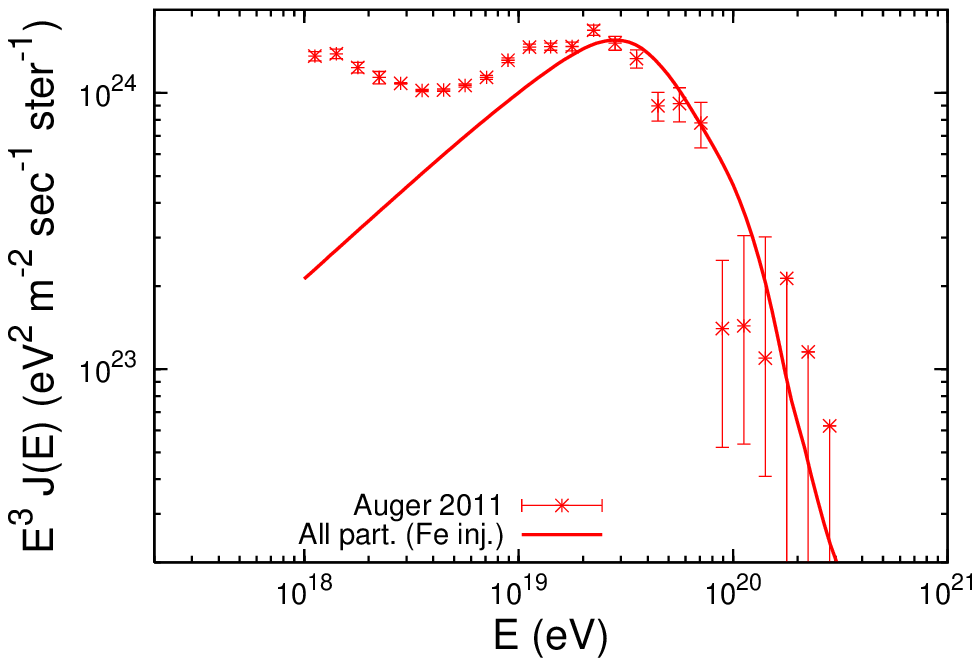}
\caption{Comparison of the Auger 2011 energy spectrum \cite{Auger2011} with the all-particles 
spectrum computed in the present paper with $\gamma_g=2.3$ 
$E_{max}=Z_0 2\times 10^{20}$ eV and a pure Iron injection at the sources.}
\label{fig18} 
\end{center}
\end{figure}

In the rigidity-acceleration models with $E_{\max} \propto Z$ the
end of the spectrum is dominated by the heaviest nuclei, like e.g. in the 
disappointing model \cite{disappointing}. Moreover, CMB photons dominate 
the photo-disintegration process only above the merging  energy, 
$E_m \approx 1.7\times 10^{20}$~eV for Iron. However, due to the approximate 
conservation of the number of nuclei, the EBL suppression of the all-particles  
spectrum is not strong (see Fig.~\ref{fig17}). Therefore, it can be expected     
that our simplified model gives a good description of he Auger spectrum at
the highest energies. The comparison is given in Fig.~\ref{fig18} with the
computed all-particles spectrum shown by the solid line. 

To conclude, we want to emphasize again that Fig. \ref{fig18} has to be
considered only as an illustration, the aim of the present paper is to introduce, 
on theoretical grounds, new techniques to calculate analytically the UHE nuclei fluxes. 
The application of the technique developed here in detailed calculations of the spectra 
with a coherent comparison with experimental data will be presented in a forthcoming paper.

Finally, we do not discuss here the comparison with HiRes and Telescope Array
observations because they show a strongly proton-dominated mass  
composition \cite{HiRes-mass2010,TA}, therefore cannot be described 
in the framework of a pure iron injection used in this paper. 

\section{Discussion and Conclusions}
\label{sec:conclusions}

In this paper we have studied comparatively three analytic
methods to determine the diffuse spectra of ultra-high energy nuclei 
propagating through background radiations: {\em(i)} trajectory method, 
{\em (ii)} kinetic-equation method combined with trajectory
calculations (the combined method),  
and {\em(iii)} coupled kinetic equations (CKE).

We summarize first our general approach. 

We have calculated the diffuse spectra of primary and secondary nuclei, and 
secondary protons in the case of homogeneously distributed sources. 
According to the propagation theorem \cite{AB04} in this case the spectra do not
depend on the specific way of propagation, and thus one may use
rectilinear propagation, or not thinking about space propagation at all,
imaging all secondary nuclei and protons filling the space
homogeneously together with primary nuclei.

Evolution in time is the basic feature of our problem. 
A primary nucleus $A_0$ is born due to acceleration and, interacting
with the background radiation, give rise to the secondary nuclei and protons.

With the help of two coupled differential equations for the propagation of a 
primary nucleus in the backward  
time, i.e. increasing redshift, we found the evolution trajectories 
$A(z)={\mathcal A}(A,\Gamma,z_0,z)$
and $\Gamma(z)={\mathcal G}(A,\Gamma,z_0,z)$, where the first three arguments
fix the initial conditions. In the evolution equations the atomic number $A$ 
changes continuously with $z$, but we use the continuous trajectory 
$A(z)$ only to determine the integer $A$ positions on the
trajectory: $A$, $A+1$ etc. At these points we assume an
instantaneous photo-disintegration, e.g. 
$(A+1)+\gamma_{\rm CMB} \rightarrow A+N$.  The Lorentz factor between 
the points of photo-disintegration is calculated assuming
$A=const$. The trajectories are basically needed to compute 
the generation parameters $z_g$ and $\Gamma_g$, at which $A(z)$ 
reaches $A_0$. Apropos, the
formal solution of the coupled trajectory equations (\ref{eq:evolve})
with continuous $A$ results in the same $\Gamma_g$ and $z_g$ as in our 
basic method.

The important feature of the calculated $A(z)$ trajectories is their 
{\em explosive behavior} for all initial energies (as an example see 
the left panel 
of Fig. \ref{fig:evolve} ). At small $\Gamma < \Gamma_c$  
a nucleus propagates with $A=const$ until 
large red-shifts $z \sim 1$ and then 'explodes' to $A_0$ on a scale of 
photo-disintegration lifetime $\tau_a \sim 10^{5}-10^{6} $~yr, see 
Eq. (\ref{eq:tau(z)}). It results in $z_A \approx z_g$ and 
$\Gamma_A \approx \Gamma_g$, where index $A$ marks the values at
the threshold of an explosion. Therefore, the generation rate for 
$A$ nuclei approximately equals to that of primary nuclei $A_0$ :  
$Q_A(\Gamma_A, z_A) \approx Q_{A_0}(\Gamma_g, z_g)$ with 
$z_g \approx z_A$ and $\Gamma_g \approx \Gamma_A$. However, the main
attention in the paper is given to the exact calculation of 
$Q_A(\Gamma_A, z_A)$ from the number of particles conservation.

We describe first the combined method.

The space density  of each species $a=A,~p,~A_0$ and their energy 
spectra are calculated  using the kinetic equation
\beq
\frac{\partial n_a(\Gamma_a,t)}{\partial t} - 
\frac{\partial }{\partial \Gamma_a}\left [ b_a(\Gamma_a,t) n_a(\Gamma_a,t) 
\right ] + \frac{n_a(\Gamma_a,t)}{\tau_a(\Gamma_a,t)} = Q_a(\Gamma_a,t),
\label{eq:kin1}
\eeq
where $b_a=-d\Gamma_a/dt=(\beta^a_{\rm pair}+ \beta^a_{\rm ad}) \Gamma_a$ 
is the rate of the Lorentz-factor loss, $\tau_a(\Gamma_a,z)$ is the
photo-disintegration lifetime of particle $a$ (for $a=p$, $\tau_a=\infty$) 
and $Q_a(\Gamma_a,z)$ is the generation rate of particles $a$.

For secondary nuclei $A$ and protons, produced in the photo-disintegration
processes, the generation rate $Q_a(\Gamma_a,z)$
is found from the conservation of the number of particles along a trajectory,  
$dN_a=dN_g$, where $dN_a$ is the number of produced particles $a$ and 
$dN_g$ is the number of generated primaries $A_0$ at acceleration. 
The relation between the  generation rate 
of primaries $A_0$ and of secondary $A$ and $p$ is given 
by Eq.~(\ref{eq:QA}), where $z_g$ and $\Gamma_g$ are calculated 
using the evolution trajectories. The solutions of the kinetic equations 
(\ref{eq:kin1}) are found analytically. 

An important physical quantity in our calculation is the critical 
Lorentz factor $\Gamma_c^A$ at epoch $z=0$.  
It is determined by the equality of the rates for changing 
of $\Gamma$ and $A$:
$\tau_{\rm dis}^{-1}(\Gamma_c^A)= \tau^{-1}_{\rm pair}(\Gamma_c^A) +H_0$,
where $H_0$ is the Hubble constant, describing here the adiabatic
energy loss, and indices 'dis' and 'pair' are 
related to photo-disintegration
and pair-production lifetimes, respectively.
This relation provides a stability condition for the $A$ nucleus at $z=0$: 
$\Gamma < \Gamma_c^A$.  The critical Lorentz factor 
$\Gamma_c$ is a basic energy scale, which gives a key for
understanding all processes considered here and which also explains  
some features in the calculated spectra. 

Therefore, the combined method includes three elements of calculations:
the kinetic equation, which gives the density of particles $n_a (\Gamma)$,
the generation rate $Q_a(\Gamma_a, z_a)$ and the limits of integration, 
especially $z_{\rm min}^a$. The first two elements are reliable
components of calculations, with $Q_a$ reliably evaluated 
with the help of the number of particles conservation; the third element 
$z_{\rm min}$ is less reliable since it involves additional
consideration and assumptions. 

The {\it trajectory method} of $n_a(\Gamma)$ calculation is less reliable 
because of the great uncertainties in the limits of integration $z_{\rm min}$
and $z_{\rm max}$ in Eq. (\ref{eq:n-traj}).

The {\it coupled kinetic equations} (CKE) method is based on the set 
of kinetic equations (\ref{eq:coupled}), where the first one describes the 
primary nucleus $A_0$, the second $A_0-1$ {\em etc} down to the $A$ of 
interest. The solution of each preceding equation $n_{A'}(\Gamma,z)$ gives 
the generation rate for a successive nucleus $A'$ as 
$n_{A'+1}(\Gamma,z)/\tau_{A'+1}(\Gamma,z)$, and the long lifetime 
$\tau_{A'+1}(\Gamma,z)$ at small $\Gamma$ and $z$ automatically 
provides $z_{\rm min}$ in the solution of a kinetic equation, while the lifetime 
$\tau_{A'}$ provides, through the term $e^{-\eta}$, the upper limit.
The $A(z)$ trajectories are formally not involved in this method, but 
actually they are. The term $n_{A_0}/\tau_{A_0}$ describes the disappearance 
of nuclei $A_0$ in the first equation of the Eqs. (\ref{eq:coupled}), 
but the same term appears in the second equation as generation of 
$(A_0-1)$ nuclei {\em etc}. Equation (\ref{eq:nA-analytic}), which includes 
the  product of terms with all $A'$ from $A$ to $A_0$, demonstrates it more 
clearly. Indeed, this equation implies the evolution of $\Gamma(t)$ with 
$A'=const$ regulated by $\exp(-t_{\rm prop}/\tau_{A'})$. Then $A'$ disappears, 
giving rise to the production term in the successive kinetic equation for $(A'-1)$.   
However, there is an essential difference with the combined method. 
The product in  Eq.(\ref{eq:nA-analytic}) is time-ordered which, 
according to the theory of probabilities, means fluctuations taken
into account.  

As we indicated above the weakness of the combined method consists 
in the procedure of defining $z_{\rm min}$, while in the coupled kinetic 
equations this limit is a natural feature of the kinetic equation itself. 
The lower limit in the combined method is introduced by the stability
condition given by Eq.(\ref{eq:z_c}), which is quite natural. However, 
in fact one can use the other definitions of stability. As a plausible 
example we consider an alternative possibility.

We used above in the combined method the cutoff in integration over $z$
imposing the condition $Q_A(\Gamma,t)=0$ at $t \geq t_{\rm max}$ in
terms of the cosmological time $t$ (see the solution in terms of $t$ 
given by Eq. \ref{eq:nA0-time}). This condition implies that at 
$t > t_{\rm max}$~ $(A+1)$-nucleus is stable. Let us introduce now this
condition through the probability of $A+1$-decay: 
\beq
Q_A(\Gamma,t) \rightarrow \left [1-\exp\left (-\int_{t_g}^t 
\frac{dt'}{\tau_{A+1}(\Gamma',t')}\right )\right ] Q_A(\Gamma,t) . 
\label{eq:Q-modify}
\eeq
One can see that the factor introduced is the decay probability of the 
$A+1$ nucleus. It can be easily seen in the case $\tau_{A+1}=const$, 
when this factor is $1-\exp(-t_{\rm prop}/\tau_{A+1})$, where 
$t_{\rm prop}=t-t_g$ is the propagation time. The decay factor in 
Eq. (\ref{eq:Q-modify}) provides the soft cutoff of the generation 
function. However, this cutoff also differs from the one obtained in 
Eq. (\ref{eq:low-limit-coupled}). 

We compare now the numerical results in some details. 

The two methods give numerically the same spectra for all nuclei at
high energies and for primary nuclei at all energies. 
In particular both methods give high-energy fluxes of the secondary 
nuclei described by identical equations (\ref{eq:nA-asymp}) and  
(\ref{eq:n_Aasympt}) with a recovery of the spectra at the highest energies 
due to an increasing $\tau_A(\Gamma)$. 

The peaks in the spectra of the secondary nuclei located at $\Gamma_c$
in the combined method (see Fig. \ref{fig:fluxA}, right panel) coincide 
with the positions of peaks in the CKE method 
(see Fig. \ref{fig:secAp-coupled} , left panel). This is a surprising result,
because the critical Lorentz factor is not introduced in the CKE method. 
The fluxes above $\Gamma_c$ are the same, but they are
much different below, which is mainly a result of the differences 
in $z_{\rm min}$.
The peculiar behaviour of $^4He$ at low energies in Fig. \ref{fig:fluxA}
(left panel), being confirmed by explosive trajectory calculation, is
not seen in Fig. \ref{fig:secAp-coupled}.

The fluxes of the secondary nuclei obey as a rule A-hierarchy: the
heavier $A$, the larger the flux
(see Figs.~\ref{fig:fluxA} and \ref{fig:secAp-coupled}).

The main features of the secondary-proton spectra are the same in both 
methods. The $A$-associating proton spectra are the same for all $A$, and the 
total spectra are universal, i.e. do not depend on $A_0$. In the combined
model this is a direct consequence of the explosive character of the 
trajectories. At low energies ($E < 1\times 10^{19}$~eV) in CKE the 
$A$-associating proton spectra are different for various $A$ (see right panel
in  Fig. \ref{fig:secAp-coupled}). This is related to a breaking of 
accuracy of the explosive trajectory approximation at low energies.

The shape of the total proton spectrum is the same in both calculations.  
It has two spectral breaks.
The high energy steepening is the usual GZK cutoff at energy 
$E \sim 5\times 10^{19}$~eV. The low-energy steepening coincides 
with $\Gamma_c$. Below $\Gamma_c$ the adiabatic energy losses dominate 
and it explains the flat spectrum $n_p(E_p)$ there, {\em i.e.} the steep one in
terms of  $E^3 n_p(E)$ spectrum as  plotted in all figures here. 

In the second paper of this series, the EBL background 
will be also included. In this case the high energy part of the nuclei
spectra is determined by CMB, as considered in the present paper, while 
the low-energy part is affected mostly by Infrared, Visible and Ultra Violet 
radiations, which compose the EBL.  
 
\section*{Acknowledgements}

We thank Pasquale Blasi, Yurii Eroshenko and Askhat Gazizov for valuable
discussions. This work is partially funded by the contract
ASI-INAF I/088/06/0 for theoretical studies in High Energy
Astrophysics and by the Gran Sasso Center for Astroparticle Physics (CFA)
funded by European Union and Regione Abruzzo under the contract P.O. FSE
Abruzzo 2007-2013, Ob. CRO. The work of SG is additionally  funded by 
the grant of President of RF SS-3517.2010.2, VB and SG - by FASI grant 
under state contract  02.740.11.5092.

\begin{appendix}

\section{Generation rates of primary nuclei, secondary nuclei 
and secondary protons.}
\label{app:generation}
We consider an expanding universe homogeneously filled by the sources
of accelerated primary UHE nuclei $A_0$ with a generation rate per unit
of comoving volume $Q_{A_0}(\Gamma,z)$ given by 
\begin{equation}
Q_{A_0}(\Gamma,z)=\frac{(\gamma_g-2)}{m_N A_0}
{\mathcal L}_0 \Gamma^{-\gamma_g},
\label{appeq:inj}
\end{equation}
where $\gamma_g>2$ is the generation index, $m_N$ is the nucleon mass, and 
${\mathcal L}_0$ is the source {\em emissivity}, i.e. the energy generated 
per unit of comoving volume and per unit time at $z=0$. 
In Eq. (\ref{appeq:inj}) 
$\Gamma_{\rm min} \sim 1$ is assumed. In all calculations in this paper
we assume also the existence of a maximum energy of acceleration 
$E^{\rm acc}_{\rm max}=Z_0\times 10^{21}$~eV 
(or $\Gamma_{\rm max}^{\rm acc}= (Z_0/A_0)\times 10^{12}$) with 
the condition $Q_{A_0}(\Gamma_g)=0$~ 
at $\Gamma_g \geq \Gamma_{\rm max}^{\rm acc}$.    

Propagating in the space, a nucleus $A_0$ experiences evolution,
producing the secondary nuclei with different $A$ and we calculate
now their generation rate $Q_A(\Gamma,z)$ per unit of comoving volume.  

For calculation of the generation rates the evolution-trajectory formalism, 
developed in section \ref{sec:trajectories}, is used. The trajectories 
$A(z)={\mathcal A}(\Gamma,A,z_0,z)$ and 
$\Gamma(z)=\mathcal{G}(\Gamma,A,z_0,z)$ in the backward time 
allows us to calculate the generation parameters $z_g$ and $\Gamma_g$, 
when $A(z)$ reaches $A_0$.  

$A(z)$ can be calculated as  continuous quantity,
but we use the method of $A$-jump trajectories, as it is described in 
section \ref{sec:trajectories}. Namely, the trajectories are
calculated assuming in Eqs. (\ref{eq:evolve}) for evolution 
of $\Gamma$ $A=const$, until $A$ reaches $A+1$ (or $A+2$).
Transition $A \to (A+1)$ is assumed to occur instantaneously.     
Accordingly, we use in this case the jump behaviour of 
$\beta^A_{\rm dis}$. In fact both methods, continuous $A$ and jump $A$ 
approximations, give the same results. 

Since a recoil momentum in the processes of transition, e.g. 
$(A+1) \to A +N$, is negligibly small, one has approximate equality of 
Lorentz-factors of all three particles 
\beq
\Gamma_{A+1} \approx \Gamma_A \approx \Gamma_N .
\label{eq:equality}
\eeq

The generation rate of secondary nuclei $A$ can be found from
conservation of number of particles. Consider an {\em allowed} 
trajectory which connects $(\Gamma_A,z_A)$ state at generation $A$
with $(\Gamma_g, z_g)$ at generation of primary nucleus $A_0$. 
During time $dt_g$ at the epoch $z_g$ we have $dN_{A_0}$ primary
nuclei generated with ($\Gamma_g,\; \Gamma_g+d\Gamma_g$):
\beq
dN_{A_0} = Q_{A_0}(\Gamma_g, z_g) d\Gamma_g dt_g
\label{dN_0}.
\eeq

All these nuclei appear as $dN_A$~ $A$-nuclei with 
$(\Gamma_A,\; \Gamma_A+d\Gamma_A)$ during time $dt_A$:  
$dN_A=dN_{A_0}$. Thus, we have
\begin{equation}
Q_A(\Gamma_A,z) d\Gamma_A dt = Q_{A_0}(\Gamma_g,z_g) d\Gamma_g dt_g 
\label{appeq:QA1}
\end{equation}
where the states ($\Gamma_A,z$) and ($\Gamma_g,z_g$) are connected by the 
allowed trajectory. From Eq. (\ref{appeq:QA1}) and the time dilatation 
relation,
\beq
dt_g/dt=(1+z)/(1+z_g),
\label{eq:dilat}
\eeq
we obtain the generation rate of nuclei $A$ as:
\begin{equation}
Q_A(\Gamma_A,z)=Q_{A_0}(\Gamma_g,z_g) \frac{1+z}{1+z_g} 
\frac{d\Gamma_g}{d\Gamma_A}
\label{eq:QA}
\end{equation}
with $d\Gamma_g/d\Gamma_A$ given in appendix \ref{app:dgamma}. 

As was emphasized above, the chain of equalities (\ref{appeq:QA1}) is valid
along the allowed trajectory, which breaks at some low $A$ and $z$, where 
$\tau_A$ becomes larger than $\tau_{\Gamma}$. 
If at $z=z_c$ $\gamma + (A+1) \rightarrow A+N$ is forbidden  
for all $\Gamma <\Gamma_c(z_c)$, then  $Q_A(\Gamma_A,z_A)=0$ at 
$z_A \leq z_c$ and $\Gamma_A \leq \Gamma_c(z_A)$ (see also 
subsection \ref{sec:stability} ). Starting from the critical redshift 
$z_c$ and below it $A$-nuclei are not produced and $(A+1)$-nuclei 
propagate remaining undestroyed. 

Note, that for derivation we do not need stationary solution, and
due to condition of homogeneity we do not need to discuss the  spatial 
displacement of a particle. 

Coming finally to the generation rate of nucleons produced in $(A+1) \to A+N$ 
transition, $Q_p^{A+1}(\Gamma_p,z)$,  one may
notice that this generation rate is the same that one for $A$ nuclei,
since the Lorentz factor of both particles are equal and they are produce
simultaneously. 
\beq
Q^{A+1}_p(\Gamma_A,z_A)=Q_{A_0}(\Gamma_g,z_g) \frac{1+z_A}{1+z_g} 
\frac{d\Gamma_g}{d\Gamma_A} ,
\label{appeq:QpA}
\eeq
where $z_A$ and $z_g$ is the redshifts of $A$ and $A_0$  production,
and $\Gamma_A$ is the Lorentz factor of nucleus $A$ and nucleon. 
We do not distinguish nucleon and proton because UHE neutron decays  
fast, on the time scale of this problem, to proton. 

\section{Ratio of energy intervals at epochs of production and observation }
\label{app:dgamma}
In this appendix we derive the ratio between Lorentz-factor intervals 
$d\Gamma_g/d\Gamma$ for a nucleus with variable atomic mass $A(z)$ 
propagating along the evolution 
trajectory $\Gamma_g(z)={\mathcal G}(A,\Gamma,z_0,z)$, where the first three 
indices describe the initial conditions and $z$ is the running 
redshift. The meaning of the 
generation index $g$ here is more general than the epoch of 
$A(z)$ evolution to $A_0$, though it includes this case, too. 
$\Gamma_g(z)$ means here the Lorentz factor that a nucleus has at 
the running epoch of evolution $z$ including $z_g$. 

We use here the Lorentz-factor loss for a nucleus $A$ due to pair production 
as $b_{\rm pair}^A(\Gamma,z)=-d\Gamma/dt$, and express it through 
$b_0^p(\Gamma)=-d\Gamma/dt$ for proton at $z=0$. At epoch $z$ we have 
\beq
b_{\rm pair}^A(\Gamma,z)=\frac{Z^2(z)}{A(z)} (1+z)^2 b_0^p [(1+z)\Gamma],
\label{eq:b-pair}
\eeq
We introduce $\bar{k}=\langle Z^2/A^2\rangle$, assuming it be
constant along the evolution trajectory $A(z)$, and thus using 
$Z^2/A=\bar{k} A(z)$. 

The Lorentz factor of nucleus $A$ along the trajectory can be
presented in a general form as 
\beq
\Gamma_g(t_1,t_2)=\Gamma+ \int_{t_1}^{t_2}dt'
\left [ \left (\frac{d\Gamma}{dt'}\right)_{\rm ad} +
\left (\frac{d\Gamma}{dt'} \right )_{\rm pair} \right]
\label{eq:Gamma-t}
\eeq
Changing the variable $t$ to $z$ and using $dt=-dz/[(1+z)H(z)]$,
we obtain 
\beq
\Gamma_g(z_0,z) = \Gamma+ \int_{z_0}^z \frac{dz'}{1+z'} \Gamma_g(z') +
\bar{k}\int_{z_0}^z dz'\frac{1+z'}{H(z')}\; A(z')\; b_0^p [(1+z')\Gamma_g(z')],
\label{eq:Gamma-z}
\eeq
Differentiating Eq. (\ref{eq:Gamma-z}) in respect to $\Gamma$ and
using 
\beq
\frac{db_0^p(\Gamma')}{d\Gamma}=\frac{\partial b_0^p(\Gamma')}
{\partial \Gamma'}\frac{d\Gamma'}{d\Gamma}, 
\nonumber
\eeq
one finds for the ratio of Lorentz-factor intervals 
$y(z) \equiv d\Gamma_g(z)/d\Gamma$: 
\beq
y(z)=1+\int_{z_0}^z\frac{dz'}{1+z'} y(z') + \bar{k}\int_{z_0}^z dz'
\frac{(1+z')^2 A(z')}{H(z')}y(z')\left (\frac{db_0^p(\Gamma')}{d\Gamma'}
\right )_{\Gamma'=(1+z')\Gamma_g(z')}. 
\label{eq:y(z)} 
\eeq 
Differentiating Eq. (\ref{eq:y(z)}) in respect to $z$ we obtain 
a differential equation for $y(z)$
\beq
\frac{1}{y(z)}\frac{dy(z)}{dz}=\frac{1}{1+z}+\frac{\bar{k}}{H_0}
\frac{(1+z)^2 A(z)}{\sqrt{\Omega_m(1+z)^3+\Omega_{\Lambda}}}
\left (\frac{db_0^p(\Gamma')}{d\Gamma'}\right )_{\Gamma'=(1+z)\Gamma_g(z)}.
\label{eq-z,diff} 
\eeq 
The solution of Eq. (\ref{eq-z,diff}) can be easily found as 
\beq
y(z)\equiv \frac{d\Gamma_g(z)}{d\Gamma (z_0)}=\frac{(1+z)}{(1+z_0)}
\exp\left[\frac{\bar{k}}{H_0}
\int_{z_0}^z
dz' \frac{(1+z')^2 A(z')}{\sqrt{\Omega_m(1+z')^3+\Omega_{\Lambda}}}
\left(\frac{db_0^p(\Gamma')}{d\Gamma'}\right)_{\Gamma'=(1+z')\Gamma_g(z')}
\right].
\label{eq:dE_g/dE} 
\eeq 

Equation (\ref{eq:dE_g/dE}) gives the ratio $d\Gamma_g/d\Gamma$, 
where $d\Gamma$ is the Lorentz-factor interval at the initial state 
$z_0$ (with most
important case $z_0=0$), when a nucleus has fixed $A$ and Lorentz
factor $\Gamma$, and $d\Gamma_g$ is the interval on the evolution
trajectory at redshift $z$ when the atomic number is $A(z)$ and the Lorentz 
factor is $\Gamma_g(z)$.

In the applications we often need $d\Gamma_g/d\Gamma$ ratio for a fixed
$A$. The most important case is the primary nuclei $A_0$. The ratio 
for fixed $A$ follows trivially from Eq. (\ref{eq:dE_g/dE}) 
substituting there $\bar{k}A(z')$ by constant value $Z^2/A$. For the
proton case this value is 1. 

\section{Solution to the kinetic equation}
\label{app:solution}
In this appendix we derive the solution of the kinetic equation for 
secondary nuclei from which the solutions for primary nuclei and
secondary protons can be easily obtained. 

We consider the secondary nuclei $A$ being produced homogeneously in the space
with the rate $Q_A(\Gamma,t)$ and then propagate as a nucleus species
with the fixed (unchanged) $A$ until it is photo-disintegrated.
The kinetic equation reads 
\begin{equation}
\frac{\partial n_A(\Gamma_A,t)}{\partial t} - 
\frac{\partial}{\partial \Gamma} [b_A(\Gamma,t) n_A(\Gamma,t)] +
\frac{n_A(\Gamma,t)}{\tau_A(\Gamma,t)} = Q_A(\Gamma,t) ,
\label{eq:app-kin1}
\end{equation}
where $b_A(\Gamma,t)$ is the rate of Lorentz factor loss given by 
\beq
b_A(\Gamma,t)=-\frac{d\Gamma}{dt}= \Gamma H(z)+ \frac{Z^2}{A}
b_{\rm pair}^p(\Gamma,z),
\label{eq:app-b_A}
\eeq
with $H(z)$ and $b_{\rm pair}^p$ being the Hubble parameter at redshift
$z$ and pair-production loss for proton, respectively. The time of 
photo-disintegration is given by 
$\tau_A^{-1}=A\beta_{\rm dis}^A$ or by $\tau_A^{-1}=dA/dt$, 
see Eq. (\ref{eq:losdis}).

The characteristic equation for the kinetic equation (\ref{eq:app-kin1})
reads
\beq
d\Gamma/dt=-b_A(\Gamma,t).
\label{eq:char-eq}
\eeq
With $\Gamma(t)$  taken on the characteristic in Eq. (\ref{eq:app-kin1}), 
the term 
$b_A(\Gamma,t) \partial n_A(\Gamma,t)/\partial \Gamma $
disappears and the kinetic equation (\ref{eq:app-kin1}) takes the form 
\begin{equation}
\frac{\partial n_A(\Gamma_A,t)}{\partial t} + n_A(\Gamma,t)
\left [-\frac{\partial b_{\rm pair}^A(\Gamma,t)}{\partial \Gamma}-
\frac{\partial b_{\rm ad}(\Gamma,t)}{\partial \Gamma} + \tau_A^{-1}(\Gamma,t) 
\right ] = Q_A(\Gamma_A,t) 
\label{eq:app-kin2}
\end{equation}

Using the notation 
\begin{equation}
\begin{array}{lll}
P_1(z) &\equiv& \partial b_{\rm ad}(z) /\partial \Gamma = H(z)\\
\nonumber
P_2(z) &\equiv& \partial b_{\rm pair}^A (\Gamma,z) /\partial \Gamma
= \frac{Z^2}{A}(1+z)^3 \left (\partial b_0^p(\Gamma')/
\partial \Gamma' \right )_{\Gamma'=(1+z)\Gamma} \;\; ,
\label{eq:app-notation}
\end{array}
\end{equation}
and taking into account that with $\Gamma$ on the characteristic, time $t$ 
becomes the only variable, one obtains the solution of 
Eq. (\ref{eq:app-kin2}) as 
\begin{equation}
n_A(t)=\int_{t_g}^{t} dt' Q_A(t') \exp\left [ -\int_{t'}^{t} dt'' 
\left (-P_1(t'') -P_2(t'') + \tau_A^{-1}(t'') \right ) \right ] .
\label{eq:app-solut1}
\end{equation}
Changing the integration variable $t$ to $z$ and using 
\beq
dt = - \frac{dz}{(1+z)H(z)}=-\frac{1}{H_0} \frac{dz}
{(1+z)\sqrt{\Omega_m(1+z)^3+\Omega_\Lambda}} ,
\label{eq:app-dt}
\eeq
the solution can be written as 
$$
n_A(\Gamma,z=0)=\int_0^{z_{\rm max}}dz'\frac{Q_A[\Gamma'(\Gamma,z')]}
{(1+z')H(z')} \times  
$$
\beq
\exp \left [\int_0^{z'} dz''\frac{P_1(z'')}{(1+z'')H(z'')}\right ] 
\exp \left [\int_0^{z'} dz''\frac{P_2(z'')}{(1+z'')H(z'')}\right ]
\exp \left [- \int_{t'}^{t_0} \frac{dt''}{\tau_A(\Gamma,t'')}\right ]
\label{eq:app-solut2}
\eeq
We keep the last integration over $t''$ to make clear the physical 
meaning of this integral as a suppression factor for the survival time of 
nucleus $A$. The upper limit $t_0$ is the age of the Universe.

Putting $P_1(z'')$ and $P_2(z'')$ from Eq. (\ref{eq:app-notation}) 
into Eq. (\ref{eq:app-solut2}), we find that the product of the first
two exponents gives the ratio of energy intervals calculated 
in appendix \ref{app:dgamma}: 
\beq 
(1+z')\exp \left [\frac{Z^2}{A}\frac{1}{H_0}\int_0^{z'} dz''
\frac{(1+z'')^2}{\sqrt{\Omega_m(1+z'')^3+\Omega_{\Lambda}}}
\left (\frac{\partial b_0^p(\Gamma'')}{\partial \Gamma''} 
\right)_{\Gamma''=(1+z'')\Gamma_g(z'')} \right ] = 
\frac{d\Gamma_g^A(z')}{d\Gamma^A}.
\label{eq:app-dgammaA}
\eeq
Finally, we have 
\beq
n_A(\Gamma,z=0)=\int_0^{z_{\rm max}} dz' \frac{Q_A[\Gamma'(\Gamma,z')]}
{(1+z')H(z')} \frac{d\Gamma'}{d\Gamma}\; e^{-\eta(\Gamma',z')},
\label{eq:app-final}
\eeq
where 
\beq
\eta(\Gamma',z')=\int_{t'}^{t_0}\frac{dt''}{\tau_A(\Gamma'',t'')}=
\int_0^{z'} dz'' \frac{1}{(1+z'')H(z'')} \frac{1}{\tau_A(\Gamma'',z'')}.
\label{eq:app-eta}
\eeq
In fact, in Eq. (\ref{eq:app-final}) one can put $z_{\rm max} \to \infty$
as upper limit, since it is regulated by the factor 
$\exp[-\eta(\Gamma',z')]$. 

The equations (\ref{eq:app-final}) and (\ref{eq:app-eta}) are valid 
for primary nuclei ($A_0, Z_0$) and secondary protons; in the latter case one
should put  $\tau_A \to \infty$ and $\eta \to 0$. 

\section{Comparison of the secondary nuclei and secondary protons fluxes}
\label{app:comparison}
We perform here the analytic comparison of the fluxes of secondary 
protons and secondary nuclei at the same Lorentz factor $\Gamma$ in 
the most general form valid for kinetic equations and demonstrate 
that fluxes of the secondary protons are always higher.

The secondary nucleus $A$ and $A$-associating secondary nucleon $N$,  
born in the same decay $(A+1) \rightarrow A+N$, are twin brothers: 
they are born at the same redshift $z$ with the same Lorentz factor
$\Gamma$ and with the same rate of production 
$Q_A(\Gamma,z)=Q_p^A(\Gamma,z)$ (we do not distinguish between 
neutron and proton because of the fast decay of the former). But one 
of the brothers, $A$-nucleus, lives short time $\tau_A$ and its flux 
is lower. We estimate here this effect analytically.

The flux of the $A$-associating secondary protons is discussed in 
sections \ref{sec:prot} and \ref{sec:coupled-kinetic} and in the
most general form is given as 
\beq
 n_p^A(\Gamma,0)=\int_0^{z_p^{\rm max}}\frac{dz'}{(1+z')H(z')}
\frac{n_{A+1}(\Gamma',z')}{\tau_{A+1}(\Gamma',z')}
\left (\frac{d\Gamma'}{d\Gamma}\right )_p ,
\label{eq:p-gen}
\eeq
where $z_0=0$ is the initial condition, the generation term 
$Q_p^A(\Gamma,z)=Q_A(\Gamma,z)$ is written 
in the most general form, and  $\Gamma'(z')=G_p(\Gamma,0,z')$ is the proton 
trajectory, on which $d\Gamma'/d\Gamma$ ratio is given by  
\beq
\left (\frac{d\Gamma'(z')}{d\Gamma}\right )_p=(1+z')\exp \left [\int_0^{z'}
\frac{dz''(1+z'')^2}{H(z'')}
\left (\frac{db_0^p(\tilde{\Gamma})}{d\tilde{\Gamma}}\right )_
{\tilde{\Gamma}=(1+z'')\Gamma''}\right ]   
\label{eq:dGamma-ratio-p}
\eeq
Flux of the $A$-nuclei is calculated as  
\beq
n_A(\Gamma,0)=\int_0^{z_A^{\rm max}}\frac{dz'}{(1+z')H(z')}
\frac{n_{A+1}(\Gamma',z')}{\tau_{A+1}(\Gamma',z')}
\left (\frac{d\Gamma'}{d\Gamma}\right )_A e^{-\eta(\Gamma'.z')} , 
\label{eq:A-gen}
\eeq
with 
\beq
\left (\frac{d\Gamma'(z')}{d\Gamma}\right)_A=(1+z')\exp \left[\frac{A}{4}
\int_0^{z'}\frac{dz''(1+z'')^2}{H(z'')}
\left (\frac{db_0^p(\tilde{\Gamma})}{d\tilde{\Gamma}}\right )_
{\tilde{\Gamma}=(1+z'')\Gamma''}\right ] ,   
\label{eq:dGamma-ratio-A}
\eeq
where we assumed $<Z/A>=1/2$,

One may observe similarity of expressions for secondary protons and
nuclei. It is easy to deduce inequality $n_p(\Gamma,0) > n_A(\Gamma,0)$ 
caused by two main reasons. The first one is given by $z_p^{\rm max}$ being 
considerably larger than $z_A^{\rm max}$, since the latter is limited by the 
decay lifetime $\tau_A$ (see factor $e^{-\eta}$ in Eq. \ref{eq:A-gen}). 
The second reason is connected with larger energy loss of nuclei to 
pair production $\beta_A(\Gamma)=(A/4)\beta_p(\Gamma)$. It results in 
$\Gamma'_A > \Gamma'_p$ at the same $z'$ in the integrand of 
Eq. (\ref{eq:A-gen}), and thus to suppression of this integral. 
The larger $d\Gamma'/d\Gamma$ for nuclei compensates partly the
discussed flux inequality. Since this inequality is valid for every
$A$, the total flux of the secondary protons also exceeds that of
nuclei: 
\beq
\Sigma_A n_p^A(\Gamma,0) > \Sigma_A n_A(\Gamma,0) .
\label{eq:secondary-equality}
\eeq
At equal energies the nuclei flux should be shifted by factor $A$ to 
higher energies and in practical cases it exceeds the secondary proton 
flux.  

\end{appendix}

\end{document}